\documentclass[a4paper,12pt]{article}
\usepackage{amsmath, amsthm, amsfonts, amssymb}
\usepackage{graphicx}
\usepackage[mathscr]{eucal}
\theoremstyle{plain}
\newtheorem{theorem}{Theorem}[section]

\newtheorem{lemma}{Lemma}[section]

\theoremstyle{remark}
\newtheorem{remark}{Remark}[section]

\numberwithin{equation}{section}

\def\e{\varepsilon}

\def\l{\lambda}

\def\k{\varkappa}
\def\E{\mbox{\rm e}}
\def\a{\alpha}
\def\b{\beta}
\def\t{\widetilde}
\def\si{\sigma}

\def\d{\delta}

\def\vs{\varsigma}

\def\vp{\varphi}

\def\h{\widehat}
\def\Odr{\mathcal{O}}
\def\H{W_2}
\def\Hloc{W_{2,loc}}

\def\di{\,\mathrm{d}}
\def\Th{\Theta}
\def\I{\mathrm{I}}
\def\iu{\mathrm{i}}

\DeclareMathOperator{\RE}{Re}

\DeclareMathOperator{\spec}{\sigma}
\DeclareMathOperator{\discspec}{\sigma_{disc}}
\DeclareMathOperator{\essspec}{\sigma_{ess}}
\DeclareMathOperator{\dist}{dist}
\DeclareMathOperator{\supp}{supp}
\DeclareMathOperator{\sgn}{sgn}

\newcommand{\sect}[1]
{\addtocounter{section}{1}
\medskip

\begin{center}
{\textbf{\large\arabic{section}. #1}}
\end{center}
\setcounter{equation}{0}\setcounter{theorem}{0}\setcounter{lemma}{0}
\setcounter{remark}{0}\smallskip }

\newcommand{\sectn}[1]
{
\medskip
\begin{center} {\textbf{\large #1}}
\end{center}
\setcounter{equation}{0}\setcounter{theorem}{0}\setcounter{lemma}{0}
\setcounter{remark}{0}\smallskip }

\begin{document}

\allowdisplaybreaks

\begin{center}

\textbf{\Large On the spectrum of a Schr\"odinger operator
perturbed by a fast oscillating potential}

\bigskip
\bigskip

{\large D.I. Borisov}

\begin{quote}
\emph{Bashkir State Pedagogical University, October rev. st.,
3a,  Ufa, Russia, 450000. E-mail: \texttt{borisovdi@yandex.ru}
\\ URL: \texttt{http://borisovdi.narod.ru/} }
\end{quote}

\end{center}

\begin{abstract}
We study the spectrum of a one-dimensional Schr\"odinger
operator perturbed by a fast oscillating potential. The
oscillation period is a small parameter. The essential spectrum
is found in an explicit form. The existence and multiplicity of
the discrete spectrum are studied. The complete asymptotics
expansions for the eigenvalues and the associated eigenfunctions
are constructed.
\end{abstract}

\sectn{Introduction}

Many books and papers are devoted to studying the asymptotic
properties of the boundary value problems with fast oscillating
perturbations (see, for instance, the monographs
\cite{Mh}-\cite{PChSh} and the references therein). Much
attention is paid to the investigating the spectral properties
of such problems imposed in bounded domains (see, for instance,
\cite[Ch. 4, \S 10]{BP}, \cite[Ch. I\!I\!I]{OIS}, \cite[Ch.
XI]{ZhKO}). The spectral properties of differential operators in
unbounded domains with fast oscillating perturbations are also
of great interest, since the operators of this kind arise in
many applications. As examples we mention the mathematical model
of the photonic crystals being materials with high-contrast
structure (see \cite{JMW}), as well as a semiclassical model of
electron dynamics in metals (see \cite{AM}). The rigorous
mathematical study of these models was carried out in many
works. Not aiming to list all these papers, we cite only the
survey \cite{Ku} and paper \cite{Zh} on photonic crystals and
survey \cite{Bu} and article \cite{Ma} on the semiclassical
model of electron dynamics. We also mention the works
\cite{B}-\cite{S2} where basing on the methods of the spectral
theory for unbounded operators the authors developed an abstract
scheme for homogenization of the periodic differential operators
in unbounded domains with fast oscillating periodic
perturbations. At the same time, spectral properties of
differential operators with fast oscillating perturbations in
unbounded domains are not clearly understood and even simply
formulated problems are not touched yet. The present paper is
devoted to one of such examples. Namely, we study the spectrum
of the operator
\begin{equation*}%\l%abel{1.1}
H_\e:=-\frac{d^2}{dx^2}+V(x)+a\left(\frac{x}{\e}\right)
\end{equation*}
in $L_2(\mathbb{R})$ with domain $\H^2(\mathbb{R})$. Here $V$ is
a real infinitely differentiable compactly supported function,
$a$ is a real 1-periodic continuous function, $\e$ is a small
positive parameter. The operator $H_\e$ is regarded as a
perturbation of the operator
\begin{equation*} H_0:=-\frac{d^2}{dx^2}+V(x)
\end{equation*}
in $L_2(\mathbb{R})$ with domain $\H^2(\mathbb{R})$ by the
potential $a\left(\frac{ x}{\e}\right)$. The operator $H_0$ is
one of the simplest operators considered in the spectral theory
of unbounded operators, while $a\left(\frac{x}{\e}\right)$ is
one the simplest examples of fast oscillating perturbation. We
find the essential spectrum of the operator $H_\e$ in an
explicit form. The existence, number and multiplicity of the
eigenvalues of the operator $H_\e$ are studied and complete
asymptotics expansions for these eigenvalues and the associated
eigenfunctions are constructed.

We note that a similar problem on the spectrum of the operator
$-\frac{\displaystyle d^2}{\displaystyle
d\xi^2}+p(\xi)+q(\e\xi)$ in $L_2(\mathbb{R})$ was considered
recently in \cite{Ma}. Here $\e$ is a small positive parameter,
$p$ and $q$ are periodic and rapidly decaying function,
respectively. The change of variables $x=\e\xi$ transforms this
problem to the problem on the spectrum of the operator
$-\e^2\frac{\displaystyle d^2} {\displaystyle
dx^2}+p\left(\frac{x}{\e}\right)+q(x)$, where the potential
$p\left(\frac{x}{\e}\right)$ becomes a fast oscillating
function. Under the row of quite severe constraints one of those
was an analyticity of the potential $q$ the leading terms of the
asymptotics expansions for the eigenvalues lying in a some
subinterval of a given lacuna were calculated.

Let us briefly describe the structure of the paper. We formulate
the main results in the next section. The second section is
devoted to the describing the essential spectrum of the operator
$H_\e$. In the third section we study the existence, number,
multiplicity, and convergence of the eigenvalues located in the
semi-infinite lacuna of the essential spectrum. The complete
asymptotics expansions for these eigenvalues and the associated
eigenfunctions are constructed in the fourth and fifth sections.
The existence, number, and multiplicity of the eigenvalues of
the operator $H_\e$ located in interior lacunas of the essential
spectrum are studied in the sixth section. In the seventh
section we construct the complete asymptotics expansions for
these eigenvalues and the associated eigenfunctions.

The results of this paper have been partially announced in
\cite{Bo}.

\sect{Main results}

The aim of this work is to study the structure and the
asymptotic properties of the spectrum of the operator $H_\e$ as
$\e\to0$.

Throughout the work the functions $a$ and $V$ are assumed to be
nonzero. We also suppose that the function $a$ obeys the
equality
\begin{equation}\label{1.0}
\int\limits_0^1 a(\xi)\di\xi=0.
\end{equation}
This equality can be always achieved by shifting the spectrum of
the operator $H_\e$ and adding an appropriate constant to the
function $a$. We denote a spectrum, discrete spectrum, and
essential spectrum by symbols $\spec(\cdot)$,
$\discspec(\cdot)$, and $\essspec(\cdot)$, respectively.
Clearly,  the operators $H_\e$ and $H_0$ are self-adjoint.

Let us formulate the main results.

\begin{theorem}\label{th1.1}
The essential spectrum of the operator $H_\e$ is given by the
equality
\begin{equation*}%\l%abel{1.2}
\essspec(H_\e)=\bigcup_{n=0}^{\infty}
\big[\mu_n^+(\e^2),\mu_{n+1}^-(\e^2)\big].
\end{equation*}
The functions $\mu_n^\pm(\cdot)$ are meromorphic and have the
form
\begin{align}
&\mu_0^+(t)=\sum\limits_{i=1}^\infty \mu_{0,i}^+ t^i,\label{1.3}
\\
&\mu_n^\pm(t)=\frac{\pi^2 n^2}{t}+\sum\limits_{i=0}^\infty
\mu_{n,i}^\pm t^i.\label{1.4}
\end{align}
The coefficients of these series are determined in accordance
with (\ref{2.18}), (\ref{2.107}), (\ref{2.29}), (\ref{2.32})
and, in particular,
\begin{align}
&\mu_{0,1}^+=-\int\limits_0^1\left(\int_0^\xi
a(\eta)\di\eta+\int_0^1
a(\eta)\eta\di\eta\right)^2\di\xi,\label{1.5}
\\
&\mu_{n,0}^\pm=\pm\sqrt{a_n^2+b_n^2},\label{1.6}
\end{align}
where
\begin{equation*}
a_n=\int\limits_0^1 a(\xi)\cos 2\pi n \xi\di\xi,\quad
b_n=\int\limits_0^1 a(\xi)\sin 2\pi n \xi\di\xi.
\end{equation*}
The lacuna $(\mu_n^-,\mu_n^+)$ in the essential spectrum of the
operator $H_\e$ is absent, if the hypothesis of
item~\ref{lm2.6it2} of Lemma~\ref{lm2.6} holds true.
\end{theorem}

According to the theorem given, the essential spectrum of the
operator $H_\e$ has a band structure and consists of an infinite
number of the segments
$\big[\mu_n^+(\e^2),\mu_{n+1}^-(\e^2)\big]$. Each of these
segments has a length of order $\Odr(\e^{-2})$. Lacunas
$\big(\mu_n^-(\e^2),\mu_n^+(\e^2)\big)$, $n\geqslant 1$, in the
essential spectrum ''run'' to infinity as $\e\to0$ keeping the
length bounded uniformly on $\e$. The finiteness condition of
$n$-th lacuna is an inequality $a_n^2+b_n^2\not=0$ (see
(\ref{1.4}), (\ref{1.6})); otherwise the length of the lacuna
tends to zero as $\e\to0$. The border $\mu_0^+$ of the essential
spectrum is located at the left of zero and has the order
$\Odr(\e^2)$, since $\mu_{0,1}^+<0$.

The operator $H_\e$ can have a discrete spectrum in the lacunas
of the essential one. The next part of the results describes the
discrete spectrum of the operator $H_\e$ in a semi-infinite
lacuna $(-\infty,\mu_0^+(\e^2))$. In order to formulate them we
will employ additional notations.

It is known that the essential spectrum of the operator $H_0$
coincides with the positive semiaxis $[0,+\infty)$, while a
discrete spectrum $H_0$ is either empty or consists of a finite
number of simple eigenvalues (see \cite[\S 30, Theorem 8, \S 31,
Theorem 25]{Gl}). We denote these eigenvalues by $\l^{(n)}_0$,
$n=-K,\ldots,-1$, where $K\geqslant 0$ (the case $K=0$
corresponds to the empty discrete spectrum of the operator
$H_0$). The eigenvalues $\l_0^{(n)}$ are taken in an ascending
order: $\l_0^{(-K)}<\l_0^{(-K+1)}<\ldots<\l_0^{(-1)}<0$. The
associated orthonormalized  in $L_2(\mathbb{R})$ eigenfunctions
are denoted by $\psi_0^{(n)}$. Let $x_0$ be a fixed number such
that $\supp V\subset (-x_0,x_0)$.

\begin{theorem}\label{th1.2}
Let the problem
\begin{equation}\label{1.7}
\left(-\frac{d^2}{dx^2}+V\right)\psi_0^{(0)}=0,\quad x\in
\mathbb{R},\qquad \psi_0^{(0)}(x)=\psi_0^{(0)}(\pm x_0),\quad
\pm x\geqslant x_0,
\end{equation}
have no nontrivial solution in $W_{2,loc}^2(\mathbb{R})$. Then
for all sufficiently small $\e$ the operator $H_\e$ has exactly
$K$ eigenvalues $\l_\e^{(n)}$, $n=-K,\ldots,-1$, in the
semi-infinite lacuna $(-\infty,\mu_0^+(\e^2))$, each of them
being simple and satisfying the asymptotics expansion:
\begin{gather}
\l_\e^{(n)}=\l_0^{(n)}+\sum\limits_{i=2}^\infty
\e^i\l_i^{(n)},\label{1.8}
\\
\l_2^{(n)}=\mu_{0,1}^+,\quad \l_3^{(n)}=0,\quad
\l_4^{(n)}=\mu_{0,2}^+
-4\int\limits_\mathbb{R}\Big|\frac{d\psi_0^{(n)}}{dx}\Big|^2\di
x\int\limits_0^1 \left|{L}_0[a](\xi)\right|^2\di\xi,
 \label{1.9}
\end{gather}
where the operator ${L}_0$ is defined in Lemma~\ref{lm2.3}.
\end{theorem}

\begin{theorem}\label{th1.2a}
Let the problem (\ref{1.7}) have a nontrivial solution in
$W_{2,loc}^2(\mathbb{R})$. Then for all sufficiently small $\e$
the operator $H_\e$ has exactly $(K+1)$ eigenvalues
$\l_0^{(n)}$, $n=-K,\ldots,0$, in the semi-infinite lacuna
$(-\infty,\mu_0^+)$. The eigenvalues $\l_\e^{(n)}$,
$n=-K,\ldots,-1$, are simple and satisfy the asymptotics
(\ref{1.8}), (\ref{1.9}). The eigenvalue $\l_\e^{(0)}$ is simple
and satisfies the asymptotics:
\begin{gather}
\l_\e^{(0)}=\sum\limits_{i=2}^\infty \e^i\l_i^{(0)},\label{1.10}
\\
\begin{gathered}
\l_{2j}^{(0)}=\mu_{0,j}^+,\quad\l_{2j+1}^{(0)}=0,\quad j=1,2,3,
\\
\l_8^{(0)}=\mu_{0,4}^+
-16\left(\int\limits_\mathbb{R}\Big|\frac{d\psi_0^{(0)}}{dx}\Big|^2\di
x \int\limits_0^1 \left|{L}_0[a](\xi)\right|^2\di\xi\right)^2,
\end{gathered}
\label{1.11}
\end{gather}
where
\begin{equation}\label{1.12}
\b_\pm=\psi_0^{(0)}(\pm x_0),\quad\b_+^2+\b_-^2=1.
\end{equation}
\end{theorem}

\begin{remark}\label{rm1.1}
Observe, if exists, a nontrivial solution to the problem
(\ref{1.7}) is unique up to a multiplicative constant and is
infinitely differentiable. The normalization condition
(\ref{1.12}) defines this solution uniquely. We also stress that
$\b_\pm\not=0$, since each of the equalities $\b_\pm=0$
contradicts to the non-triviality of the function
$\psi_0^{(0)}$.
\end{remark}

Thus, Theorems~\ref{th1.2},~\ref{th1.2a} maintain that the
number of the eigenvalues of the operator $H_\e$ in the
semi-infinite lacuna coincides or is greater by one the number
of the eigenvalues of the operator $H_0$. The perturbation
$a\left(\frac{x}{\e}\right)$ shifts the eigenvalues of the
operator $H_0$ to the left, the value of the shift coincides
with the shift of the essential spectrum in leading terms. At
the same time, the shift of the eigenvalues of the operator
$H_0$ is less than the shift of the essential spectrum, since
the difference $(\l_\e^{(n)}-\l_0^{(n)}-\mu_0^+(\e^2))$ is
negative and of order $\Odr(\e^4)$.

Apart from the eigenvalues, converging to the eigenvalues of the
operator $H_0$ as $\e\to0$, in the semi-infinite lacuna the
operator $H_\e$ can have an additional eigenvalue, emerging from
the border of the essential spectrum. The existence of this
eigenvalue is equivalent to the existence of the nontrivial
solution to the problem (\ref{1.7}). It is clear that the
existence of this solution is independent on the choice of the
point $x_0$ and is determined by the potential $V$ only. In the
case of the eigenvalue emerging from the essential spectrum, the
distance from this eigenvalue to the border of the essential
spectrum is of order $\Odr(\e^8)$. Thus, the potential
$a\left(\frac{x}{\e}\right)$ behaves as a negative perturbation
with respect to the eigenvalues in the semi-infinite lacuna and
the border of the essential spectrum. We notice that earlier a
similar phenomenon was revealed in \cite{Si,Kl} for the
eigenvalue emerging from the essential spectrum in studying the
operator $-\frac{\displaystyle d^2}{\displaystyle dx^2}+\e
V(x)$, where $V$ is a compactly supported or sufficiently
rapidly decaying function with zero mean value.

The remaining part of the results is devoted to the discrete
spectrum of the operator $H_\e$ in the interior lacunas. We will
deal only with lacunas $\big(\mu_n^-(\e^2),\mu_n^+(\e^2)\big)$,
$n\geqslant 1$, the condition $a_n^2+b_n^2\not=0$ holds for,
what is equivalent to the finiteness of the lacuna's length.

\begin{theorem}\label{th1.3a}
Let at least one of the numbers $a_n$ and $b_n$ be nonzero for
some $n\geqslant 1$. Then the operator $H_\e$ has at most two
eigenvalues in the lacuna
$\left(\mu_n^-(\e^2),\mu_n^+(\e^2)\right)$. Each of these
eigenvalues is simple and satisfies the equality
\begin{equation}\label{1.30}
\l_\e-\mu(\e^2)=o(1),\quad \e\to0,
\end{equation}
where $\mu=\mu_n^-$ or $\mu=\mu_n^+$. If the lacuna
$\big(\mu_n^-(\e^2),\mu_n^+(\e^2)\big)$ contains two eigenvalues
of the operator $H_\e$, then one of them satisfies the equality
(\ref{1.30}) with $\mu=\mu_n^-$, while the other does with
$\mu=\mu_n^+$.
\end{theorem}

Let the hypothesis of this theorem holds true. If exists, the
eigenvalue of the operator $H_\e$ in the lacuna
$\left(\mu_n^-(\e^2),\mu_n^+(\e^2)\right)$ obeying (\ref{1.30})
with $\mu=\mu_n^-$ will be denoted by $\l_{\e,-}^{(n)}$.
Similarly, if exists, the eigenvalue of the operator $H_\e$ in
the lacuna $\left(\mu_n^-(\e^2),\mu_n^+(\e^2)\right)$ obeying
(\ref{1.30}) with $\mu=\mu_n^+$ will be denoted by
$\l_{\e,+}^{(n)}$.

\begin{theorem}\label{th1.3b}
Let at least one of the numbers $a_n$ and $b_n$ be nonzero for
some $n\geqslant 1$. Then
\begin{enumerate}
\def\theenumi{(\arabic{enumi})}
\item\label{it1th1.3b} The operator $H_\e$ has an eigenvalue
$\l_{\e,-}^{(n)}$, if and only if  $\int\limits_\mathbb{R}
V(x)\di x\geqslant 0$.

\item\label{it2th1.3b} The operator $H_\e$ has an eigenvalue
$\l_{\e,+}^{(n)}$, if and only if
\begin{equation}\label{1.16}
\int\limits_{\mathbb{R}} \phi_n^+\left(\frac{x}{\e}, \e^2\right)
\left(\I+\e V T_{14}^+(\e)\right)^{-1}V(x)\phi_n^+
\left(\frac{x}{\e},\e^2\right)\di x<0.
\end{equation}
Here the function $\phi_n^+$ is defined by the formulas
(\ref{2.9c}), (\ref{2.103}), (\ref{2.107}), (\ref{2.31}) and
Lemma~\ref{lm2.7}, while the operator $T_{14}^+$ is given by the
equality (\ref{6.21a}).
\end{enumerate}

\end{theorem}

\begin{theorem}\label{th1.4}
Let at least one of the numbers $a_n$ and $b_n$ be nonzero for
some $n\geqslant 1$ and $\int\limits_\mathbb{R} V(x)\di
x\geqslant 0$. Then the asymptotics of the eigenvalue
$\l_{\e,-}^{(n)}$ is as follows:
\begin{gather}
\l_{\e,-}^{(n)}=\frac{\pi^2 n^2}{\e^2}+\sum\limits_{i=0}^\infty
\e^i\l_{i,-}^{(n)},\label{1.17}
\\
\l_{0,-}^{(n)}=\mu_{n,0}^-,\quad \l_{1,-}^{(n)}=0,\quad
\l_{2,-}^{(n)}=\mu_{n,1}^-  +\frac{2\pi^2 n^2
\left(\tau_2^-\right)^2}{\sqrt{a_n^2+b_n^2}}, \label{1.18}
\end{gather}
where $\tau_2^-$ is from (\ref{7.12}). If
$\int\limits_\mathbb{R} V(x)\di x=0$, then
\begin{equation}\label{1.19}
\l_{2,-}^{(n)}=\mu_{n,1}^-,\quad\l_{3,-}^{(n)}=\l_{5,-}^{(n)}=0,\quad
\l_{4,-}^{(n)}=\mu_{n,2}^-, \quad\l_{6,-}^{(n)}=\mu_{n,3}^- +
\frac{2\pi^2 n^2 \left(\tau_4^-\right)^2}{\sqrt{a_n^2+b_n^2}},
\end{equation}
where $\tau_4^-$ is defined in (\ref{7.30a}).
\end{theorem}

\begin{theorem}\label{th1.5}
Let at least one of the numbers $a_n$ and $b_n$ be nonzero for
some $n\geqslant 1$. Let there exists a natural number
$\mathfrak{n}_+$ so that the numbers $\tau_i^+$, defined in
accordance with (\ref{7.12}), (\ref{7.29a}), (\ref{7.30a}) and
Lemma~\ref{lm7.2}, meet the relations: $\tau_i^+=0$, $i\leqslant
\mathfrak{n}_+ -1$, $\tau_{\mathfrak{n}_+}\not=0$. Then the
eigenvalue $\l_{\e,+}^{(n)}$ exists, if and only if
$\tau_{\mathfrak{n}_+}>0$. In this case the asymptotics of the
eigenvalue $\l_{\e,+}^{(n)}$ has the form:
\begin{gather}
\l_{\e,+}^{(n)}=\frac{\pi^2 n^2}{\e^2}+\sum\limits_{i=0}^\infty
\e^i\l_{i,+}^{(n)}\label{1.20}
\\
\l_{2j,+}^{(n)}=\mu_{n,j}^+,\quad \l_{2j+1,+}^{(n)}=0,\quad
j\leqslant\mathfrak{n}_+-2,\quad
\l_{2\mathfrak{n}_+-2,+}^{(n)}=\mu_{n,\mathfrak{n}_+}^+ -
\frac{2\pi^2
n^2\left(\tau_{\mathfrak{n}_+}^+\right)^2}{\sqrt{a_n^2+b_n^2}}.
\label{1.21}
\end{gather}

\end{theorem}

\begin{remark}\label{rm1.2}
In the fourth, fifth, and seventh sections we give the algorithm
determining all the coefficients of the series (\ref{1.8}),
(\ref{1.10}), (\ref{1.17}), (\ref{1.20}). We do not give these
formulas in
Theorems~\ref{th1.2},~\ref{th1.2a},~\ref{th1.4},~\ref{th1.5},
since they include a great amount of additional notations.
Because of the same reason we do not give the asymptotics
expansions for the eigenfunctions of the operator $H_\e$ but
formulate corresponding statements in the end of the fourth,
fifth, and seventh sections (see
Theorems~\ref{th4.1},~\ref{th5.1},~\ref{th7.1}).
\end{remark}

In accordance with item~\ref{it1th1.3b} of Theorem~\ref{th1.3b}
the eigenvalue $\l_{\e,-}^{(n)}$ exists, if and only if
$\int\limits_\mathbb{R} V(x)\di x\geqslant 0$, while due to
Theorem~\ref{th1.5} and formula (\ref{7.12}) for $\tau_2^+$ the
eigenvalue $\l_{\e,+}^{(n)}$ exists, if
$\int\limits_\mathbb{R}V(x)\di x<0$, and is absent, if
$\int\limits_\mathbb{R}V(x)\di x>0$. Hence, the finite lacuna
$\big(\mu_n^-(\e^2),\mu_n^+(\e^2)\big)$ contains at least one
eigenvalue of the operator $H_\e$. Moreover, in the case
$\int\limits_\mathbb{R} V(x)\di x\not=0$ the finite lacuna
contains exactly one eigenvalue located near one of the edges of
the lacuna subject to the sign of $\int\limits_\mathbb{R}
V(x)\di x$ and differs from this edge by a quantity of order
$\Odr(\e^2)$. If $\int\limits_\mathbb{R} V(x)\di x=0$, then the
eigenvalue $\l_{\e,-}^{(n)}$ exists in the finite lacuna and is
distant from the left edge by a quantity of order $\Odr(\e^6)$.
The existence of the eigenvalue $\l_{\e,+}^{(n)}$ is determined
by the numbers $\tau_i^+$ constructed in the seventh section
(see (\ref{7.21}), (\ref{7.30a})). In particular, it follows
from Theorems~\ref{th1.5} and (\ref{7.30a}) that the eigenvalue
$\l_{\e,-}^{(n)}$ exists, if
\begin{equation}\label{1.31}
2\int\limits_\mathbb{R} V^2(x)\di
x<\sqrt{a_n^2+b_n^2}\int\limits_\mathbb{R} \bigg(
\int\limits_\mathbb{R} \sgn(x-t) V(t)\di t\bigg)^2 \di x
\end{equation}
and is absent, if  the opposite inequality takes place. Clearly,
given a potential $V$, one can always achieve a prescribed sign
in this inequality by a suitable choice of $a$. Thus, in the
case $\int\limits_\mathbb{R} V(x)\di x=0$ the lacuna
$\big(\mu_n^-(\e^2),\mu_n^+(\e^2)\big)$ can contain both one and
two eigenvalues of the operator $H_\e$. We also note that, if
exists, the distance from the eigenvalue $\l_{\e,+}^{(n)}$ to
the right edge of the lacuna is of order $\Odr(\e^{2m})$, where
$m\geqslant 3$, and the case $m>3$ can be realized by choosing
the potentials $a$ and $V$ so that the left-hand side  in
(\ref{1.31}) equals to the right-hand side. Hence, in contrast
to the semi-infinite lacuna, in a finite lacuna the potential
$a\left(\frac{x}{\e}\right)$ behaves as a perturbation of
variable sign. We should also notice, that the statements of
Theorems~\ref{th1.3a}-\ref{th1.5} are in agreement with the
results of works \cite{RB}-\cite{GS}. In these papers the
operator $-\frac{\displaystyle d^2}{\displaystyle
dx^2}+p(x)+q(x)$ was considered where $p\in
L_{1,loc}(\mathbb{R})$ is a periodic function, and $q$ obeys the
condition $\int\limits_\mathbb{R} (1+|x|)|q(x)|\di x<\infty$. It
was shown that the lacunas of the essential spectrum contain the
finite number of the eigenvalues, a distant (in number) lacuna
contain at most two eigenvalues, and in the case
$\int\limits_\mathbb{R} q(x)\di x\not=0$ a distant lacuna
contains exactly one eigenvalue.

\sect{Essential spectrum}

This section is devoted to the proof of Theorem~\ref{th1.1}. An
operator
\begin{equation*}
\t H_\e:=-\frac{d^2}{dx^2}+a\left(\frac{x}{\e}\right)
\end{equation*}
in $L_2(\mathbb{R})$ with domain $\H^2(\mathbb{R})$ is
self-adjoint. It is easy to check that the multiplication
operator by $V(x)$ is $\t H_\e$-compact. By Theorem~5.35 from
\cite[Ch. IV, \S 5.6]{K} it implies the equality
$\essspec(H_\e)=\essspec(\t H_\e)$. Hence, in order to prove
Theorem~\ref{th1.1} it is sufficient to study the structure of
the essential spectrum of the operator $\t H_\e$. Due to the
periodicity of the potential $a\left(\frac{x}{\e}\right)$ and in
accordance with \cite[Ch. 2, \S 2.8]{Sh} the essential spectrum
of the operator $\t H_\e$ is as follows:
\begin{equation}\label{2.1}
\essspec(\t H_\e)=\bigcup_{n=0}^\infty
\big[\mu_n^+,\mu_{n+1}^-\big],
\end{equation}
where $\mu_0^+$ and $\mu_n^\pm$ are eigenvalues of the boundary
value problems
\begin{align*}
&\left(-\frac{d^2}{dx^2}+a\left(\frac{x}{\e}\right)\right)\phi=
\mu\phi,\quad x\in[0,\e],%\l%abel{2.3}
\\
&\phi(0)-\phi(\e)=0,\quad
\frac{d\phi}{dx}(0)-\frac{d\phi}{dx}(\e)=0, \quad \text{if $n$
is even},
%\l%abel{2.4}
\\
&\phi(0)+\phi(\e)=0,\quad
\frac{d\phi}{dx}(0)+\frac{d\phi}{dx}(\e)=0, \quad \text{if $n$
is odd}.
\end{align*}
The solution to these problems are sought in the space
$C^2[0,\e]$. Making change of variable $\xi=x/\e$, we obtain
that the quantities $M_n^\pm:=\e^2\mu_n^\pm$ are eigenvalues of
the boundary value problems for the equation
\begin{equation}\label{2.6}
\left(-\frac{d^2}{d\xi^2}+\e^2a(\xi)\right)\phi=M\phi,\quad
\xi\in[0,1],
\end{equation}
subject to boundary conditions
\begin{equation}\label{2.7}
\phi(0)-\phi(1)=0,\quad
\frac{d\phi}{d\xi}(0)-\frac{d\phi}{d\xi}(1)=0,
\end{equation}
if $n$ is even, and
\begin{equation}\label{2.8}
\phi(0)+\phi(1)=0,\quad
\frac{d\phi}{d\xi}(0)+\frac{d\phi}{d\xi}(1)=0,
\end{equation}
if $n$ is odd. Here $\phi=\phi(\xi)$.

\begin{lemma}\label{lm2.1}
The eigenvalues of the problems (\ref{2.6}), (\ref{2.7}) and
(\ref{2.6}), (\ref{2.8}) are holomorphic on $\e^2$. The
associated eigenfunctions orthonormalized in $L_2(0,1)$ can be
chosen as holomorphic on $\e^2$ in the norm of $C^2[0,1]$.
\end{lemma}
\begin{proof}
By $\mathcal{W}^+$ (respectively, by $\mathcal{W}^-$) we denote
the subset of functions belonging to $\H^2(0,1)$ and satisfying
the boundary conditions (\ref{2.7}) (respectively, (\ref{2.8})).
Using the estimate
\begin{equation*}
|u(0)|+|u(1)|+|u'(0)|+|u'(1)|\le C\|u\|_{\H^2(0,1)}
\end{equation*}
it is easy to check that the sets $\mathcal{W}^\pm$ are Hilbert
spaces. An operator $-\frac{\displaystyle d^2}{\displaystyle
d\xi^2}+\d a(\xi)$ in $L_2(0,1)$ with domain $\mathcal{W}^+$
(respectively, $\mathcal{W}^-$) will be indicated as $A^+(\d)$
(respectively, $A^-(\d)$). A complex parameter $\d$ is supposed
to belong to a small neighborhood of an interval $(-\d_0,\d_0)$,
where $\d_0$ is a sufficiently small fixed number. This
neighbourhood is assumed to be symmetric with respect to the
real axis. Following the definition, one checks that the
operators $A^\pm(\d)$ are self-adjoint holomorphic families of
the type (A) (see the definition in \cite[Ch. VII, \S\S 2.1,
3.1]{K}). It is also easy to make sure that the resolvents of
the operators $A^\pm(0)$ are compact, what by Theorem~2.4 from
\cite[Ch. VII, \S 2.1]{K} yields that the operators $A^\pm(\d)$
have the compact resolvent as well. The described properties of
the operators $A^\pm(\d)$ allow to apply Theorem~3.9 from
\cite[Ch. VII, \S 3.5]{K} to these operators, which implies the
holomorphy on $\d\in(-\d_0,\d_0)$ of the eigenvalues of the
operators $A^\pm(\d)$ and the associated orthonormalized in
$L_2(0,1)$ eigenfunctions (in norm of the same space). Treating
these eigenfunctions as solutions of the boundary value problems
for the equation
\begin{equation}\label{2.9}
\left(-\frac{d^2}{d\xi^2}+1\right)\phi=\phi-\d a\phi+M\phi
\end{equation}
with boundary conditions (\ref{2.7}) or (\ref{2.8}), we deduce
that the eigenfunctions of the operators $A^\pm(\d)$ are
holomorphic in the norm of $\H^2(0,1)$. Due to the inclusion
$C^1[0,1]\subset\H^2(0,1)$ it follows that the eigenfunctions
are holomorphic in norm of $C^1[0,1]$. Treating these
eigenfunctions as solutions to the boundary value problems for
the equation (\ref{2.9}) once again, we conclude that they
belong to $C^2[0,1]$ and are holomorphic on $\d$ in the norm of
this space.
\end{proof}

According to the proven lemma, the eigenvalues $M_n^\pm(\e^2)$
read as follows
\begin{equation}
\label{2.9a} M_n^\pm=M_n^\pm(\e^2)=\sum\limits_{i=0}^\infty
\e^{2i}\mu_{n,i-1}^\pm,
\end{equation}
where $\mu_{n,i-1}^\pm$ are some numbers. The numbers
$\mu_{n,-1}^\pm$ are eigenvalues of the problems (\ref{2.6}),
(\ref{2.7}) and (\ref{2.6}), (\ref{2.8}) with $\e=0$. One can
easily check that $\mu_{0,-1}^+=0$, $\mu_{n,-1}^\pm=\pi^2 n^2$,
$n\geqslant 1$. Let us determine other coefficients of the
series (\ref{2.9a}). We begin with the case $n=0$. The
eigenvalue $\mu_{0,-1}^+$ of the problem (\ref{2.6}),
(\ref{2.7}) is simple, and the associated eigenfunction takes
the form: $\phi_{0,0}^+(\xi)\equiv1$. By Lemma~\ref{lm2.1} it
follows that the eigenvalue $M_0^+(\e^2)$ is simple as well.

Hereinafter the symbol $(\cdot,\cdot)_{X}$ denotes the inner
product in a Hilbert space $X$.

\begin{lemma}\label{lm2.2}
The eigenfunction $\phi_{0}^+$ associated with $M_0^+(\e^2)$ can
be chosen as holomorphic on $\e^2$ in the norm of $C^2[0,1]$:
\begin{equation}\label{2.9b}
\phi_0^+(\xi,\e^2)=\sum\limits_{i=0}^\infty
\e^{2i}\phi_{0,i}^+(\xi),
\end{equation}
where the functions $\phi_{0,i}^+$ meet the equalities
\begin{equation}
\int\limits_0^1\phi_{0,i}^+(\xi)\di\xi=0,\quad
i\geqslant1.\label{2.100}
\end{equation}
\end{lemma}

\begin{proof}
By Lemma~\ref{lm2.1} the eigenfunction associated with $M_0^+$
can be chosen as normalized in $L_2(\mathbb{R})$ and holomorphic
on $\e^2$ in the norm of $C^2[0,1]$. We denote such function by
$\t\phi_0^+=\t\phi_0^+(\xi,\e^2)$. Without loss of generality we
suppose that $\t\phi_0^+(\xi,0)\equiv \phi_{0,0}^+(\xi)$.
Bearing in mind the properties of $\t\phi_0^+$, it is easy to
check that the function
$\phi_0(\xi,\e^2):=\big(\t\phi_0^+,\phi_{0,0}^+\big)_{L_2(0,1)}^{-1}
\t\phi_0(\xi,\e^2)$ satisfies the lemma.
\end{proof}

In what follows the function $\phi_0^+$ is assumed to be chosen
in accordance with Lemma~\ref{lm2.2}. We substitute the series
(\ref{2.9a}) and (\ref{2.9b}) into the boundary value problem
(\ref{2.6}), (\ref{2.7}) and evaluate the coefficients of the
same powers of $\e$. In view of holomorphy of the functions
$M_0^+$ and $\phi_0^+$ on $\e^2$ such substitution is not a
formal procedure, that is why we obtain that the problem
(\ref{2.6}), (\ref{2.7}) for $M_0^+$ and $\phi_0^+$ is
equivalent to the following recurrent system of boundary value
problems:
\begin{equation}\label{2.13}
\begin{aligned}
&-\frac{d^2}{d\xi^2}\phi_{0,i}^+=-a\phi_{0,i-1}^+
+\sum\limits_{j=1}^i \mu_{0,j-1}^+\phi_{0,i-j}^+,\quad
\xi\in[0,1],
\\
&\phi_{0,i}^+(0)-\phi_{0,i}^+(1)=0,\quad
\frac{d\phi_{0,i}^+}{d\xi}(0)-\frac{d\phi_{0,i}^+}{d\xi}(1)=0,\quad
i\geqslant 1.
\end{aligned}
\end{equation}
It is sufficient to find out the solutions of these problems
satisfying the conditions (\ref{2.100}) in order to determine
the coefficients of the series (\ref{2.9a}), (\ref{2.9b}) for
$n=0$.

By direct calculations one can check the following statement.
\begin{lemma}\label{lm2.3}
Let $f\in C[0,1]$. The boundary value problem
\begin{equation*}
-\frac{d^2u}{d\xi^2}=f,\quad \xi\in[0,1],\qquad
u(0)-u(1)=0,\quad \frac{du}{d\xi}(0)-\frac{du}{d\xi}(1)=0
\end{equation*}
has a unique solution $u\in C^2[0,1]$ obeying the equality
$\int\limits_0^1u(\xi)\di\xi=0$, if and only if $\int\limits_0^1
f(\xi)\di\xi=0$. The solution $u$ is given  by the formula
\begin{equation*}
u(\xi)={L}_0[f](\xi):=-\int\limits_0^\xi
(\xi-\eta)f(\eta)\di\eta
+\frac{1}{2}\int\limits_0^1(\eta^2-\eta-2\xi\eta)f(\eta)\di\eta.
\end{equation*}
\end{lemma}

Solving sequentially the boundary value problems (\ref{2.13}) by
Lemma~\ref{lm2.3} and taking  into account (\ref{1.0}),
(\ref{2.100}), we get
\begin{equation}
\begin{aligned}
& \mu_{0,0}^+=0,\quad
\mu_{0,i}^+=\int\limits_0^1a\phi_{0,i}^+\di\xi,\quad
i\geqslant 1,%\l%abel{2.17}
\\
& \phi_{0,1}^+=-{L}_0[a],\quad
\phi_{0,i}^+={L}_0\left[-a\phi_{0,i-1}^+ +\sum\limits_{j=2}^i
\mu_{0,j-1}^+\phi_{0,i-j}^+\right],\quad i\geqslant2.\nonumber
\end{aligned}\label{2.18}
\end{equation}
where the coefficients $\mu_{0,i}^+$ are determined by the
solvability conditions of the boundary value problems
(\ref{2.13}). We calculate $\mu_{0,1}^+$:
\begin{equation*}
\mu_{0,1}^+=\int\limits_0^1 a\phi_{0,1}^+\di\xi=\int\limits_0^1
\phi_{0,1}^+\frac{d^2}{d\xi^2}\phi_{0,1}^+\di\xi=-\int\limits_0^1
\left(\frac{d}{d\xi}\phi_{0,1}^+\right)^2\di\xi,
\end{equation*}
what in view of the definition of the operator ${L}_0$ implies
the formula (\ref{1.5}). The conclusion of Theorem~\ref{th1.1}
on $\mu_0^+$ is complete.

We proceed to the calculating the coefficients $\mu_{n,i}^\pm$
for $n>0$. The {eigen\-func\-tions} $\phi_{n,0}^\pm$ associated
with $\mu_{n,-1}^\pm$ will be indicated as $\phi_{n,0}^\pm$ and
are chosen as orthogonal in $L_2(0,1)$ in the form:
\begin{equation}\label{2.20}
\phi_{n,0}^+(\xi)=\sqrt{2}\cos(\pi n \xi+\a),\quad
\phi_{n,0}^-(\xi)=\sqrt{2}\sin(\pi n \xi+\a),
\end{equation}
where $\a\in(-\pi/2,\pi/2]$ is a some number depending on $n$.
The functions $M_n^\pm(\e^2)$ begin holomorphic on $\e^2$,
without loss of generality we assume that
\begin{equation}
M_n^-(\e^2)\leqslant M_n^+(\e^2).\label{2.106}
\end{equation}

\begin{lemma}\label{lm2.4}
The eigenfunctions $\phi_{n}^\pm$ associated with $M_n^\pm$ and
the number $\a$ in (\ref{2.20}) can be chosen so that the
functions $\phi_n^\pm$ are orthogonal in $L_2(0,1)$ and
holomorphic on $\e^2$ in the norm of $C^2[0,1]$:
\begin{equation}\label{2.9c}
\phi_n^\pm(\xi,\e^2)=\sum\limits_{i=0}^\infty
\e^{2i}\phi_{n,i}^\pm(\xi),
\end{equation}
where the functions $\phi_{n,i}^\pm$ obey the equalities
\begin{equation}\label{2.9d}
\left(\phi_{n,i}^\pm,\phi_{n,0}^\pm\right)_{L_2(0,1)}=0, \quad
i\geqslant1.
\end{equation}
\end{lemma}
\begin{proof}
Employing Lemma~\ref{lm2.1}, we chose the eigenfunctions
$\t\phi_n^\pm=\t\phi_n^\pm(\xi,\e^2)$ associated with $M_n^\pm$
as orthonormalized in $L_2(0,1)$ and holomorphic on $\e^2$ in
$C^2[0,1]$. Since $\t\phi_n^\pm(\xi,0)$ are the eigenfunctions
of the problem (\ref{2.6}), (\ref{2.7}) or (\ref{2.6}),
(\ref{2.8}) associated with $\mu_{n,-1}^\pm$ and orthonormalized
in $L_2(0,1)$, it follows that the number $\a$ in (\ref{2.20})
can chosen so that the equalities
$\t\phi_n^\pm(\xi,0)\equiv\phi_{n,0}^\pm(\xi)$ take place. Now
it is easy to check that the functions $\phi_n^\pm(\xi,\e^2):=
\big(\phi_n^\pm,\phi_{n,0}^\pm\big)_{L_2(0,1)}^{-1}
\t\phi_n^\pm(\xi,\e^2)$ meet the conclusion of the lemma.
\end{proof}

Hereafter the functions $\phi_n^\pm$ are assumed to be chosen in
accordance with {Lem\-ma}~\ref{lm2.4}. As in the case $n=0$, in
order to determine the coefficients of the series (\ref{2.9a}),
(\ref{2.9c}) one should substitute these series into one of the
boundary value problems (\ref{2.6}), (\ref{2.7}) and
(\ref{2.6}), (\ref{2.8}) (subject to the parity of $n$) and
collect the coefficients of the same powers of $\e$, what leads
us to the following boundary value problems:
\begin{equation}
\begin{aligned}
&-\left(\frac{d^2}{d\xi^2}+\pi^2n^2\right)\phi_{n,i}^\pm
=-a(\xi)\phi_{n,i-1}^\pm+\sum\limits_{j=1}^i
\mu_{n,j-1}^\pm\phi_{n,i-j}^\pm,\quad \xi\in[0,1],
\\
&\phi_{n,i}^\pm(0)+(-1)^{n+1}\phi_{n,i}^\pm(1)=0,\quad
\frac{d\phi_{n,i}^\pm}{d\xi}(0)+(-1)^{n+1}
\frac{d\phi_{n,i}^\pm}{d\xi}(1)=0,\quad i\geqslant 1.
\end{aligned}\label{2.21}
\end{equation}

In what follows we will make use of the following auxiliary
statements.

\begin{lemma}\label{lm2.5}
Let $f\in C[0,1]$. The boundary value problem
\begin{equation}\label{2.101}
\begin{gathered}
-\left(\frac{d^2}{d\xi^2}+\pi^2 n^2\right)u=f,\quad \xi\in[0,1],
\\
u(0)+(-1)^{n+1}u(1)=0,\quad
\frac{du}{d\xi}(0)+(-1)^{n+1}\frac{du}{d\xi}(1)=0
\end{gathered}
\end{equation}
has a unique solution $u\in C^2[0,1]$ orthogonal to the function
$\phi_{n,0}^\pm$ in $L_2(0,1)$, if and only if
$\left(f,\phi_{n,0}^\pm\right)_{L_2(0,1)}=0$. The solution $u$
is given by the formula:
\begin{equation*}
u(\xi)={L}_n[f](\xi):=-\frac{1}{\pi n}\int\limits_0^\xi \sin\pi
n(\xi-\eta)f(\eta)\di\eta-\frac{1}{\pi n}\int\limits_0^1
\eta\sin\pi n(\xi-\eta)f(\eta)\di\eta.
\end{equation*}
\end{lemma}
The validity of the lemma is checked by direct calculations.

\begin{lemma}\label{lm2.8}
Let $p$, $s$ be some natural numbers, $A_i$, $B_{i,j}$ be
arbitrary number sequences. The equality
\begin{equation*}
\sum\limits_{i=s}^{p-s}\sum\limits_{j=0}^{i-s} A_j B_{p-i,i-j}=
\sum\limits_{i=s}^{p-s}\sum\limits_{j=0}^{i-s} A_j B_{i-j,p-i}
\end{equation*}
holds true
\end{lemma}

\begin{proof}
The validity of the lemma follows from the chain of equalities:
\begin{align*}
&\sum\limits_{i=s}^{p-s}\sum\limits_{j=0}^{i-s} A_j B_{p-i,i-j}=
\sum\limits_{j=0}^{p-2s}\sum\limits_{i=j+s}^{p-s} A_j
B_{p-i,i-j}=
\\
& =\sum\limits_{j=0}^{p-2s}\sum\limits_{\t i=j+s}^{p-s} A_j
B_{\t i-j,p-\t i}=\sum\limits_{\t
i=s}^{p-s}\sum\limits_{j=0}^{\t i-s} A_j B_{\t i-j,p-\t i},
\end{align*}
where the change of summation index $\t i=p-i+j$ has been
carried out.
\end{proof}

We set:
\begin{align*}
&\Phi_{n,0}^+(\xi):=\sqrt{2}\cos(\pi n\xi),\quad
\Phi_{n,0}^-(\xi):=\sqrt{2}\sin(\pi n\xi),\quad
M_{n,0}^{\pm,+}:=M_{n,0}^{\pm,-}:=0,
\\
&\Phi_{n,i}^\pm:={L}_n\left[-a\Phi_{n,i-1}^\pm+
\sum\limits_{j=1}^i M_{n,j}^{\pm,\pm}\Phi_{n,i-j}^\pm\right],
\\
&M_{n,i}^{\pm,+}:=
\left(a\Phi_{n,i-1}^\pm,\Phi_{n,0}^+\right)_{L_2(0,1)},\quad
M_{n,i}^{\pm,-}:=
\left(a\Phi_{n,i-1}^\pm,\Phi_{n,0}^-\right)_{L_2(0,1)}.
\end{align*}

\begin{lemma}\label{lm2.6}
One of two situations takes place:
\begin{enumerate}
\def\theenumi{(\arabic{enumi})}
\item\label{lm2.6it1}
There exists a number $N\geqslant1$ such that the equalities
\begin{equation}\label{2.22}
M_{n,i}^{+,-}=M_{n,i}^{-,+}=M_{n,i}^{+,+}-M_{n,i}^{-,-}=0
\end{equation}
hold true for $i\leqslant N-1$ and at least one of three numbers
$M_{n,N}^{+,-}$, $M_{n,N}^{-,+}$,
$\left(M_{n,N}^{+,+}-M_{n,N}^{-,-}\right)$ is nonzero. In this
case the functions $\Phi_{n,i}^\pm$, $1\leqslant i\leqslant N$
satisfy the boundary conditions in (\ref{2.101}), are orthogonal
to $\Phi_{n,0}^+$ and $\Phi_{n,0}^-$ in $L_2(0,1)$, and the
equality $M_{n,N}^{+,-}=M_{n,N}^{-,+}$ holds true.

\item\label{lm2.6it2}

The equalities (\ref{2.22}) are valid for each $i\geqslant 1$.
In this case $\mu_n^-(\e^2)\equiv \mu_n^+(\e^2)$ and the lacuna
$\big(\mu_n^-(\e^2),\mu_n^+(\e^2)\big)$ in the essential
spectrum of the operator $H_\e$ is absent.
\end{enumerate}
\end{lemma}

\begin{proof}
Obviously, the hypothesis of one of the cases \ref{lm2.6it1} and
\ref{lm2.6it2} always takes place and these cases exclude each
other. Suppose the hypothesis of the first case holds. Assume
that the equalities (\ref{2.22}) are valid $i\leqslant m$, and
the functions $\Phi_{n,i}^\pm$, $1\leqslant i\leqslant m$, are
orthogonal to $\Phi_{n,0}^+$ and $\Phi_{n,0}^-$ in $L_2(0,1)$.
In this case the functions
$f^\pm:=-a\Phi_{n,m}^\pm+\sum\limits_{j=1}^{m+1}
M_{n,j}^{\pm,\pm}\Phi_{m-j+1}^\pm$ satisfy the hypothesis of
Lemma~\ref{lm2.5}, and $\Phi_{n,m+1}^\pm$ are the corresponding
solutions of the boundary value problem (\ref{2.101}). It
follows that the functions $\Phi_{n,m+1}^\pm$ obey the boundary
conditions in (\ref{2.101}) and are orthogonal to $\Phi_{n,0}^+$
and $\Phi_{n,0}^-$ in $L_2(0,1)$. Therefore, all the functions
$\Phi_{n,i}^\pm$, $1\leqslant i\leqslant N-1$, satisfy the
boundary conditions in (\ref{2.101}), and are orthogonal to
$\Phi_{n,0}^+$ and $\Phi_{n,0}^-$ in $L_2(0,1)$.

Let us prove the equality $M_{n,N}^{+,-}=M_{n,N}^{-,+}$. If
$N=1$, then this equality is obvious. Let $N\geqslant 2$.
Bearing in mind the definition and the proven properties of the
functions $\Phi_{n,i}^\pm$, and integrating by parts, we get
($1\leqslant i\leqslant N-1$, $0\leqslant j\leqslant N-2$):
\begin{align*}
&\int\limits_0^1 a\Phi_{n,i}^+\Phi_{n,j}^-\di\xi=
\int\limits_0^1
\Phi_{n,i}^+\left(\frac{d^2}{d\xi^2}+\pi^2n^2\right)
\Phi_{n,j+1}^-\di\xi+\sum\limits_{q=1}^{j+1}
M_{n,q}^{-,-}\int\limits_0^1 \Phi_{n,i}^+
\Phi_{n,j-q+1}^-\di\xi=
\\
&=\int\limits_0^1 a\Phi_{n,i-1}^+\Phi_{n,j+1}^-\di\xi-
\sum\limits_{q=1}^{i-1} M_{n,q}^{+,+}\int\limits_0^1
\Phi_{n,i-q}^+ \Phi_{n,j+1}^-\di\xi+\sum\limits_{q=1}^{j}
M_{n,q}^{-,-}\int\limits_0^1 \Phi_{n,i}^+
\Phi_{n,j-q+1}^-\di\xi,
\end{align*}
what together with (\ref{2.22}) imply
\begin{align*}
\int\limits_0^1 a\Phi_{n,N-1}^+\Phi_{n,0}^-\di\xi
=&\sum\limits_{j=1}^{N-2} \sum\limits_{q=1}^{j}
M_{n,q}^{+,+}\int\limits_0^1 \Phi_{n,N-j-1}^+
\Phi_{n,j-q+1}^-\di\xi-
\\
&-\sum\limits_{i=2}^{N-1}\sum\limits_{q=1}^{i-1}
M_{n,q}^{+,+}\int\limits_0^1 \Phi_{n,i-q}^+
\Phi_{n,N-i}^-\di\xi+\int\limits_0^1
a\Phi_{n,0}^+\Phi_{n,N-1}^-\di\xi.
\end{align*}
Changing the summation index  $j\mapsto j+1$ in the first
summand in the right hand side of the equality obtained and
applying Lemma~\ref{lm2.8} with $p=N$, $s=1$,
$A_j=M_{n,j}^{+,+}$, $j\geqslant 0$, $B_{i,j}=\int\limits_0^1
\Phi_{n,i}^+\Phi_{n,j}^-\di\xi$ to the second summand, we arrive
at the equality $M_{n,N}^{+,-}=M_{n,N}^{-,+}$.

Suppose that the case~\ref{lm2.6it2} takes place, and the
equalities (\ref{2.22}) hold for each $i\geqslant1$. Then it
follows from the definition of $\Phi_{n,i}^\pm$ that the
functions
\begin{equation*}
\phi_{n,i}^+=\Phi_{n,i}^+\cos\a-\Phi_{n,i}^-\sin\a,\quad
\phi_{n,i}^-=\Phi_{n,i}^-\sin\a+\Phi_{n,i}^+\cos\a
\end{equation*}
are solutions to the boundary value problems (\ref{2.21}) with
$\mu_{n,i}^\pm=M_{n,i+1}^{\pm,\pm}$. The relations (\ref{2.22})
yield the equalities $\mu_{n,i}^-=\mu_{n,i}^+$, what implies
that $\mu_n^-(\e^2)\equiv\mu_n^+(\e^2)$.
\end{proof}

Everywhere till the end of this section we suppose that the
case~\ref{lm2.6it1} of Lemma~\ref{lm2.6} takes place. Using
Lemmas~\ref{lm2.5},~\ref{lm2.6} and the definition of the
functions $\Phi_{n,i}^\pm$ and numbers $M_{n,i}^\pm$, by direct
calculations one can easily check that the general solutions to
the problems (\ref{2.21}) for $i\leqslant N-1$ obeying the
conditions (\ref{2.9d}) and numbers $\mu_{n,i}^\pm$ with
$i\leqslant N-2$ are of the form:
\begin{align}
&
\phi_{n,i}^\pm=\t\phi_{n,i}^\pm+\sum\limits_{j=1}^{i}c_{n,j}^\pm
\phi_{n,i-j}^\mp,\quad i\leqslant N-1,\label{2.103}
\\
&\t\phi_{n,i}^+=\Phi_{n,i}^+\cos\a-\Phi_{n,i}^-\sin\a, \quad
\t\phi_{n,i}^-=\Phi_{n,i}^+\sin\a+\Phi_{n,i}^-\cos\a,\quad
0\leqslant i\leqslant N-1,\nonumber
\\
&
\mu_{n,i}^+=M_{n,i+1}^{+,+}=M_{n,i+1}^{-,-}=\mu_{n,i}^{-},\quad
0\leqslant i\leqslant N-2, \label{2.107}
\end{align}
where $c_{n,i}^\pm$ are some numbers. This fact together with
Lemma~\ref{lm2.5}, the item~\ref{lm2.6it1} of Lemma~\ref{lm2.6},
and the definition of the numbers $M_{n,i}^{\pm,\pm}$ imply that
the solvability conditions of the boundary value problem
(\ref{2.21}) for $\phi_{n,N}^\pm$ are as follows:
\begin{equation}\label{2.30}
\begin{gathered}
\mu_{n,N-1}^\pm=
\big(a\t\phi_{n,N-1}^\pm,\phi_{n,0}^\pm\big)_{L_2(0,1)},
\\
\big(a\t\phi_{n,N-1}^+,\phi_{n,0}^-\big)_{L_2(0,1)}=
\big(a\t\phi_{n,N-1}^-,\phi_{n,0}^+\big)_{L_2(0,1)}=0,
\end{gathered}
\end{equation}
what yields:
\begin{gather}
\mu_{n,N-1}^\pm=\frac{M_{n,N}^{+,+}+M_{n,N}^{-,-}\pm
\left(\left(M_{n,N}^{+,+}-M_{n,N}^{-,-}\right)\cos 2\a-2
M_{n,N}^{+,-}\sin 2\a\right)}{2},\nonumber%\l%abel{2.25}
\\
\left(M_{n,N}^{+,+}-M_{n,N}^{-,-}\right)\sin
2\a+2M_{n,N}^{+,-}\cos 2\a=0.\label{2.108}
\end{gather}
According to the hypothesis of item~\ref{lm2.6it1} of
Lemma~\ref{lm2.6}, one of the numbers $M_{n,N}^{+,-}$,
$\left(M_{n,N}^{+,+}-M_{n,N}^{-,-}\right)$ is nonzero, this is
why the equality (\ref{2.108}) regarded as an equation for $\a$
has exactly two roots differing by $\pi$ in the interval
$(-\pi/2,\pi/2]$. Using (\ref{2.108}), we check that for each of
these root the equality
\begin{equation*}
\left(\left(M_{n,N}^{+,+}-M_{n,N}^{-,-}\right)\cos 2\a-2
M_{n,N}^{+,-}\sin
2\a\right)^2=\left(M_{n,N}^{+,+}-M_{n,N}^{-,-}\right)^2+
4\left(M_{n,N}^{+,-}\right)^2
\end{equation*}
holds. Taking into account this equality, it is not difficult to
make sure that
\begin{equation*}
\left(M_{n,N}^{+,+}-M_{n,N}^{-,-}\right)\cos 2\a-2
M_{n,N}^{+,-}\sin
2\a=\sqrt{\left(M_{n,N}^{+,+}-M_{n,N}^{-,-}\right)^2+
4\left(M_{n,N}^{+,-}\right)^2}>0
\end{equation*}
for one of the roots of the equation (\ref{2.108}). We choose
$\a\in(-\pi/2,\pi/2]$ so that the last equality is true. This
choice is explained by the fact that in this case the inequality
$\mu_{n,N-1}^-<\mu_{n,N-1}^+$ is valid, what in virtue of
(\ref{2.107}) corresponds to the ordering  (\ref{2.106}):
\begin{equation}\label{2.29}
\mu_{n,N-1}^\pm=\frac{M_{n,N}^{+,+}+M_{n,N}^{-,-}\pm
\sqrt{\left(M_{n,N}^{+,+}-M_{n,N}^{-,-}\right)^2+
4\left(M_{n,N}^{+,-}\right)^2}}{2}.
\end{equation}
Bearing in mind (\ref{2.103}) and the boundary value problems
for $\phi_{n,N}^\pm$, we deduce that the functions
$\phi_{n,N}^\pm$ have the form:
\begin{equation}\label{2.31}
\phi_{n,N}^\pm=\t\phi_{n,N}^\pm+\sum\limits_{j=1}^N
c_{n,j}^\pm\t\phi_{n,N-j}^\mp, \quad
\t\phi_{n,N}^\pm={L}_n\left[-a\t\phi_{n,N-1}^\pm+
\sum\limits_{j=1}^{N} \mu_{n,j-1}^\pm \t\phi_{n,N-j}^\pm\right].
\end{equation}

\begin{lemma}\label{lm2.7}
The functions $\phi_{n,i}^\pm$ for $i\geqslant N$ and numbers
$\mu_{n,i}^\pm$ for $i\geqslant N-1$ are determined by the
equalities:
\begin{align}
&\phi_{n,i}^\pm=\t\phi_{n,i}^\pm+ \sum\limits_{j=i-N+1}^i
c^\pm_{n,j} \t\phi_{n,i-j}^\mp,\quad
\mu_{n,i}^\pm=\big(a\t\phi_{n,i}^\pm,
\phi_{n,0}^\pm\big)_{L_2(0,1)}, \label{2.32}
\\
&\t\phi_{n,i}^\pm={L}_n\Bigg[-a\left(\t\phi_{n,i-1}^\pm+
c_{n,i-N}^\pm\t\phi_{n,N-1}^\mp\right)+ \sum\limits_{j=N+1}^i
\mu_{n,j-1}^\pm\phi_{n,i-j}^\pm+\nonumber
\\
&\hphantom{\t\phi_{n,i}^\pm={L}_n\Bigg[} +\sum\limits_{j=1}^N
\mu_{n,j-1}^\pm\bigg(\t\phi_{n,i-j}^\pm+\sum\limits_{p=\max\{i-j-N+1,1\}}^{i-N}
c_{n,p}^\pm\t\phi^\mp_{n,i-j-p}\bigg)\Bigg],\nonumber
\\
&c^\pm_{n,i-N}=\frac{1}{\mu_{n,N-1}^\pm-\mu_{n,N-1}^\mp}\left(
\left(a\t\phi_{n,i-1}^\pm,\phi_{n,0}^\mp\right)_{L_2(0,1)}-
\sum\limits_{j=N+1}^{i-1}\mu_{n,j-1}^\pm
c_{n,i-j}^\pm\right).\nonumber
\end{align}
The functions $\t\phi_{n,i}^\pm$ are orthogonal to
$\phi_{n,0}^+$ and $\phi_{n,0}^-$ in $L_2(0,1)$.
\end{lemma}

\begin{proof}
We prove the lemma by induction. The conclusion of the lemma for
$\phi_{n,N}^\pm$, $\mu^\pm_{n,N-1}$ and $c_{n,0}^\pm$ follows
from the formulas (\ref{2.30}), (\ref{2.31}), if we put
$c_{n,0}^\pm=0$. Suppose the lemma is valid for $\phi_{n,i}^\pm$
and $c_{n,i-N}^\pm$ as $i\leqslant m$ and for $\mu_{n,i}^\pm$ as
$i\leqslant m-1$. We denote by $f_{m+1}^\pm$ the right hand
sides of the equations in (\ref{2.21}) for $i=m+1$. By the
induction assumption these functions are of the form:
\begin{align*}
f^\pm_{m+1}=&-a\left(\t\phi_{n,m}^\pm-\sum\limits_{j=m-N+1}^m
c_{n,j}^\pm\t\phi_{n,m-j}^\mp\right)+\sum\limits_{j=N+1}^{m+1}
\mu_{n,j-1}^\pm\phi_{n,m-j+1}^\pm+
\\
&+\sum\limits_{j=1}^N \mu_{n,j-1}^\pm\left(\t\phi_{n,m-j+1}^\pm+
\sum\limits_{p=\max\{m-j-N+2,1\}}^{m-N}
c_{n,p}^\pm\t\phi_{n,m-j-p+1}^\mp\right)+
\\
&+\sum\limits_{j=m-N+1}^m c_{n,j}^\pm\sum\limits_{p=1}^{m-j+1}
\mu_{n,p-1}^\pm\t\phi^\mp_{n,m-j-p+1}.
\end{align*}
Now we write out the solvability conditions of the boundary
value problems (\ref{2.21}), taking into account the induction
assumption, item~\ref{lm2.6it1} of Lemma~\ref{lm2.6} and the
equalities (\ref{2.103}), (\ref{2.30}), (\ref{2.29}), and
$\mu_{n,i}^+=\mu_{n,i}^-$, $i\leqslant N-2$:
\begin{gather*}
\mu_{n,m}^\pm-
\big(a\t\phi_{n,m}^\pm,\phi_{n,0}^\pm\big)_{L_2(0,1)}=0,
\\
c_{n,m-N+1}^\pm\left(\mu_{n,N-1}^\pm-\mu_{n,N-1}^\mp\right)-
\big(a\t\phi_{n,m}^\pm,\phi_{n,0}^\mp\big)_{L_2(0,1)}+
\sum\limits_{j=N+1}^m \mu_{n,j-1}^\pm c^\pm_{n,m-j+1}=0.
\end{gather*}
The equalities obtained prove the statement of the lemma for
$\mu_{n,m}^\pm$ and $c_{n,N-m+1}^\pm$. The statement of the
lemma for $\phi_{n,m+1}^\pm$ follows from the form of the
functions $f_{m+1}^\pm$ and the definition of the operator
${L}_n$ and the functions $\phi_{n,i}^\pm$, $i\leqslant N-1$.
\end{proof}

Let us prove the formula (\ref{1.6}). We calculate
$M_{n,1}^{\pm,\pm}$:
\begin{align*}
&M_{n,1}^{+,-}=M_{n,1}^{-,+}=2\int\limits_0^1 a(\xi)\sin\pi
n\xi\cos\pi n\xi\di\xi=b_n,
\\
&M_{n,1}^{+,+}=2\int\limits_0^1 a(\xi)\cos^2\pi
n\xi\di\xi=a_n,\quad M_{n,1}^{-,-}=2\int\limits_0^1
a(\xi)\sin^2\pi n\xi\di\xi=-a_n.
\end{align*}
If at least one of the numbers $a_n$, $b_n$ is nonzero, then
$N=1$. In this case the formula (\ref{1.6}) follows immediately
from (\ref{2.29}). If $a_n=b_n=0$, then $N\geqslant 2$ and
$M_{n,1}^{+,+}=M_{n,1}^{-,-}=0$, what by (\ref{2.107}) implies
the formula (\ref{1.6}).

\begin{remark}\label{rm2.2}
Notice, the formula (\ref{1.6}) follows also from the results of
\cite[Ch. XXI, \S 11]{T}, where the spectrum
$-\frac{\displaystyle d^2}{\displaystyle dx^2}+\e q(x)$ was
studied for a periodic function $q(x)$ and the leading terms for
the asymptotics expansions of the edges of the essential
spectrum of this operator were constructed. At the same time, in
the book cited the complete asymptotics expansions were not
constructed and the holomorphic dependence of these edges on
small parameter was not proved as well.
\end{remark}

\sect{Existence and convergence of the discrete {spect\-rum} in
the semi-infinite lacuna}

In this section we study the existence, number, multiplicity,
and convergence of the eigenvalues of the operator $H_\e$
located in the semi-infinite lacuna
$\big(-\infty,\mu_0^+(\e^2)\big)$.

By $\mathfrak{C}$ we denote the set of all finite intervals on
the real axis. We begin with an auxiliary lemma.

\begin{lemma}\label{lm3.2}
Let $\mathfrak{L}$ be an arbitrary compact set in the complex
plane such that for all sufficiently small $\e$ the equalities
\begin{equation}\label{3.1}
\mathfrak{L}\cap \discspec(H_0)=\emptyset,\quad \mathfrak{L}\cap
\essspec(H_\e)=\emptyset
\end{equation}
hold. Then for all sufficiently small $\e$ and
$\l\in\mathfrak{L}$ the resolvent $(H_\e-\l)^{-1}$ is well
defined and is a bounded linear operator from $L_2(\mathbb{R})$
into $\H^2(\mathbb{R})$:
\begin{equation}\label{3.2}
\|(H_\e-\l)^{-1}f\|_{\H^2(\mathbb{R})}\le
C\|f\|_{L_2(\mathbb{R})},
\end{equation}
where the constant $C$ is independent on $\e$, $f\in
L_2(\mathbb{R})$, $\l\in \mathfrak{L}$. For each function $f\in
L_2(\mathbb{R})$ an uniform on $\l\in \mathfrak{L}$ convergence
\begin{equation}\label{3.3}
(H_\e-\l)^{-1}f\to(H_0-\l)^{-1}f,\quad \e\to0,
\end{equation}
holds. The convergence is weak in $\H^2(\mathbb{R})$ and strong
in $\H^1(Q)$ for each $Q\in \mathfrak{C}$.
\end{lemma}

\begin{proof}
First we prove that for all sufficiently small $\e$ for each
function $u\in \H^2(\mathbb{R})$ the inequality
\begin{equation}\label{3.4}
\|u\|_{\H^2(\mathbb{R})}\le C\|f\|_{L_2(\mathbb{R})}
\end{equation}
holds true, where $f=(H_\e-\l)u$, and the constant $C$ is
independent on $u$, $\e$, and $\l\in \mathfrak{L}$. We argue by
contradiction. Suppose that there exist sequences $\e_p\to0$,
$\l_p\in \mathfrak{L}$, $u_p\in \H^2(\mathbb{R})$ such that
\begin{equation}\label{3.5}
\|u_p\|_{\H^2(\mathbb{R})}\geqslant
p\|f_p\|_{L_2(\mathbb{R})},\quad f_p:=(H_{\e_p}-\l_p)u_p.
\end{equation}
Without loss of generality we assume that $\l_p\to\l_*\in
\mathfrak{L}$ and $\|u_p\|_{L_2(\mathbb{R})}=1$. Taking into
account the normalization of the functions $u_p$, we obtain:
\begin{align*}
\big(f_p,u_p\big)_{L_2(\mathbb{R})}&=
\big((H_{\e_p}-\l_p)u_p,u_p\big)_{L_2(\mathbb{R})}=
\\
&= \left\|\frac{du_p}{dx}\right\|^2_{L_2(\mathbb{R})}-\l_p
+\left(\left(V(x)
+a\left(\frac{x}{\e}\right)\right)u_p,u_p\right)_{L_2(\mathbb{R})},
\end{align*}
what implies
\begin{equation*}
\|u_p\|_{\H^2(\mathbb{R})}\leqslant\|f_p\|_{L_2(\mathbb{R})}+
C\|u_p\|_{\H^1(\mathbb{R})}\leqslant
C\left(\|f_p\|_{L_2(\mathbb{R})}+1\right),
\end{equation*}
where the constant $C$ is independent on $p$. The estimate
obtained and (\ref{3.5}) yield the uniform on $p$ inequality
\begin{equation}\label{3.6}
\|u_p\|_{\H^2(\mathbb{R})}\leqslant C.
\end{equation}
Substituting the inequality obtained into (\ref{3.5}), we get:
\begin{equation}\label{3.7}
\|f_p\|_{L_2(\mathbb{R})}\to0.
\end{equation}
It follows from (\ref{3.6}) that the sequence $u_p$ contains a
subsequence (which is denoted by $u_p$ as well) converging to a
function $u_*\in \H^2(\mathbb{R})$ weakly in $\H^2(\mathbb{R})$.
For each interval $Q\in \mathfrak{C}$ the operator of embedding
$\H^2(\mathbb{R})$ into $\H^1(Q)$ is compact. Since the compact
operator maps weakly {con\-ver\-ging} sequence into strongly
{con\-ver\-ging} one and the weak limit is mapped into strong
one (see, for instance, \cite[Ch. VI, \S 1, Theorem 1]{LS}), it
follows that the sequence $u_{p}$ converges to $u_*$ strongly in
$\H^1(Q)$ for each $Q\in\mathfrak{C}$.

Obviously, the functions $u_p$ satisfy the equality $(\t
H_{\e_p}-\l_p)u_p=f_p-V u_p$, what by (\ref{2.1}), (\ref{3.1})
and formula (3.16) from \cite[Ch. 5, \S 3.5]{K} yield
\begin{equation}\label{3.100}
\begin{aligned}
1=&\|u_p\|_{L_2(\mathbb{R})}=\|(\t H_{\e_p}-\l_p)^{-1}(f_p-V
u_p)\|_{L_2(\mathbb{R})}\leqslant \\
&\leqslant\frac{\left(\|f_p\|_{L_2(\mathbb{R})}+
\|Vu_p\|_{L_2(\mathbb{R})}\right)}{\dist(\l_p,\si(\t
H_\e))}\leqslant \frac{\left(\|f_p\|_{L_2(\mathbb{R})}+
\|Vu_p\|_{L_2(\mathbb{R})}\right)}{\dist(\mathfrak{L},\essspec(
H_\e))}.
\end{aligned}
\end{equation}
The last inequality by (\ref{3.1}), (\ref{3.7}) leads us to an
uniform on $p$ estimate
\begin{equation*}
\|Vu_p\|_{L_2(\mathbb{R})}\geqslant C>0,
\end{equation*}
what by $V$ being compactly supported and convergence of $u_p$
to $u_*$ in $\H^1(\supp V)$ implies: $u_*\not\equiv0$.

For a test function $\vs\in C_0^\infty(\mathbb{R})$ the relation
\begin{equation*}
\big((H_{\e_p}-\l_p)u_p,\vs\big)_{L_2(\mathbb{R})}=
(f_p,\vs)_{L_2(\mathbb{R})}
\end{equation*}
gives rise to the equality
\begin{equation*}%\l%abel{3.9}
-\left(\frac{du_p}{dx},\frac{d\vs}{dx}\right)_{L_2(\mathbb{R})}
-\l_p(u_p,\vs)_{L_2(\mathbb{R})}+ (Vu_p,\vs)_{L_2(\mathbb{R})}+
\left(a\left(\frac{x}{\e_p}\right)u_p,\vs\right)_{L_2(\mathbb{R})}
=(f_p,\vs)_{L_2(\mathbb{R})}.
\end{equation*}
By $\vs$ being compactly supported, strong in $\H^1(\supp\vs)$
convergence $u_p\to u_*$, the equality (\ref{1.0}) and Lemma~4.1
from \cite[Ch. V, \S 4]{SP} we deduce that the last summand in
the left hand side of the equality obtained tends to zero as
$\e\to0$. Bearing in mind this fact, (\ref{3.7}) and weak in
$\H^1(\mathbb{R})$ convergence $u_p\to u_*$, we pass to limit in
the last equality as $p\to+\infty$, what results in
\begin{equation*}
-\left(\frac{du_*}{dx},\frac{d\vs}{dx}\right)_{L_2(\mathbb{R})}
-\l_*(u_*,\vs)_{L_2(\mathbb{R})}+
(Vu_*,\vs)_{L_2(\mathbb{R})}=0.
\end{equation*}
The equality obtained implies that  $u_*$ satisfies the equation
$(H_0-\l_*)u_*=0$. Due to the inequality $u_*\not\equiv0$
established above this equation means that $\l_*\in
\mathfrak{L}$ is an eigenvalue of the operator $H_0$, what
contradicts to (\ref{3.1}). The estimate (\ref{3.4}) is proven.

The results of \cite[Ch. VII, \S 7]{LS} and the estimate
(\ref{3.4}) imply the boundedness of the resolvent
$(H_\e-\l)^{-1}: L_2(\mathbb{R})\to \H^2(\mathbb{R})$ as well as
the estimate (\ref{3.2}).

Employing the estimate (\ref{3.2}) instead of  (\ref{3.6}), by
arguments similar to the proof of the estimate (\ref{3.2}) it is
easy to show that for all sequences $\e_p\to0$, $\l_p\to\l_*$,
$\l_p\in \mathfrak{L}$ the function $(H_{\e_p}-\l_p)^{-1}f$
converges to $(H_0-\l_*)^{-1}f$ weakly in $\H^2(\mathbb{R})$ and
strongly in $\H^1(Q)$ for each $Q\in \mathfrak{C}$. Bearing in
mind this convergence and the analyticity of the function
$(H_0-\l)^{-1}f$ on $\l\in\mathfrak{L}$ in the norm of
$\H^2(\mathbb{R})$, by arguing by contradiction one proves the
convergence (\ref{3.3}).
\end{proof}

\begin{remark}\label{rm3.2}
The ideas employed in the proof of Lemma~\ref{lm3.2} are
borrowed from the proof of Theorem~2.1 in the work \cite{G3}.
\end{remark}

We denote $B_\d(\l_0):=\{\l\in \mathbb{C}: |\l-\l_0|<\d\}$.

\begin{lemma}\label{lm3.3}
Let a number $\d_0<0$ be such that
$\discspec(H_0)=\{\l_0^{(-K)},\ldots,\l_0^{(-1)}\}
\subset(-\infty,\d_0]$. Then for all sufficiently small $\e$ a
half-interval $(-\infty,\d_0]$ contains exactly $K$ eigenvalues
$\l_\e^{(n)}$, $n=-K,\ldots,-1$ of the operator $H_\e$. Each of
these eigenvalues is simple and the convergences
$\l_\e^{(n)}\to\l_0^{(n)}$, $n=-K,\ldots,-1$, hold. The
associated orthonormalized in $L_2(\mathbb{R})$ eigenfunctions
can be chosen so that the convergences
$\psi_\e^{(n)}\to\psi_0^{(n)}$, $n=-K,\ldots,-1$, weak in
$\H^2(\mathbb{R})$ and strong in $\H^1(Q)$ for each $Q\in
\mathfrak{C}$ hold.
\end{lemma}

\begin{proof}
According to \cite{RB}, for each value $\e>0$ the number of the
eigenvalues of the operator $H_\e$ located in the semi-infinite
lacuna $(-\infty,\mu_0^+(\e^2))$ is finite. The discrete
spectrum of the operator $H_\e$ is semi-bounded from below:
\begin{equation*}
(-\infty,0)\cap\discspec(H_\e)\subset[c,0],\quad
c=\min\limits_{[0,1]} a(\xi)+\min\limits_{\mathbb{R}}V(x).
\end{equation*}
Let $\t{\mathfrak{L}}$ be some fixed bounded neighbourhood in
the complex plane of the segment $[c,\d_0]$ of the real axis and
$\t{\mathfrak{L}}\cap\essspec{H_\e}=\emptyset$ for all
sufficiently small $\e$. We denote by $\mathfrak{L}$ the closure
of the set
$\widetilde{\mathfrak{L}}\setminus\bigcup\limits_{n=-K}^{-1}
B_\d(\l_0^{(n)})$, where $\d$ is an arbitrary small number. Then
the compact set $\mathfrak{L}$ obeys the hypothesis of
Lemma~\ref{lm3.2},  and, therefore, for all sufficiently small
$\e$ the resolvent $(H_\e-\l)^{-1}$ is bounded uniformly on
$\l\in \mathfrak{L}$, this is why the set $\mathfrak{L}$
contains no eigenvalues of the operator $H_\e$ for all
sufficiently small $\e$. Since $\d$ is arbitrary, it follows
that the eigenvalues of the operator $H_\e$ located in
$(-\infty,\d_0]$ converge to the eigenvalues of the operator
$H_0$.

We fix $n$ and choose $\d$ so that the equality
$\overline{B_\d(\l_0^{(n)})}\cap \si(H_0)=\{\l_0^{(n)}\}$ is
true. Then due to (\ref{3.3}) with $f=\psi_0^{(n)}$ the
convergence follows
\begin{equation}\label{3.9a}
\frac{1}{2\pi\iu}\int\limits_{\partial B_\d(\l_0^{(n)})}
(H_\e-\l)^{-1}\psi_0^{(n)}\to-\psi_0^{(n)},\quad \e\to0,
\end{equation}
which is weak in $\H^2(\mathbb{R})$ and strong in $\H^1(Q)$ for
each $Q\in \mathfrak{C}$. According to \cite[Ch. V, \S 3.5]{K},
the isolated eigenvalues of the operator $H_\e$ are simple poles
for the resolvent $(H_\e-\l)^{-1}: L_2(\mathbb{R})\to
L_2(\mathbb{R})$. Therefore, being considered as an operator
from $L_2(\mathbb{R})$ into $L_2(Q)$, where $Q\in \mathfrak{C}$,
the resolvent $(H_\e-\l)^{-1}$ has simple poles as well. Taking
into account that the right hand side (\ref{3.9a}) is nonzero,
we conclude, that a disk $B_\d(\l_0^{(n)})$ contains at least
one eigenvalue of the operator $H_\e$ for all sufficiently small
$\e$. This is why at least one eigenvalue of the operator $H_\e$
converges to each eigenvalue of the operator $H_0$.

Let $\l_{\e,j}$, $j=1,\ldots,m_n(\e)$ be the eigenvalues of the
operator $H_\e$, converging to $\l_0^{(n)}$ as $\e\to0$. Assume
that on some sequence $\e_p\to0$ at least two eigenvalues
$\l_{\e,1}$ and $\l_{\e,2}$ taken accounting multiplicity
converge to the eigenvalue $\l_0^{(n)}$. We indicate by
$\psi_{\e,1}$ and $\psi_{\e,2}$ the associated eigenfunctions
orthonormalized in $L_2(\mathbb{R})$. Similarly to the proof of
Lemma~\ref{lm3.2} it is easy to show that selecting a
subsequence of $\{\e_p\}$, if needed, the functions
$\psi_{\e_p,i}$ can be assumed to converge to $\psi_{0,i}$
weakly in  $\H^2(\mathbb{R})$ and strongly in $\H^1(Q)$ for each
$Q\in \mathfrak{C}$, where $\psi_{0,i}$ are eigenfunctions
associated with $\l_0^{(n)}$, and $\psi_{0,i}\not\equiv0$. Since
$\l_0^{(n)}$ is a simple eigenvalue, it follows that
$\psi_{0,i}=c_i\psi_0^{(n)}$.The function
$\t\psi_{\e_p}:=c_2\psi_{\e_p,1}-c_1\psi_{\e_p,2}$, obviously,
converges to zero strongly in $\H^1(Q)$ for each $Q\in
\mathfrak{C}$. On the other hand, the function $\t\psi_{\e_p}$
satisfies the equation
\begin{equation*}
(\t
H_{\e_p}-\l_0^{(n)})\t\psi_{\e_p}=c_2(\l_{\e_p,1}-\l_0^{(n)})
\psi_{\e_p,1}-
c_1(\l_{\e_p,2}-\l_0^{(n)})\psi_{\e_p,2}-V\t\psi_{\e_p},
\end{equation*}
what by analogy with (\ref{3.100}) implies:
\begin{equation*}
0<c_1^2+c_2^2\leqslant\frac{|c_2||\l_{\e_p,1}-\l_0^{(n)}|+
|c_1||\l_{\e_p,2}-\l_0^{(n)}|+\|V\t\psi_{\e_p}\|_{L_2(\mathbb{R})}}
{\mu_0^+(\e^2)-\l_0^{(n)}}.
\end{equation*}
The last inequality contradicts to the convergence
$\t\psi_{\e_p}\to0$ in $L_2(\supp V)$. Therefore, only one
eigenvalue of the operator $H_\e$ converges to the eigenvalue
$\l_0^{(n)}$ and this eigenvalue of the operator $H_\e$ is
simple.

Let us prove the convergence for the eigenfunctions. According
to \cite[Ch. V, \S 3.5]{K}, the convergence
$\l_\e^{(n)}\to\l_0^{(n)}$ implies that for all sufficiently
small $\e$ the equality
\begin{equation*}
\frac{1}{2\pi\iu}\int\limits_{\partial B_\d(\l_0^{(n)})}
\left(H_\e-\l\right)^{-1}\psi_0^{(n)}\di\l=
-\big(\psi_0^{(n)},\psi_\e^{(n)}\big)_{L_2(\mathbb{R})}
\psi_\e^{(n)},
\end{equation*}
holds true, where $\psi_\e^{(n)}$ is the normalized in
$L_2(\mathbb{R})$ eigenfunction associated with $\l_\e^{(n)}$,
and $\d$ is the same as in (\ref{3.9a}). By (\ref{3.9a}) it
follows the convergence
\begin{equation*}
\big(\psi_0^{(n)},\psi_\e^{(n)}\big)_{L_2(\mathbb{R})}
\psi_\e^{(n)}\to\psi_0^{(n)},\quad \e\to0,
\end{equation*}
which is weak in $\H^2(\mathbb{R})$ and strong in  $\H^1(Q)$ for
each $Q\in \mathfrak{C}$. A weak  $\H^2(\mathbb{R})$ convergence
yielding a weak $L_2(\mathbb{R})$ convergence in, we multiply
the last convergence by $\psi_0^{(n)}$ in $L_2(\mathbb{R})$ and
get:
\begin{equation*}
\big(\psi_0^{(n)},\psi_\e^{(n)}\big)_{L_2(\mathbb{R})}^2 \to
1,\quad \e\to0.
\end{equation*}
The last two convergences imply that the eigenfunction
$\psi_\e^{(n)}\sgn\big(\psi_0^{(n)},\psi_\e^{(n)}\big)_{L_2(\mathbb{R})}$
meets the conclusion of the lemma.
\end{proof}

In constructing the asymptotics expansions for the eigenvalues
$\l_\e^{(n)}$, $n=-K,\ldots,-1$, we will employ an auxiliary
statement, and it is convenient to formulate it in the present
section.

\begin{lemma}\label{lm3.4}
For $\l$ close to $\l_0^{(n)}$ each function $f\in
L_2(\mathbb{R})$ meets the representation:
\begin{equation}\label{3.10}
(H_\e-\l)^{-1}f=-\frac{\psi_\e^{(n)}}{\l-\l_\e^{(n)}}
\left(f,\psi_\e^{(n)}\right)_{L_2(\mathbb{R})}+\t u(x,\l,\e),
\end{equation}
where the function $\t u(x,\l,\e)$ satisfies the estimate
\begin{equation}\label{3.11}
\|\t u\|_{\H^2(\mathbb{R})}\leqslant C\|f\|_{L_2(\mathbb{R})}
\end{equation}
with the constant $C$ independent on $\e$, $\l$, and $f$.
\end{lemma}

\begin{proof}
The representation (\ref{3.10}) with the function $\t u$
holomorphic on $\l$ in the norm of $L_2(\mathbb{R})$ follows
from \cite[Ch. V, \S 3.5]{K}. Let $\d$ be the same as in
(\ref{3.9a}). The function $\t u$ takes the form
\begin{equation*}
\t u=(H_\e-\l)^{-1}\t f,\quad \t
f=f-(f,\psi_\e^{(n)})_{L_2(\mathbb{R})}\psi_\e^{(n)}.
\end{equation*}
By the holomorphy of $\t u$ on $\l$ in the norm of
$L_2(\mathbb{R})$ it follows the holomorphy of $\t u$ in the
norm of $\H^2(\mathbb{R})$. A compact set $\partial B_\d(\l_0)$
obeys the hypothesis of Lemma~\ref{lm3.2}, this is why the
representation given for $\t u$ implies the estimate
(\ref{3.11}) uniform on $\l\in
\partial B_\d(\l_0)$. By the maximum principle for holomorphic
functions the function $\t u$ satisfies the estimate
(\ref{3.11}) for $\l\in B_\d(\l_0)$ as well.
\end{proof}

Besides the eigenvalues $\l_\e^{(n)}$, $n=-K,\ldots,-1$, in the
semi-infinite lacuna $(-\infty,\mu_0^+(\e^2))$ the operator
$H_\e$ can also have the eigenvalues not converging to the
eigenvalues of $H_0$. As it follows from Lemma~\ref{lm3.3}, such
eigenvalues must converge to zero as $\e\to0$. The remaining
part of this section is devoted to the proof of the fact that
the operator $H_\e$ can have at most one such eigenvalue and, if
exists, such eigenvalue is simple. We will also show that the
necessary condition for such eigenvalue to exist is the
existence of the nontrivial solution to the problem (\ref{1.7}).

First we prove the following auxiliary lemmas.

\begin{lemma}\label{lm3.5}
The solutions $\vp_{i}=\vp_{i}(\xi,\e^2)$ of the equation
(\ref{2.6}) with $M=\e^2\mu_0^+(\e^2)$ subject to the initial
conditions
\begin{equation*}
\vp_{1}(0,\e^2)=1,\quad \vp'_{1}(0,\e^2)=0,\quad
\vp_{2}(0,\e^2)=0,\quad \vp'_{2}(0,\e^2)=1,
\end{equation*}
are of the form
\begin{equation*}%\l%abel{3.13}
\vp_{1}=1+\sum\limits_{j=1}^\infty \e^{2j}
{P}^j(\e^2)[1],\quad\vp_{2}=\xi+\sum\limits_{j=1}^\infty \e^{2j}
{P}^j(\e^2)[\xi]
\end{equation*}
for all sufficiently small $\e$. Here the operator ${P}(\e^2)$
is defined by the equality
\begin{equation*}
{P}(\e^2)[f](\xi,\e^2):=\int\limits_0^\xi(\xi-\eta)
(a(\eta)-\mu_0^+(\e^2))f(\eta)\di\eta
\end{equation*}
and is a linear bounded operator from $C[0,1]$ into $C^2[0,1]$
holomorphic on $\e^2$. The functions $\vp_{i}$ are holomorphic
on $\e^2$ in the norm of $C^2[0,1]$.
\end{lemma}

\begin{remark}\label{rm3.3}
The notations ${P}^j(\e^2)[1]$ and ${P}^j(\e^2)[\xi]$ in the
lemma mean the applying of the $j$-th power of the operator
${P}(\e^2)$ to the functions $f(\xi)\equiv 1$ and
$f(\xi)\equiv\xi$.
\end{remark}

\begin{proof}
The maintained properties of the operator ${P}$ are obvious. The
Cauchy problems for the functions $\vp_{i}$ are easily reduced
to the integral equations
\begin{equation*}
\vp_{1}-\e^2{P}(\e^2)[\vp_{1}]=1,\quad
\vp_{2}-\e^2{P}(\e^2)[\vp_{2}]=\xi,
\end{equation*}
what implies the required series for the functions $\vp_{i}$ and
the holomorphy of this functions on $\e^2$.
\end{proof}

\begin{lemma}\label{lm3.6}
The solutions $\Th_{i}=\Th_{i}(\xi,\e^2,k^2)$ of the equation
(\ref{2.6}) with $M=\e^2(\mu_0^+(\e^2)-k^2)$, where $k$ is a
small complex parameter, subject to the initial conditions
\begin{equation*}
\Th_{1}(0,\e^2,k^2)=1,\quad \Th'_{1}(0,\e^2,k^2)=0,\quad
\Th_{2}(0,\e^2,k^2)=0,\quad \Th'_{2}(0,\e^2,k^2)=1,
\end{equation*}
are of the form
\begin{equation*}
\Th_{i}=\vp_{i}+\sum\limits_{j=1}^\infty \e^{2j} k^{2j}
\widetilde{{P}}^j(\e^2)[\vp_{i}],
\end{equation*}
for all sufficiently small $\e$. Here the operator
$\widetilde{{P}}(\e^2)$ is defined by the equality
\begin{equation*}
\widetilde{{P}}(\e^2)[f](\xi,\e^2):=\int\limits_0^\xi
\left(\vp_{2}(\xi,\e^2)\vp_{1}(\eta,\e^2)-
\vp_{2}(\eta,\e^2)\vp_{1}(\xi,\e^2)\right)f(\eta)\di\eta
\end{equation*}
and is a linear bounded operator from $C[0,1]$ into $C^2[0,1]$
holomorphic on $\e^2$. The functions $\Th_{i}$ are holomorphic
on $\e^2$ in the norm of $C^2[0,1]$.
\end{lemma}

\begin{proof}
The boundedness and holomorphy of the operator $\widetilde{{P}}$
follows from Lemma~\ref{lm3.5}. Reducing the equation
(\ref{2.6}) to an integral one $\Th_{i}-\e^2 k^2
\widetilde{{P}}(\e^2)[\Th_{i}]=\vp_{i}$, one can easily deduce
the statement of the lemma on the functions $\Th_{i}$.
\end{proof}

The equation (\ref{2.6}) having 1-periodic coefficients, the
Floquet theory is applicable to it. Let us calculate the
multipliers corresponding to this equation with
$M=\e^2\left(\mu_0^+(\e^2)-k^2\right)$. We introduce the
function
\begin{equation*}
{D}={D}(\e^2,k^2):=\Th_{1}(1,\e^2,k^2)+ \Th'_{2}(1,\e^2,k^2).
\end{equation*}
Due to the formulas (8.13) from \cite[Ch. 2, \S 2.8]{Sh} the
multipliers $\k_\pm=\k_\pm(\e,k)$ are of the form
\begin{equation}\label{3.14}
\k_\pm=\k^{\pm1},\quad \k=\frac{{D}+\sqrt{{D}^2-4}}{2}.
\end{equation}
Since $\mu_0^+$ is an edge of the essential spectrum of the
operator $H_\e$ and the associated solution $\phi_0^+$ of the
equation (\ref{2.6}) satisfies the periodic boundary condition
in (\ref{2.7}), in accordance with \cite[Ch. 2, \S 2.8]{Sh} the
identity
\begin{equation}\label{3.14e}
\vp_{1}(1,\e^2)+\vp'_{2}(1,\e^2)\equiv 2
\end{equation}
holds. Taking into account this identity, by
Lemmas~\ref{lm3.5},~\ref{lm3.6} we obtain
\begin{align*}
&{D}(\e^2,k^2)=2+\e^2 k^2\t D(\e^2)+\e^4 k^2
\widehat{D}(\e^2,k^2),
\\
&\t D(\e^2):=\left(\t P(\e^2)[\vp_1](\xi,\e^2)+\frac{d}{d\xi}\t
P(\e^2)[\vp_2](\xi,\e^2)\right)\Bigg|_{\xi=1}=1+\Odr(\e^2),
\end{align*}
where  $\widetilde{{D}}$, $\widehat{D}$ are holomorphic
functions. Substituting these equalities into (\ref{3.14}), we
get
\begin{equation}\label{3.14a}
\k(\e,k)=1+\e k+\e^3k\t\k^{(1)}(\e,k)+\e^2k^2\t\k^{(2)}(\e,k),
\end{equation}
where $\t\k^{(i)}(\e,k)$ are holomorphic on $\e$ and $k$
functions. By Floquet-Lyapunov theorem the equation (\ref{2.6})
with $M=\e^2(\mu_0^+(\e^2)-k^2)$ has the solutions of the form
\begin{equation}\label{3.14b}
\Th^\pm=\Th^\pm(\xi,\e,k)=\E^{\mp\xi\ln\k(\e,k)}
\Th_{per}^\pm(\xi,\e,k),
\end{equation}
where $\Th_{per}^\pm$ are 1-periodic on $\xi$ functions. We
indicate by $W_t(f(t),g(t))$ the wronskian of functions $f(t)$
and $g(t)$. Bearing in mind the equalities
\begin{equation}
\begin{aligned}
&\Th_{i}(\xi+m,\e^2,k^2)=\Th_{i}(m,\e^2,k^2)
\Th_{1}(\xi,\e^2,k^2)+ \Th'_{i}(m,\e^2,k^2)
\Th_{2}(\xi,\e^2,k^2),\quad m\in \mathbb{Z},
\\
&W_\xi\left(\Th_{1}(\xi,\e^2,k^2),\Th_{2}(\xi,\e^2,k^2)\right)
\equiv1,
\\ &
\k(\e,k)+\k^{-1}(\e,k)=\Th_{1}(1,\e^2,k^2)+\Th'_{2}(1,\e^2,k^2),
\end{aligned}\label{3.16a}
\end{equation}
it is not difficult to check that the functions $\Th^\pm$ can be
chosen as follows:
\begin{equation}\label{3.15}
\Th^\pm(\xi,\e,k)=\Th_{2}(1,\e^2,k^2)\Th_{1}(\xi,\e^2,k^2)+
\left(
\k^{\mp1}(\e,k)-\Th_{1}(1,\e^2,k^2)\right)\Th_{2}(\xi,\e^2,k^2).
\end{equation}
By  (\ref{3.14a}), (\ref{3.14b}), (\ref{3.15}), and
Lemmas~\ref{lm3.5},~\ref{lm3.6} the functions
$\Th^\pm_{per}(\xi,\e,k)$ satisfy the equalities
\begin{equation}\label{3.22}
\Th_{per}^\pm(\xi,\e,k)=1+\e^3k\Th_{per}^{\pm,1}(\xi,\e,k)+\e^2
k^2\Th_{per}^{\pm,2}(\xi,\e,k),
\end{equation}
where $\Th_{per}^{\pm,i}$ are 1-periodic on $\xi$ functions
holomorphic on $\e$ and $k$ in the norm of $C^2[0,1]$.

Let us consider the equation
\begin{equation}\label{3.16}
\left(-\frac{d^2}{dx^2}+V(x)+a\left(\frac{x}{\e}\right)-\l
\right)u=f,\quad
x\in\mathbb{R},
\end{equation}
where $f\in L_2(\mathbb{R})$ is a compactly supported function,
and $\supp f\subseteq[-x_0,x_0]$. We choose the parameter $\l$
in this equation as $\l=\mu_0^+(\e^2)-k^2$, where $k$ is a small
complex parameter. We seek the solutions of the equation
(\ref{3.16}) with such $\l$ in the class $\Hloc^2(\mathbb{R})$
and impose the constraint
\begin{equation}\label{3.17}
u(x,\e,k)=c_\pm(\e,k)
\Th^\pm\left(\frac{x}{\e},\e,k\right),\quad \pm x\geqslant x_0.
\end{equation}
These equalities can be replaced by the boundary conditions
\begin{equation}\label{3.18}
\frac{d}{dx}\left(\frac{u}{\Th^\pm
\left(\frac{x}{\e},\e,k\right)}\right)=0,\quad x=\pm x_0.
\end{equation}
Indeed, the solution of the problem (\ref{3.16}), (\ref{3.17})
with $\l=\mu_0^+(\e^2)-k^2$ satisfies, obviously, the boundary
value problem (\ref{3.16}), (\ref{3.18}). The solution of the
problem (\ref{3.16}), (\ref{3.18}) with $\l=\mu_0^+(\e^2)-k^2$
extended by the equality
\begin{equation}\label{3.18a}
u(x,\e,k):=\frac{u(\pm x_0,\e,k)}
{\Th^\pm\left(\pm\frac{x_0}{\e},\e,k\right)}
\Th^\pm\left(\frac{x}{\e},\e,k\right),\quad \pm x\geqslant x_0,
\end{equation}
is a solution to the problem (\ref{3.16}), (\ref{3.17}). We note
that the relations (\ref{3.14a}), (\ref{3.14b}), (\ref{3.22})
imply that $\Th^\pm\left(\pm\frac{x_0}{\e},k,\e\right)\not=0$.

In the problem (\ref{3.16}), (\ref{3.18}) we make a substitution
\begin{equation}\label{3.18b}
u(x,\e,k)=U(x,\e,k){\vp} \left(\frac{x}{\e},\e^2\right),
\end{equation}
where ${\vp}(\xi,\e^2):=\vp_{2}(1,\e^2)\vp_{1}(\xi,\e^2)+
(1-\vp_{1}(1,\e^2))\vp_{2}(\xi,\e^2)$ is 1-periodic on $\xi$
function. This substitution is well defined, since due to
Lemma~\ref{lm3.5} the function ${\vp}$ is holomorphic on $\e^2$
in the norm of  $C^2[0,1]$, and ${\vp}(\xi,0)\equiv1$. As a
result of the substitution we arrive at the following boundary
value problem for $U$:
\begin{equation}\label{3.20}
\begin{gathered}
\left(-\frac{d^2}{dx^2}-2\e^{-1}
\frac{\vp'\left(\frac{x}{\e},\e^2\right)}
{{\vp}\left(\frac{x}{\e},\e^2\right)}
\frac{d}{dx}+V(x)+k^2\right)U=F,\quad x\in(-x_0,x_0),
\\
\frac{dU}{dx}-U\frac{d}{dx}
\ln\frac{{\vp}\left(\frac{x}{\e},\e^2\right)}
{\Th^\pm\left(\frac{x}{\e},\e,k\right)}=0,\quad x=\pm x_0,
\end{gathered}
\end{equation}
where $F=F(x,\e)=f(x)/{\vp}\left(\frac{x}{\e},\e^2\right)$. Let
us study the solvability of the problem (\ref{3.20}). In order
to do this we will employ the following
\begin{lemma}\label{lm3.7}
The statements
\begin{enumerate}
\def\theenumi{(\arabic{enumi})}
\item\label{it1lm3.7} For all sufficiently small $\e$
an uniform on $\e$ and $x$ estimate
\begin{equation*}
\left|\frac{\vp'\left(\frac{x}{\e},\e^2\right)}
{{\vp}\left(\frac{x}{\e},\e^2\right)}\right|\le C\e^2
\end{equation*}
holds.

\item\label{it2lm3.7} The functions
\begin{equation*}
\rho_\pm=\rho_\pm(\e,k):=-\frac{d}{dx}\ln
\frac{{\vp}\left(\frac{x}{\e},\e^2\right)}{\Th^\pm
\left(\frac{x}{\e},\e,k\right)}\Bigg|_{x=\pm x_0}
\end{equation*}
can be represented as $\rho_\pm(\e,k)=\mp k+\e\t\rho_\pm(\e,k)$,
where $\t\rho_\pm(\e,k)$ are holomorphic on $k$ functions
bounded uniformly on $\e$ and $k$ together with their
derivatives $\frac{\displaystyle d\t\rho_\pm}{\displaystyle
dk}$.
\end{enumerate}
are valid.
\end{lemma}

\begin{proof}
Item~\ref{it1lm3.7} follows directly from the definition of the
function $\vp$ and Lemma~\ref{lm3.5}. From (\ref{3.14b}) we
deduce
\begin{equation*}
\frac{d}{dx}\ln\Th^\pm(\xi,\e,k)=\mp\frac{\ln\k(\e,k)}{\e}+
\frac{d}{dx}\ln\Th_{per}^\pm\left(\frac{x}{\e},\e,k\right),
\end{equation*}
what by (\ref{3.14a}), (\ref{3.22}) and item~\ref{it1lm3.7}
implies item~\ref{it2lm3.7}.
\end{proof}

We will treat a solution of the problem (\ref{3.20}) in a
generalized sense, namely, as a solution to the integral
equation
\begin{equation}\label{3.23}
\begin{aligned}
&\left(\frac{dU}{dx},\frac{d\vs}{dx}\right)_{L_2(-x_0,x_0)}
-2\e^{-1} \left(\frac{\vp'\left(\frac{x}{\e},\e^2\right)}
{{\vp}\left(\frac{x}{\e},\e^2\right)}
\frac{dU}{dx},\vs\right)_{L_2(-x_0,x_0)}+
\\
&+\left(VU,\vs\right)_{L_2(-x_0,x_0)}+
k^2\left(U,\vs\right)_{L_2(-x_0,x_0)}+
\rho_+(x_0)U(x_0,\e,k)\overline{\vs(x_0)}-
\\
&-\rho_-(-x_0)U(-x_0,\e,k)\overline{\vs(-x_0)}=
\left(f,\vs\right)_{L_2(-x_0,x_0)}
\end{aligned}
\end{equation}
for each $\vs\in \H^1(-x_0,x_0)$. The line in this equation
indicates the complex conjugation. Due to Lemma~$1'$ from
\cite[Ch. IV, \S 1]{M} there exists a linear bounded operator
$T_1: L_2(-x_0,x_0)\to \H^1(-x_0,x_0)$ such that for each
$\vs\in\H^1(-x_0,x_0)$ the equality
\begin{equation}\label{3.24}
(v,\vs)_{L_2(-x_0,x_0)}=(T_1v,\vs)_{\H^1(-x_0,x_0)}
\end{equation}
holds. The restriction of the operator $T_1$ on $\H^1(-x_0,x_0)$
is compact and self-adjoint.

\begin{lemma}\label{lm3.8}
There exist linear bounded operators $T_2(\e)$, $T_3^\pm:
\H^1(-x_0,x_0)\to \H^1(-x_0,x_0)$ such that the operator
$T_2(\e)$ is bounded uniformly on $\e$, the operators $T_3^\pm$
are compact, and the equalities
\begin{gather*}
\e^{-2}\left(\frac{\vp'\left(\frac{x}{\e},\e^2\right)}
{{\vp}\left(\frac{x}{\e},\e^2\right)}
\frac{dU}{dx},\vs\right)_{L_2(-x_0,x_0)}=
(T_2(\e)U,\vs)_{\H^1(-x_0,x_0)},
\\
U(\pm x_0)\overline{\vs(\pm x_0)}=(T_3^\pm U,
\vs)_{\H^1(-x_0,x_0)}
\end{gather*}
hold.
\end{lemma}

\begin{proof}
We set
\begin{equation*}
T_2(\e)U:=\e^{-2}T_1\frac{\vp'\left(\frac{x}{\e},\e^2\right)}
{\vp\left(\frac{x}{\e},\e^2\right)}\frac{dU}{dx}, \quad T_3^\pm
U:=\frac{U(\pm x_0)\cosh(x\pm x_0)}{\sinh 2 x_0}.
\end{equation*}
The conclusion of the lemma on the operator $T_2(\e)$ follows
from item~\ref{it1lm3.7} of Lemma~\ref{lm3.7}, (\ref{3.24}), and
the boundedness of the operator $T_1$. The maintained properties
of the operators $T_3^\pm$ are checked by direct calculations.
\end{proof}

Due to (\ref{3.24}) and the proven lemma the equation
(\ref{3.23}) is equivalent to an operator equation in
$\H^1(-x_0,x_0)$:
\begin{equation}\label{3.25}
(\I-2\e T_2(\e)+T_1V+(k^2-1)T_1+\rho_+(\e,k)T_3^+-
\rho_-(\e,k)T_3^-)U=T_1F,
\end{equation}
where $\I$ is the identity mapping, and $V$ is the operator of
multiplication by the function $V(x)$. First we study the
solvability of this equation for $\e=0$ (we put $T_2(0):=0$).
Denote $T_4(k):=T_1V+(k^2-1)T_1-k(T_3^-+T_3^+)$.

\begin{lemma}\label{lm3.9}
If there exists no nontrivial solution of the problem
(\ref{1.7}), then for all sufficiently small $k$ the operator
$(\I+T_4(k))$ has an inverse operator bounded uniformly on $k$.
If there exists the nontrivial solution of the problem
(\ref{1.7}), then for all sufficiently small $k$ the equality
\begin{equation*}
(\I+T_4(k))^{-1}g=-\frac{\psi_0^{(0)}}{k}
\big(g,\psi_0^{(0)}\big)_{\H^1(-x_0,x_0)} +T_5(k)g
\end{equation*}
holds true, where $g\in \H^1(-x_0,x_0)$, $T_5(k):
\H^1(-x_0,x_0)\to\H^1(-x_0,x_0)$ is a linear bounded operator
holomorphic on $k$, and the function $\psi_0^{(0)}$ obeys the
normalization condition (\ref{1.12}).
\end{lemma}

\begin{proof}
Let the problem (\ref{1.7}) have no nontrivial solution. The
operator $T_4(0)$ being compact, due to Fredholm theorems the
invertibility of the operator $(\I+T_4(0))$ is equivalent to the
absence of nontrivial solutions to the equation
\begin{equation}\label{3.25a}
(\I+T_4(0))U=0.
\end{equation}
It is easy to check that each solution of such equation is a
generalized solution to a boundary value problem
\begin{equation*}
\left(-\frac{d^2}{dx^2}+V\right)U=0,\quad x\in(-x_0,x_0),\qquad
\frac{dU}{dx}=0,\quad x=\pm x_0.
\end{equation*}
This problem is equivalent to the problem (\ref{1.7}), which has
no nontrivial solution. Therefore, the operator $(\I+T_4(0))$
has a bounded inverse operator, what implies that for all
sufficiently small $k$  the operator $(\I+T_4(k))$ has an
inverse operator bounded uniformly on $k$.

Assume that the problem (\ref{1.7}) has a nontrivial solution.
The operator equation $(\I+T_4(k))U=0$ describes generalized
solutions to the following boundary value problem:
\begin{equation*}
\left(-\frac{d^2}{dx^2}+V-k^2\right)U=0,\quad x\in(-x_0,x_0),
\qquad \left(\frac{d}{dx}\pm k\right)U=0,\quad x=\pm x_0.
\end{equation*}
For $k>0$ nontrivial solutions $U$ of this problem extended by
the rule $U(x,k):=U(\pm x_0,k)\E^{\mp k(x\mp x_0)}$, $\pm
x\geqslant x_0$, are eigenfunctions of the operator  $H_0$
associated with eigenvalues $\l=-k^2$. The negative spectrum of
the operator $H_0$ being discrete, there exists a number $k_*>0$
such that $\l=-k_*^2$ is not an eigenvalue of the operator
$H_0$, i.e., the operator $(\I+T_4(k_*))^{-1}$ is invertible.
Bearing in mind this fact  as well as the compactness of the
operator $T_4(k)$, in virtue of Theorem~7.1 from \cite[Ch. XV,
\S 7]{SP} the operator $(\I+T_4(k))^{-1}$ is meromorphic on $k$.
The function $\psi_0^{(0)}$ is a solution of the equation
(\ref{3.25a}), what implies that the operator $(\I+T_4(k))^{-1}$
has a pole at zero:
\begin{equation*}%\l%abel{3.26}
U=(\I+T_4(k))^{-1}g=\frac{U_0}{k^m}+k^{-m+1}T_5(k)g,
\end{equation*}
where $m\geqslant1$ is the order of the pole, $T_5(k):
\H^1(-x_0,x_0)\to \H^1(-x_0,x_0)$ is a bounded linear operator
holomorphic on $k$. This equality and the definition of the
operator $T_4(k)$ yield that
\begin{equation}\label{3.27}
g=\frac{(\I+T_4(0))U_0}{k^m}+\frac{
(\I+T_4(0))T_5(0)g-(T_3^++T_3^-)U_0}{k^{m-1}}+\Odr(k^{-m+2}),
\end{equation}
i.e., $U_0$ is a solution of the equation (\ref{3.25a}):
$U_0=C(g)\psi_0^{(0)}$, where $C(g)$ is a some linear nonzero
functional. Let us multiply the equation (\ref{3.27}) by
$\psi_0^{(0)}$ in $\H^1(-x_0,x_0)$ and take into account the
definition of the operators $T_3^\pm$ and the normalization
condition (\ref{1.12}). This procedure results in
\begin{equation*}
\big(g,\psi_0^{(0)}\big)_{\H^1(-x_0,x_0)}=\frac{
\big((\I+T_4(0))T_5(0)g, \psi_0^{(0)}\big)_{\H^1(-x_0,x_0)}
-C(g)}{k^{m-1}}+\Odr(k^{-m+2}).
\end{equation*}
Employing the definition of the operator $T_1$, it is not
difficult to check that the operator $T_4(0)$ is self-adjoint,
what implies that
\begin{equation*}
\big((\I+T_4(0))T_5(0)g, \psi_0^{(0)}\big)_{\H^1(-x_0,x_0)}=
\big(T_5(0)g, (\I+T_4(0))\psi_0^{(0)}\big)_{\H^1(-x_0,x_0)}=0.
\end{equation*}
Since $C(g)$ is not identically zero, it follows from last two
equalities that $m=1$ and
$C(g)=-\big(g,\psi_0^{(0)}\big)_{\H^1(-x_0,x_0)}$.
\end{proof}

\begin{lemma}\label{lm3.12}
Let there exists no nontrivial solution of the problem
(\ref{1.7}). Then the operator $H_\e$ has no eigenvalues tending
to zero as $\e\to0$.
\end{lemma}

\begin{proof}
It follows from Lemmas~\ref{lm3.8},~\ref{lm3.9} and
item~\ref{it2lm3.7} of Lemma~\ref{lm3.7} that the operator in
the right hand side of the equation (\ref{3.25}) is boundedly
invertible, i.e., the equation (\ref{3.25}) with $F=0$ has no
nontrivial solutions for all sufficiently small $\e$ and $k$.
Thus, the problem (\ref{3.20}), and, therefore, the problem
(\ref{3.16}), (\ref{3.17}) have no nontrivial solution as $F=0$
for all sufficiently small $\e$ and $k$. This implies the
absence of small eigenvalues of the operator $H_\e$.
\end{proof}

We proceed to the case of presence of the nontrivial solution to
the problem (\ref{1.7}). This solution is assumed to meet the
equality (\ref{1.12}). In studying this case we will employ the
approach suggested in \cite{G1} (see also \cite{BEG},
\cite{BE}).

Applying the operator $(\I+T_4(k))^{-1}$ to the equation
(\ref{3.25}), by Lemma~\ref{lm3.9} and item~\ref{it2lm3.7} we
obtain:
\begin{equation}\label{3.29}
\begin{aligned}
U-\frac{\e}{k}\psi_0^{(0)}&
\big(T_6(\e,k)U,\psi_0^{(0)}\big)_{\H^1(-x_0,x_0)} +\e
T_5(k)T_6(\e,k)U=\\ &=-\frac{1}{k}\psi_0^{(0)}
\big(T_1F,\psi_0^{(0)}\big)_{\H^1(-x_0,x_0)}+T_5(k)T_1F,
\end{aligned}
\end{equation}
where
$T_6(\e,k):=-2T_2(\e)+\t\rho_+(\e,k)T_3^+-\t\rho_-(\e,k)T_3^-$.
Item~\ref{it2lm3.7} of Lemma~\ref{lm3.7} and Lemma~\ref{lm3.8}
imply the uniform on $\e$ and  $k$ boundedness of the operator
$T_6(\e,k)$. In view of holomorphy of the operator $T_5(k)$  the
bounded invertibility of the operator $(\I+\e T_5(k)T_6(\e,k))$
follows for all sufficiently small $\e$ and $k$. We denote
$T_7(\e,k):=(\I+\e T_5(k)T_6(\e,k))^{-1}$. Clearly, the uniform
on $k$ convergence
\begin{equation}\label{3.29a}
T_7(\e,k)\xrightarrow[\e\to0]{}\I
\end{equation}
holds true. Applying $T_7(\e,k)$ to the equation (\ref{3.29})
and bearing in mind the equality (\ref{3.24}), we get
\begin{equation}\label{3.30}
\begin{aligned}
&U-\frac{\e}{k}\big(T_6(\e,k)U,\psi_0^{(0)}\big)_{\H^1(-x_0,x_0)}
T_7(\e,k)\psi_0^{(0)}=
\\
&= -\frac{1}{k}\big(F,\psi_0^{(0)}\big)_{L_2(-x_0,x_0)}
T_7(\e,k)\psi_0^{(0)}+T_7(\e,k)T_5(k)T_1F.
\end{aligned}
\end{equation}
Now we apply the operator $T_6(\e,k)$ to this equation, and then
calculate the inner product with $\psi_0^{(0)}$ in
$\H^1(-x_0,x_0)$ and multiply by $k$:
\begin{gather}
\begin{aligned}
(k-\e \mathfrak{g}(\e,k))
&\big(T_6(\e,k)U,\psi_0^{(0)}\big)_{\H^1(-x_0,x_0)}=
-\mathfrak{g}(\e,k)\big(F,\psi_0^{(0)}\big)_{L_2(-x_0,x_0)} +
\\
&\hphantom{\Big(T_6(\e,k)U,}
+k\big(T_6(\e,k)T_7(\e,k)T_5(k)T_1F,
\psi_0^{(0)}\big)_{\H^1(-x_0,x_0)},
\end{aligned}\label{3.31}
\\
\mathfrak{g}(\e,k):=\big(T_6(\e,k)T_7(\e,k)\psi_0^{(0)},
\psi_0^{(0)}\big)_{\H^1(-x_0,x_0)}.\nonumber
\end{gather}
A nontrivial solution of the equation (\ref{3.25}) with $F=0$
satisfy the equation (\ref{3.31}) with $F=0$. Moreover, for such
solutions the inner product
$\big(T_6(\e,k)U,\psi_0^{(0)}\big)_{\H^1(-x_0,x_0)}$ is nonzero,
otherwise the equation (\ref{3.30}) would imply that $U=0$.
Thus, the equation (\ref{3.25}) with $F=0$ can have nontrivial
solution only for $k$ those satisfy the equation
\begin{equation}\label{3.32}
k=\e \mathfrak{g}(\e,k).
\end{equation}
As it follows from (\ref{3.30}) with $F=0$, the nontrivial
solutions of the equation (\ref{3.25}) associated with such
roots $k=k_\e$ are unique up to a multiplicative constant and
takes the form
\begin{equation}\label{3.33}
U_\e=T_7(\e,k_\e)\psi_0^{(0)}.
\end{equation}
Moreover, $U_\e\not=0$, since by (\ref{3.29a}) the convergence
\begin{equation*}
U_\e=T_7(\e,k_\e)\psi_0^{(0)}\to\psi_0^{(0)}\not=0
\end{equation*}
is valid in $\H^1(-x_0,x_0)$. Hence, the equation (\ref{3.32})
determines $k$ those nontrivial solutions of the equation
(\ref{3.25}) with $F=0$ exist for.

The function $\mathfrak{g}(\e,k)$ is holomorphic on $k$ and
bounded uniformly on $\e$ and $k$. Owing to this reason by
choosing $\e$ the right hand side of (\ref{3.32}) can be made as
small as needed for $|k|=\d$, where $\d$ is a small fixed
number. By Rouche theorem it follows that in the disc $\{k:
|k|\leqslant \d \}$ the function $k\mapsto (k-\e
\mathfrak{g}(\e,k))$ has as many zeros as the function $k\mapsto
k$. Therefore, the equation (\ref{3.32}) has exactly one root
$k_\e$, and $k_\e\to0$ as $\e\to0$. The associated nontrivial
solution of the equation (\ref{3.25}) with $F=0$ is given by the
formula (\ref{3.33}). This nontrivial solution is an element of
$L_2(\mathbb{R})$, if and only if $\RE k_\e>0$ (see
(\ref{3.14a}), (\ref{3.14b}), (\ref{3.17}), (\ref{3.18b})), this
is why only in this case the operator $H_\e$ has an eigenvalue
$\l_\e=\mu_0^+(\e^2)-k_\e^2$ tending to zero as $\e\to0$. Thus,
we have proved
\begin{lemma}\label{lm3.10}
The operator $H_\e$ has the eigenvalue converging to zero as
$\e\to0$, if and only if there exists the nontrivial solution to
the problem (\ref{1.7}), and the solution $k_\e$ of the equation
(\ref{3.32}) meets the inequality $\RE k_\e>0$. If exists, this
eigenvalue is simple and of the form
$\l_\e^{(0)}=\mu_0^+(\e^2)-k_\e^2$. An associated eigenfunction
is given by the equality
\begin{equation*}%\l%abel{3.38a}
\psi_\e^{(0)}(x)=\left\{
\begin{aligned}
&{\vp}\left(\frac{x}{\e},\e^2\right)U_\e(x), & |x|&<x_0,
\\
&\frac{{\vp}\left(\pm\frac{x_0}{\e},\e^2\right)U_\e(\pm
x_0)}{\Th^\pm \left(\pm\frac{x_0}{\e},\e,k_\e\right)}\Th^\pm
\left(\frac{x}{\e},\e,k_\e\right), & \pm x&\geqslant x_0,
\end{aligned} \right.
\end{equation*}
and the convergence
\begin{equation}\label{3.38b}
\psi_\e^{(0)}\xrightarrow[\e\to0]{}\psi_0^{(0)}
\end{equation}
holds true in $\H^1(Q)$ for each $Q\in \mathfrak{C}$.
\end{lemma}

In the fifth section we will show that the solution of the
equation (\ref{3.32}) satisfies the inequality $\RE k_\e>0$,
what we will need the following auxiliary statement for.

\begin{lemma}\label{lm3.11}
For $k$ close to zero each function $F\in L_2(-x_0,x_0)$ can be
represented as
\begin{equation}\label{3.33a}
(\I-2\e T_2(\e)+T_1V+(k^2-1)T_1+\rho_+ T_3^+
-\rho_-T_3^-)^{-1}T_1F=\frac{T_8(\e,k)F}{k-k_\e}U_\e+T_9(\e,k)F,
\end{equation}
where $T_8(\e,k): L_2(-x_0,x_0)\to \mathbb{C}$, $T_9(\e,k):
L_2(-x_0,x_0)\to \H^1(-x_0,x_0)$ are linear functional and
operator bounded uniformly on $\e$ and $k$.
\end{lemma}

\begin{proof}
Let us complete the solving of the equation (\ref{3.25}). We
express the inner product
$\big(T_6(\e,k)U,\psi_0^{(0)}\big)_{\H^1(-x_0,x_0)}$ from
(\ref{3.31}) and substitute the result into (\ref{3.30}):
\begin{align}\label{3.34}
&U=\frac{T_{10}(\e,k)F}{k-\e \mathfrak{g}(\e,k)}
T_7(\e,k)\psi_0^{(0)}+T_7(\e,k) T_5(k)T_1F,
\\
&T_{10}(\e,k)F:=\big(T_6(\e,k)T_7(\e,k)T_5(k)T_1F,
\psi_0^{(0)}\big)_{\H^1(-x_0,x_0)}-
\big(F,\psi_0^{(0)}\big)_{L_2(-x_0,x_0)},\nonumber
\end{align}
By the equation (\ref{3.32}) we deduce:
$k-\e\mathfrak{g}(\e,k)=(k-k_\e)\left(1+\e
\widetilde{\mathfrak{g}}(\e,k)\right)$,
$\widetilde{\mathfrak{g}}(\e,k)=
\frac{\displaystyle\mathfrak{g}(\e,k_\e)-\mathfrak{g}(\e,k)}{\displaystyle
k-k_\e}$. Item~\ref{it2lm3.7} of Lemma~\ref{lm3.7}, holomorphy
of the operator $T_5(k)$, and the definition of the operators
$T_6(\e,k)$, $T_7(\e,k)$, and the function $\mathfrak{g}(\e,k)$
imply that the derivative $\frac{\displaystyle d}{\displaystyle
dk}\mathfrak{g}(\e,k)$ is bounded uniformly on $\e$ and $k$.
This fact by Hadamard lemma yields a uniform on $\e$ and $k$
estimate: $\left|\widetilde{\mathfrak{g}}(\e,k)\right|\leqslant
C$. In the same way one can prove an uniform on $\e$ and $k$
estimate:
\begin{equation}
\|T_7(\e,k)\psi_0^{(0)}-U_\e\|_{\H^1(-x_0,x_0)}=
\|(T_7(\e,k)-T_7(\e,k_\e))\psi_0^{(0)}\|_{\H^1(-x_0,x_0)}\leqslant
C|k-k_\e|.\label{3.43a}
\end{equation}
Last two estimates and (\ref{3.34}) lead us to the
representation (\ref{3.33a}), where
\begin{align*}
&T_8(\e,k)=\frac{T_{10}(\e,k)F}{1+\e
\widetilde{\mathfrak{g}}(\e,k)},
\\
&T_9(\e,k)=\frac{(T_7(\e,k)-T_7(\e,k_\e))\psi_0^{(0)}}{k-\e
\mathfrak{g}(\e,k)}T_{10}(\e,k)+T_7(\e,k) T_5(k)T_1.
\end{align*}
\end{proof}

\sect{Asymptotics for the finite eigenvalues in the
semi-infinite lacuna}

In the present section we construct the asymptotics expansions
for the eigenvalues $\l_\e^{(n)}$, $n=-K,\ldots,-1$, of the
operator $H_\e$. The asymptotics expansions for the associated
eigenfunctions will be constructed as well. First we construct
the asymptotics expansions formally, and then we justify them
rigorously.

The asymptotics for an eigenvalue $\l_\e^{(n)}$ is constructed
as the series (\ref{1.8}). In constructing the asymptotics for
the associated eigenfunction we employ the two-scale asymptotics
expansions method (asymptotic method of homogenization)
\cite{BP}, \cite{ZhKO}, \cite{PChSh}. Following this method, the
asymptotics for the eigenfunction is sought as:
\begin{equation}\label{4.1}
\psi_\e^{(n)}(x)=\psi_0^{(n)}(x)+\sum\limits_{i=2}^\infty\e^i
\psi_i^{(n)}(x,\xi),
\end{equation}
where, we remind, $\xi=x/\e$. The functions $\psi_i$ are sought
as 1-periodic on second argument. The eigenfunction, associated
with $\l_\e^{(n)}$ is an element of $L_2(\mathbb{R})$, this is
why for the functions $\psi_i$ we impose an additional
restriction: $\psi_i(x,x/\e)\in L_2(\mathbb{R})$.

The aim of the formal constructing is to determine the
coefficients of the series (\ref{1.8}) and (\ref{4.1}).
Throughout the constructing we will omit the superscript $(n)$
in the notations $\l_i^{(n)}$ and $\psi_i^{(n)}$, at the same
time having in mind that all the functions appearing in
constructing depend on $n$ as well.

The eigenfunction associated with $\l_\e^{(n)}$ satisfies the
equation (\ref{3.16}) with $f=0$, $\l=\l_\e^{(n)}$. Because of
this we substitute the series (\ref{1.8}), (\ref{4.1}) into the
equation (\ref{3.16}) with $f=0$, collect the coefficients of
the same powers of $\e$ and equate them to zero. It gives the
following equations:
\begin{equation}\label{4.3}
\begin{aligned}
-\frac{\partial^2}{\partial
\xi^2}\psi_{i+2}=&\left(\frac{\partial^2}{\partial
x^2}-a-V\right)\psi_i+2\frac{\partial^2}{\partial
x\partial\xi}\psi_{i+1}+
\\
&+ \sum\limits_{j=0}^{i-2}\l_j \psi_{i-j}+\l_i\psi_0,\quad
(x,\xi)\in \mathbb{R}^2,\quad i\geqslant0.
\end{aligned}
\end{equation}
where $a=a(\xi)$, $V=V(x)$, $\l_1=0$,
$\psi_0(x,\xi):=\psi_0(x)$, $\psi_1(x,\xi):=0$. Following the
two-scale asymptotics expansions method, the variables $x$ and
$\xi$ in these equations are treated to be independent.
According Lemma~\ref{lm2.3}, the equations (\ref{4.3}) have
1-periodic on $\xi$ solutions, if their right hand sides meet
the %following
 solvability conditions:
\begin{equation}\label{4.4}
\left(-\frac{d^2}{dx^2}+ V-\l_0\right)\int\limits_0^1
\psi_i\di\xi=-\int\limits_0^1a\psi_i\di\xi+
\sum\limits_{j=2}^{i-2}\l_j
\int\limits_0^1\psi_{i-j}\di\xi+\l_i\psi_0,\quad x\in
\mathbb{R}.
\end{equation}
Here we have also taken into account that the function
$\psi_{i+1}$ is supposed to be 1-periodic on $\xi$. We denote
$\H^\infty(\mathbb{R}):=\bigcap\limits_{p=1}^\infty
\H^p(\mathbb{R})$. By standard embedding theorems we have
$\H^\infty(\mathbb{R})\subset C^\infty(\mathbb{R})$ (see, for
instance, \cite[Ch. I\!I\!I, \S 6]{M}). Obviously, $\psi_0\in
\H^\infty(\mathbb{R})$.

In order to solve the equations (\ref{4.3}), (\ref{4.4}) we will
make use of the following auxiliary lemma.

\begin{lemma}\label{lm4.1}
An equation
\begin{equation}\label{4.5}
\left(-\frac{d^2}{dx^2}+V-\l_0\right)u=f,\quad x\in \mathbb{R},
\end{equation}
where $f\in L_2(\mathbb{R})$ is solvable in the space
$\H^2(\mathbb{R})$, if and only if
$(f,\psi_0)_{L_2(\mathbb{R})}=0$. In this case a solution of the
equation (\ref{4.5}) is unique up to an additive term $C\psi_0$,
$C=const$. There exists unique solution $u\in \H^2(\mathbb{R})$
orthogonal to $\psi_0$ in $L_2(\mathbb{R})$. If $f\in
\H^\infty(\mathbb{R})$, then $u\in \H^\infty(\mathbb{R})$.
\end{lemma}

\begin{proof}
The resolvent $(H_0-\l)^{-1}:
L_2(\mathbb{R})\to\H^2(\mathbb{R})$ is meromorphic on $\l$ in a
small neighbourhood of $\l_0$ and in accordance with \cite[Ch.V,
\S 3.5]{K} it has a simple pole at the point $\l_0$, the residue
at this pole being a projector on $\psi_0$. Therefore, the
resolvent $(H_0-\l_0)^{-1}$ is a bounded operator on the
orthogonal complement of $\psi_0$ in $L_2(\mathbb{R})$, what
proves the condition of solvability of the equation (\ref{4.5})
in the space $\H^2(\mathbb{R})$. If $f\in \H^p(\mathbb{R})$,
then expressing $\frac{\displaystyle d^2 u}{\displaystyle dx^2}$
from the equation (\ref{4.5}), it is not difficult to show that
$u\in\H^{p+2}(\mathbb{R})$. Thus, if  $f\in
\H^\infty(\mathbb{R})$, then $u\in \H^\infty(\mathbb{R})$.
\end{proof}

Obviously, the equation (\ref{4.4}) holds for $i=0,1$. It is
easy to see that 1-periodic on $\xi$ solution to the equation
(\ref{4.3}) with $i=0$ is of the form:
\begin{equation*}
\psi_2(x,\xi)=u_{2,0}(x)+\t\psi_2(x,\xi), \quad
\t\psi_2(x,\xi)=u_{2,1}(x)\psi_{2,1}(\xi),
\end{equation*}
where $u_{2,0}$ is a some function,
\begin{equation}\label{4.7}
u_{2,1}(x)=\psi_0(x),\quad \psi_{2,1}(\xi)=-{L}_0[a](\xi).
\end{equation}
Clearly, $\psi_0\in C^\infty(\mathbb{R})$, $\psi_{2,1}\in
C^2(\mathbb{R})$. The condition $\psi_2\left(x,x/\e\right)\in
L_2(\mathbb{R})$ holds, if the same condition holds for the
function $u_{2,0}$, this is why the function $u_{2,0}$ is sought
in the space $L_2(\mathbb{R})$. Now we substitute the obtained
expression for $\psi_2$ into the equation (\ref{4.4}) with $i=2$
and take into account that due to Lemma~\ref{lm2.3} the equality
$\int\limits_0^1\psi_{2,1}(\xi)\di\xi=0$ takes place. As a
result we arrive at the following equation:
\begin{equation*}%\l%abel{4.7a}
\left(-\frac{d^2}{dx^2}+V-\l_0\right)u_{2,0}=
-\left(a,\psi_{2,1}\right)_{L_2(0,1)}\psi_0+\l_2\psi_0, \quad
x\in \mathbb{R}.
\end{equation*}
The right hand side of this equation being an element of
$L_2(\mathbb{R})$, in view of Lemma~\ref{lm4.1} the equation is
solvable in $L_2(\mathbb{R})$, if the equality
\begin{equation*}
-\left(a,\psi_{2,1}\right)_{L_2(0,1)}+\l_2=0
\end{equation*}
takes place. By (\ref{2.18}) and (\ref{4.7}) it follows the
formula (\ref{1.9}) for $\l_2$. In turn, this equality implies
that $u_{2,0}=C\psi_0$. The constant $C$ is chosen by the
orthogonality condition $(u_{2,0},\psi_0)_{L_2(\mathbb{R})}=0$,
i.e., $u_{2,0}\equiv 0$. Thus, the function $\psi_2$ has the
form
\begin{equation}\label{4.8}
\psi_2(x,\xi)=\t\psi_2(x,\xi)=u_{2,1}(x)\psi_{2,1}(\xi),
\end{equation}
where $u_{2,1}$ and $\psi_{2,1}$ are defined by the equalities
(\ref{4.7}). Other functions $\psi_i$ and numbers $\l_i$ are
given by the following lemma.

\begin{lemma}\label{lm4.2}
There exist the solutions of the equations (\ref{4.3}),
(\ref{4.4}) having the form
\begin{align}
&\psi_i(x,\xi)=u_{i,0}(x)+\t\psi_i(x,\xi),\label{4.9}
\\
&\t\psi_i(x,\xi)={L}_0[G_{i}(x,\cdot)](\xi)
=\sum\limits_{j=1}^{\mathfrak{m}_i}
u_{i,j}(x)\psi_{i,j}(\xi),\label{4.10}
\\
&G_i:=\left(\frac{\partial^2}{\partial
x^2}-a-V\right)\t\psi_{i-2}+\int\limits_0^1a\t\psi_{i-2}
\di\xi+2\frac{\partial^2}{\partial
x\partial\xi}\t\psi_{i-1}+\sum\limits_{j=0}^{i-4}
\l_j\t\psi_{i-j-2}-au_{i-2,0},\label{4.10a}
\end{align}
where $G_i=G_i(x,\xi)$, $a=a(\xi)$, $V=V(x)$,
$\t\psi_0=\t\psi_1:=0$, $u_{0,0}:=\psi_0$, $\mathfrak{m}_i$ are
some numbers, $u_{i,j}\in \H^\infty(\mathbb{R})$, $\psi_{i,j}\in
C^2(\mathbb{R})$ are 1-periodic functions obeying the equalities
\begin{equation}\label{4.11}
\int\limits_0^1\psi_{i,j}(\xi)\di\xi=0.
\end{equation}
The functions $u_{i,0}\in \H^\infty(\mathbb{R})$ are solutions
of the equations
\begin{equation}\label{4.12}
\begin{aligned}
\left(-\frac{d^2}{dx^2}+ V-\l_0\right)u_{i,0}=&
-\int\limits_0^1a\t\psi_i\di\xi+\sum\limits_{j=1}^{i-1}\l_j
u_{i-j,0}+\l_i\psi_0,\quad x\in \mathbb{R},
\end{aligned}
\end{equation}
orthogonal to $\psi_0$ in $L_2(\mathbb{R})$. The numbers $\l_i$
are given by the formulas
\begin{equation}\label{4.13}
\l_i=\int\limits_{\mathbb{R}}\int\limits_0^1a(\xi)\psi_0(x)
\t\psi_i(x,\xi)\di\xi\di x.
\end{equation}
\end{lemma}
\begin{remark}\label{rm4.1}
The notation ${L}_0[G_{i}(x,\cdot)]$ means the applying of the
operator ${L}_0$ to the function $G_{i}(x,\xi)$ as a function on
$\xi$ depending on additional parameter $x$.
\end{remark}
\begin{proof}
We prove the lemma by induction. For $i=2$ the conclusion of the
lemma follows from the formula (\ref{1.9}) of $\l_2$ and
formulas (\ref{4.7}), (\ref{4.8}) for the function
$\psi_2(x,\xi)$. Moreover, the equation (\ref{4.3}) holds for
$i=0$, and the equation (\ref{4.4}) holds for $i=0,1,2$.

Suppose the formulas (\ref{4.9})--(\ref{4.11}) hold for
$i\leqslant m+1$, the equations (\ref{4.3}), (\ref{4.4}),
(\ref{4.12}) and the formulas (\ref{4.13}) do for $i\leqslant
m-1$. Let us prove that then the equations (\ref{4.3}),
(\ref{4.4}) are solvable for $i=m$ and their solutions are
determined in accordance with (\ref{4.9})--(\ref{4.12}), and the
number $\l_m$ is given by the formula (\ref{4.13}). By the
induction assumption the functions $\psi_{i,j}$, $i\leqslant m$,
obey the equalities (\ref{4.11}). Therefore, the condition of
solvability of the equation (\ref{4.3}) with $i=m$ in the class
of 1-periodic on $\xi$ functions is the equation (\ref{4.4})
with $i=m$. We substitute the formulas (\ref{4.9}) for $\psi_i$,
$i\leqslant m$, to this equation. These formulas are valid due
to the induction assumption. Bearing in mind the relations
(\ref{4.11}), we get the equation (\ref{4.12}) for the function
$u_{m,0}$. By the induction assumption the right hand side of
this equation is an element of the space
$\H^\infty(\mathbb{R})$. The functions $u_{m,j}$, $j\geqslant
1$, belonging to $\H^2(\mathbb{R})$, it is sufficient the
inclusion $u_{m,0}\in L_2(\mathbb{R})$ to be true to meet the
restriction $\psi_m\left(x,\frac{x}{\e}\right)\in
L_2(\mathbb{R})$. The condition of solvability of the equation
(\ref{4.12}) with $i=m$ in the space $L_2(\mathbb{R})$ is given
by Lemma~\ref{lm4.1}. Writing out this condition and bearing in
mind the orthogonality of the functions $u_{i,0}$, $i\leqslant
m-1$, to the functions $\psi_0$ in $L_2(\mathbb{R})$, we arrive
at the equality (\ref{4.13}) for $\l_m$. The function
$u_{m,0}\in\H^\infty(\mathbb{R})$ is chosen as orthogonal to $
\psi_0$ in $L_2(\mathbb{R})$.

Thus, the solvability condition for the equation (\ref{4.3})
with $i=m$ holds. The right hand side of the equation
(\ref{4.3}) with $i=m$ coincides with the function $G_{m+2}$
from (\ref{4.10a}) by the induction assumption and the equation
(\ref{4.12}) for $u_{m,0}$. We represent the function $G_{m+2}$
as
\begin{align}
&G_{m+2}(x,\xi)=G^{(1)}_{m+2}(x,\xi)+G^{(2)}_{m+2}(x,\xi),\label{4.33}
\\
&G_{m+2}^{(1)}=\left(\frac{\partial^2}{\partial
x^2}-V\right)\t\psi_{m}+2\frac{\partial^2}{\partial
x\partial\xi}\t\psi_{m+1}+\sum\limits_{j=0}^{m-2}
\l_j\t\psi_{m-j}-au_{m,0},\nonumber
\\
& G^{(2)}_{m+2}=-a\t\psi_{m}+
\int\limits_0^1a\t\psi_{m}\di\xi.\nonumber
\end{align}
By the formulas (\ref{4.10}) for $\t\psi_i$, $i\leqslant m+1$,
the functions $G_{m+2}^{(i)}$, $i=1,2$, are finite series
\begin{equation}\label{4.34}
G_{m+2}^{(i)}(x,\xi)=\sum\limits_j G_{m+2,j}^{(i,1)}(x)
G_{m+2,j}^{(i,2)}(\xi),
\end{equation}
where $G_{m+2,j}^{i,1}\in\H^\infty(\mathbb{R})$. The equalities
(\ref{1.0}) and (\ref{4.11}) for $i\leqslant m+1$ and the
definition of the function $G_{m+2}^{(1)}$ imply that
\begin{equation}\label{4.35}
\int\limits_0^1 G_{m+2,j}^{(1,2)}(\xi)\di\xi=0
\end{equation}
for all $j$. The same equalities hold for the functions
$G_{m+2,j}^{(2,2)}$ as well, what follows from the definition of
the function $G_{m+2}^{(2)}$ and the formula  (\ref{4.10}) with
$i=m$. Taking into account the established properties of the
function $G_{m+2}$ and solving the equation (\ref{4.3}) with
$i=m$ by Lemma~\ref{lm2.3}, we arrive at the representations
(\ref{4.9}), (\ref{4.10}) for $\psi_{m+2}(x,\xi)$, where
$u_{m+2,0}$ is a some function, $u_{m+2,j}\in
\H^\infty(\mathbb{R})$, $j\geqslant 1$, and $\psi_{m+2,j}\in
C^2(\mathbb{R})$ are 1-periodic functions satisfying the
equalities (\ref{4.11}).
\end{proof}

Let us prove the formulas (\ref{1.9}) for $\l_3$ and $\l_4$.
Employing (\ref{4.7})--(\ref{4.10a}), it is not difficult to
check that
\begin{equation}\label{4.13a}
\mathfrak{m}_3=1,\quad
u_{3,1}=2\frac{du_{2,1}}{dx}=2\frac{d\psi_0}{dx}, \quad
\psi_{3,1}={L}_0\left[\frac{d\psi_{2,1}}{d\xi}\right].
\end{equation}
These formulas and the definition of the function $\psi_{2,1}$
imply:
\begin{equation}\label{4.13b}
\int\limits_0^1 a\psi_{3,1}\di\xi=\int\limits_0^1\psi_{3,1}
\frac{d^2\psi_{2,1}}{d\xi^2}\di\xi=\int\limits_0^1\psi_{2,1}
\frac{d^2\psi_{3,1}}{d\xi^2}\di\xi=-\int\limits_0^1\psi_{2,1}
\frac{d\psi_{2,1}}{d\xi}\di\xi=0.
\end{equation}
In view of the equality obtained and (\ref{4.13}) we conclude
that $\l_3=0$. Therefore, the equation (\ref{4.12}) for
$u_{3,0}$ is homogeneous, this is why
\begin{equation}\label{4.13c}
u_{3,0}\equiv0.
\end{equation}
From  (\ref{4.8})--(\ref{4.10a}), (\ref{4.13}) it follows that
\begin{equation}\label{4.13d}
\begin{aligned}
&\mathfrak{m}_4=2,\quad u_{4,1}=\psi_0,\quad
\psi_{4,1}={L}_0[\l_2-a\psi_{2,1}],
\\
&u_{4,2}=2\frac{dU_{3,1}}{dx}=4\frac{d^2\psi_0}{dx^2},\quad
\psi_{4,2}={L}_0\left[\frac{d\psi_{3,1}}{d\xi}\right].
\end{aligned}
\end{equation}
By (\ref{4.13}) and the normalization condition for $\psi_0$ it
implies
\begin{equation*}
\l_4=\int\limits_0^1a\psi_{4,1}\di\xi+4\int\limits_{\mathbb{R}}
\psi_0\frac{d^2\psi_0}{dx^2}\di
x\int\limits_0^1a\psi_{4,2}\di\xi.
\end{equation*}
The first summand in the right hand side of the last equality
coincides with $\mu_{0,2}^+$ (see (\ref{2.18}), (\ref{4.7}),
(\ref{4.13d})). Let us calculate the second summand:
\begin{equation}
\begin{aligned}
& 4\int\limits_\mathbb{R}\psi\frac{d^2\psi_0}{dx^2}\di
x=-4\int\limits_\mathbb{R}\left|\frac{d\psi_0}{dx}\right|^2\di
x,
\\
&\int\limits_0^1 a\psi_{4,2}\di\xi=\int\limits_0^1\psi_{4,2}
\frac{d^2\psi_{2,1}}{d\xi^2}\di\xi=-\int\limits_0^1
\psi_{2,1}\frac{d\psi_{3,1}}{d\xi}\di\xi=\int\limits_0^1
\left|\psi_{2,1}\right|^2\di\xi,
\end{aligned}\label{4.20a}
\end{equation}
what yields immediately the formula (\ref{1.9}) for $\l_4$.

Let $m\geqslant 2$. We denote
\begin{equation*}
\psi_{\e,m}(x):=\psi_0(x)+\sum\limits_{i=2}^m\e^i\psi_i
\left(x,\frac{x}{\e}\right),\quad
\l_{\e,m}:=\l_0+\sum\limits_{i=2}^{m-2}\e^i\l_i.
\end{equation*}
Lemma~\ref{lm4.2} implies the following statement.
\begin{lemma}\label{lm4.3}
The function $\psi_{\e,m}$ is an element of the space
$\H^2(\mathbb{R})\cap C^\infty(\mathbb{R})$ and converges to
$\psi_0$ in $\H^1(\mathbb{R})$ as $\e\to0$. The function
$f_{\e,m}:=(H_\e-\l_{\e,m})\psi_{\e,m}$ obeys the estimate
$\|f_{\e,m}\|_{L_2(\mathbb{R})}=\Odr(\e^{m-1})$.
\end{lemma}
Since  $f_{\e,m}\in L_2(\mathbb{R})$ and
$\psi_{\e,m}=(H_\e-\l_{\e,m})^{-1}f_{\e,m}$, it follows that
Lemma~\ref{lm3.4} and, in particular, the representation
(\ref{3.10}) are applicable to the function $\psi_{\e,m}$:
\begin{equation}\label{4.14}
\psi_{\e,m}=-\frac{\psi_\e^{(n)}}{\l-\l_\e^{(n)}}
\left(f_{\e,m},\psi_\e^{(n)}\right)_{L_2(\mathbb{R})}+\t\psi_{\e,m},
\end{equation}
where by (\ref{3.11}) and Lemma~\ref{lm4.3} the function
$\t\psi_{\e,m}$ satisfies the uniform on $\e$ estimate:
\begin{equation}\label{4.15}
\|\t\psi_{\e,m}\|_{\H^2(\mathbb{R})}\leqslant
C\|f_{\e,m}\|_{L_2(\mathbb{R})}\leqslant C\e^{m-1}.
\end{equation}
The representation (\ref{4.14}) yields
\begin{equation*}
\|\psi_{\e,m}-\t\psi_{\e,m}\|_{L_2(\mathbb{R})}\leqslant
\frac{\|f_{\e,m}\|_{L_2(\mathbb{R})}}{|\l_\e^{(n)}-\l_{\e,m}|},
\end{equation*}
what together with the estimate (\ref{4.15}) and convergence of
$\psi_{\e,m}$ to $\psi_0$ in $\H^1(\mathbb{R})$ imply
\begin{equation*}
|\l_\e^{(n)}-\l_{\e,m}|=\Odr(\e^{m-1}).
\end{equation*}
This equality proves the asymptotics (\ref{1.8}) and completes
the proof of Theorem~\ref{th1.2}.

Multiplying (\ref{4.14}) by $\psi_\e^{(n)}$ in $L_2(\mathbb{R})$
and bearing in mind the normalization of $\psi_\e^{(n)}$, we
obtain:
\begin{equation*}
-\frac{\big(f_{\e,m},\psi_\e^{(n)}\big)_{L_2(\mathbb{R})}}
{\l_{\e,m}-\l_\e^{(n)}}=
\big(\psi_{\e,m}-\t\psi_{\e,m},\psi_\e^{(n)}\big)_{L_2(\mathbb{R})}.
\end{equation*}
Lemma~\ref{lm4.2} and (\ref{4.15}) imply that for any $m_1$,
$m_2$
\begin{equation*}
\big(\psi_{\e,m_1}-\t\psi_{\e,m_1},
\psi_\e^{(n)}\big)_{L_2(\mathbb{R})}-
\big(\psi_{\e,m_2}-\t\psi_{\e,m_2},
\psi_\e^{(n)}\big)_{L_2(\mathbb{R})}=
\Odr(\e^{\min\{m_1,m_2\}+1}).
\end{equation*}
The last two relations and Lemmas~\ref{lm3.3},~\ref{lm4.3} yield
that there exists the function $c(\e)$ such that
\begin{equation}\label{4.16}
c(\e)=-\frac{\big(f_{\e,m},\psi_\e^{(n)}\big)_{L_2(\mathbb{R})}}
{\l_{\e,m}-\l_\e^{(n)}}+\Odr(\e^{m+1}), \quad c(\e)=1+o(1),\quad
\e\to0,
\end{equation}
for each $m$. The equalities (\ref{4.14})-(\ref{4.16}) lead us
to the estimates:
\begin{equation*}
\|\psi_{\e,m}-c(\e)\psi_\e^{(n)}\|_{\H^2(\mathbb{R})}=\Odr(\e^{m-1}).
\end{equation*}
Thus, we have proved
\begin{theorem}\label{th4.1}
The eigenfunction associated with the eigenvalue $\l_\e^{(n)}$,
$n=-K,\ldots,-1$ can be chosen such that in norm of
$\H^2(\mathbb{R})$ it has an asymptotics expansion (\ref{4.1}),
where the coefficients of the expansions are determined by the
equalities (\ref{4.7}), (\ref{4.8}), (\ref{4.13a}),
(\ref{4.13c}), (\ref{4.13d}), and Lemma~\ref{lm4.2}.
\end{theorem}

\sect{Asymptotics for the small eigenvalue in the semi-infinite
lacuna}

In the present section we prove Theorem~\ref{th1.2a}. The
constructing and justifying of the asymptotics expansions for
the eigenvalues $\l_\e^{(n)}$, $n=-K,\ldots,-1$, and the
associated eigenfunctions under hypothesis of
Theorem~\ref{th1.2a} coincide completely with the arguments of
the previous section. This is why in this section we study only
the eigenvalue of the operator $H_\e$ converging to zero as
$\e\to0$. In addition to the proof of Theorem~\ref{th1.2a} we
will also obtain the asymptotics expansion of the associated
eigenfunction.

Throughout this section the problem (\ref{1.7}) is assumed to
have the nontrivial solution obeying the condition (\ref{1.12}).

According to Lemma~\ref{lm3.10}, under hypothesis of
Theorem~\ref{th1.2a} the operator $H_\e$ has the eigenvalue
converging to zero, if and only if the inequality  $\RE k_\e>0$
holds for the root of the equation (\ref{3.32}). We will
construct the asymptotics expansions of the number
$\l_\e=\mu_0^+(\e^2)-k_\e^2$ and the associated nontrivial
solution to the problem (\ref{3.16}), (\ref{3.17}) with $f=0$,
$k=k_\e$, which is denoted by $\psi_\e$. Then on the base of
these asymptotics expansions we will show that $\RE k_\e>0$,
$\l_\e$ is a required eigenvalue of the operator $H_\e$, and
$\psi_\e$ is the associated eigenfunction. Like in the previous
section, first we construct the asymptotics expansions formally
and then we justify them rigorously.

The asymptotics of $\l_\e$ is constructed as the series
(\ref{1.10}). The asymptotics of the associated eigenfunction
$\psi_\e$ of the problem (\ref{3.16}), (\ref{3.17}) is
constructed on the base of two-scale asymptotics expansions
method as follows
\begin{gather}
\psi_\e(x)=h(x,\e)\left(\psi_0^{(0)}(x)+
\sum\limits_{i=2}^\infty\e^i\psi_i^{(0)}(x,\xi)\right),\label{5.3}
\\
h(x,\tau_\e)=\chi(x)+\left(1-\chi(x)\right)\E^{-\tau_\e|x|},\label{5.4}
\end{gather}
where $\chi\in C^\infty(\mathbb{R})$ is an even real cut-off
function taking values from the segment $[0,1]$, equalling to
one in some fixed neighbourhood of the support of $V$ and
vanishing as $|x|\geqslant x_0$. The symbol $\tau_\e$ in
(\ref{5.4}) denotes a function, whose asymptotics is constructed
as follows
\begin{equation}\label{5.5}
\tau_\e=\sum\limits_{i=4}^\infty \e^i \tau_i.
\end{equation}
The aim of the formal constructing is to determine the functions
$\psi_i^{(0)}$ and numbers $\l_i^{(0)}$, $\tau_i$. Everywhere in
constructing we will omit the superscript $(0)$ in the notations
$\psi_i^{(0)}$ and $\l_i^{(0)}$.

We will employ the symbol $\mathcal{V}$ for the subset of the
functions $u=u(x)$ from $C(\mathbb{R})$ satisfying the
constraint $u(x)=u(\pm x_0)$, $\pm x\geqslant x_0$. The
functions $\t\psi_i(x,\xi)$ are sought as 1-periodic on $\xi$
and belonging to the set $\mathcal{V}$ as function on $x$.

Let us explain the choice of the ansatz  (\ref{5.3}). The
function $\psi_\e$ satisfies the equalities (\ref{3.17}) with
$k=k_\e$. Because of this reason the asymptotics expansions
(\ref{5.3}) has been chosen so that it has the same structure
for $\pm x\geqslant x_0$ as the functions $\Th^\pm$ from
(\ref{3.17}). The function $h(x,\e)$ in (\ref{5.3}) models the
exponents being a factor in the functions
$\Th^\pm\left(\frac{x}{\e},\e,k_\e\right)$ (see (\ref{3.14b})).
We also stress that the function $h$ is positive. The
requirement of 1-periodicity on $\xi$ for the functions $\psi_i$
corresponds to 1-periodicity of the function $\Th_{per}^\pm$.
The same reason explains the requirement $\psi_i(\cdot,\xi)\in
\mathcal{V}$. A priori choice of inferior summation limits in
the series (\ref{1.10}), (\ref{5.3}), (\ref{5.5}) is explained
by the fact that in constructing the asymptotics given below the
coefficients of the absent powers of $\e$ are happened to be
zero. This is why we do not introduce them, what also simplify
the equations appearing in determining the other coefficients of
the series (\ref{1.10}), (\ref{5.3}), (\ref{5.5}).

We substitute the series (\ref{1.10}), (\ref{5.3}), (\ref{5.5})
into the equation (\ref{3.16}) with $f=0$, divide this equation
by $h(x,\tau_\e)$, expand it in power series on $\e$ and collect
the coefficients of the same powers of $\e$. As a result we get
the following equations:
\begin{equation}\label{5.11}
\begin{aligned}
-\frac{\partial^2}{\partial\xi^2}\psi_{i+2}=&2\frac{\partial^2}{\partial
x\partial\xi}\psi_{i+1}+\left(\frac{\partial^2}{\partial
x^2}-V-a\right)\psi_{i}+2\sum\limits_{j=4}^{i-1}h_j^{(1)}
\frac{\partial}{\partial\xi}\psi_{i-j+1}+
\\
&+ \sum\limits_{j=2}^{i}\left(2h_j^{(1)}\frac{\partial}{\partial
x}+ h_j^{(2)}+\l_j\right)\psi_{i-j}, \quad (x,\xi)\in
\mathbb{R}^2,\quad i\geqslant 0,
\end{aligned}
\end{equation}
where $V=V(x)$, $a=a(\xi)$, $\psi_1:=0$,
$h_i^{(1)}=h_i^{(2)}:=0$, $i=2,3$. The functions
$h_i^{(1)}=h_i^{(1)}(x,\boldsymbol{\tau}_i)$ and
$h_i^{(2)}=h_i^{(2)}(x,\boldsymbol{\tau}_i)$, $i\geqslant 4$,
$\boldsymbol{\tau}_i:=(\tau_4,\ldots,\tau_i)$ are the
coefficients of the power asymptotics expansion on $\e$ for the
functions $h'(x,\tau_\e)/h(x,\tau_\e)$ and
$h''(x,\tau_\e)/h(x,\tau_\e)$, respectively. The functions
$h_i^{(j)}\in C^\infty(\mathbb{R})$ can be represented in the
form
\begin{equation}\label{5.11a}
h_i^{(j)}(x,\boldsymbol{\tau}_i)=-\tau_i\frac{d^j}{dx^j}
\left(|x|(1-\chi(x))\right)+\t
h_i^{(j)}(x,\boldsymbol{\tau}_{i-1}),\quad j=1,2,
\end{equation}
where $\t h_i^{(j)}\in C^\infty(R)\cap \mathcal{V}$ and, in
particular,
\begin{equation}\label{5.12}
\t h_4^{(1)}(x)=\t h_4^{(2)}(x)=0.
\end{equation}
We also note that for $\pm x\geqslant x_0$ the equalities
\begin{equation}\label{5.12a}
h_i^{(1)}(x,\boldsymbol{\tau}_i)=\mp\tau_i,\quad
h_i^{(2)}(x,\boldsymbol{\tau}_i)=
\sum\limits_{j=4}^{i-4}\tau_j\tau_{i-j}
\end{equation}
take place. These equalities are consequences of (\ref{5.5}) and
the following relations
\begin{equation}\label{5.8a}
\frac{h'(x,\tau_\e)}{h(x,\tau_\e)}=\mp \tau_\e,\quad
\frac{h''(x,\tau_\e)}{h(x,\tau_\e)}=\tau_\e^2,\quad \pm
x\geqslant x_0.
\end{equation}

In accordance with Lemma~\ref{lm2.3} the condition of
solvability of the equation (\ref{5.11}) in the class of
1-periodic on $\xi$ functions $\xi$ is the equality
\begin{equation}\label{5.10}
\begin{aligned}
\left(-\frac{d^2}{dx^2}+V\right)&\int\limits_0^1\psi_i\di\xi=
-\int\limits_0^1 a\psi_i \di\xi+
\\
&+\sum\limits_{j=2}^{i} \left(2h_j^{(1)} \frac{d}{dx}+ h_j^{(2)}
+\l_j\right)\int\limits_0^1\psi_{i-j} \di\xi, \quad x\in
\mathbb{R},\quad i\geqslant 0,
\end{aligned}
\end{equation}
where $a=a(\xi)$. In deducing this equation we have also beared
in mind that the functions $\psi_{j}$, $j\leqslant i+1$, are
assumed to be 1-periodic on $\xi$. For $i=0,1$ the equation
(\ref{5.10}) holds due to (\ref{1.0}), (\ref{1.7}) and the
equality $\psi_1=0$. Therefore, the equation (\ref{5.11}) is
solvable for $i=0,1$. It is easy to make sure that 1-periodic on
$\xi$ solutions of the equation (\ref{5.11}) with $i=0,1$ take
the form
\begin{equation}
\begin{aligned}
&\psi_i(x,\xi)=u_{i,0}(x)+\t\psi_i(x,\xi),\quad &&
\t\psi_i(x,\xi)=u_{i,1}(x)\psi_{i,1}(\xi),\quad i=2,3,
\\
&\psi_{2,1}(\xi)=-{L}_0[a](\xi),\quad && \psi_{3,1}(\xi)=-{L}_0
\left[\frac{d\psi_{2,1}}{d\xi}\right](\xi),
\\
& u_{2,1}(x)=\psi_0(x),\quad && u_{3,1}=2\frac{d}{dx}\psi_0(x),
\end{aligned}\label{5.11b}
\end{equation}
where $u_{2,0}$, $u_{3,0}$ are some functions. Clearly,
$\psi_{2,1}, \psi_{3,1}\in C^2(\mathbb{R})$, $u_{2,1},
u_{3,1}\in C^\infty(\mathbb{R})\cap \mathcal{V}$.

The equations (\ref{5.11}), (\ref{5.10}) are solvable for
$i\geqslant 2$ as well. In order to prove this fact we will
employ the following auxiliary statement.

\begin{lemma}\label{lm5.1}
The equation
\begin{equation}\label{5.15}
\left(-\frac{d^2}{dx^2}+V\right)u=f,\quad x\in \mathbb{R},
\end{equation}
where $f\in C(\mathbb{R})$ has a solution $u\in
C^2(\mathbb{R})\cap \mathcal{V}$, if and only if $\supp
f\subseteq[-x_0,x_0]$ and the equality
$\int\limits_{\mathbb{R}}f\psi_0\di x=0$ holds. In this case the
solution of the equation (\ref{5.15}) is unique up to an
additive term $C\psi_0$, $C=const$. There exists a unique
solution of the equation (\ref{5.15}) satisfying the equality
\begin{equation}\label{5.16}
\b_-u(x_0)+\b_+u(-x_0)=0.
\end{equation}
This solution is of the form $u(x)={S}[f](x)$,
\begin{equation*}
{S}[f](x)=\psi_0(x) \int\limits_{-\infty}^x f(t)\t\psi_0(t)\di
t+\t\psi_0(x)\int\limits_x^{+\infty}f(t)\psi_0(t)\di t-
\frac{\psi_0(x)}{2}\int\limits_\mathbb{R} f(t)\t\psi_0(t)\di t,
\end{equation*}
where $\t\psi_0$ are the solutions of the equation from
(\ref{1.7}), such that $W_x(\psi_0,\t\psi_0)\equiv 1$. If $f\in
C_0^\infty(\mathbb{R})$, then $u\in C^\infty(\mathbb{R})$.
\end{lemma}

The statement of the lemma is checked by direct calculations.

\begin{lemma}\label{lm5.2}
There exist the solutions of the equations (\ref{5.11}),
(\ref{5.10}) having the form
\begin{align}
&\psi_i(x,\xi)=u_{i,0}(x)+\t\psi_i(x,\xi),\label{5.16a}
\\
&\t\psi_i(x,\xi)={L}_0[G_i(x,\cdot)](x,\xi)=
\sum\limits_{j=1}^{\mathfrak{m}_i}
u_{i,j}(x)\psi_{i,j}(\xi),\label{5.21}
\\
& u_{i,0}(x)={S}[f_i](x),\label{5.23}
\\
&f_i=\sum\limits_{j=2}^{i}
\left(2h_j^{(1)}\frac{d}{dx}+h_j^{(2)}+\l_j\right)u_{i-j,0}
-\int\limits_0^1 a\t\psi_i\di\xi, \label{5.24}
\\
&
\begin{aligned}
&G_i=2\frac{\partial^2}{\partial x\partial\xi}\t\psi_{i-1}+
\left(\frac{\partial^2}{\partial x^2}-V-a\right)\t\psi_{i-2}+
\int\limits_0^1a\t\psi_{i-2}\di\xi+
\\
&+ 2\sum\limits_{j=4}^{i-3}h_j^{(1)}
\frac{\partial}{\partial\xi}\t\psi_{i-j-1}+
\sum\limits_{j=2}^{i-4}\left(2h_j^{(1)} \frac{\partial}{\partial
x}+ h_j^{(2)} +\l_j\right)\t\psi_{i-j-2}-au_{i-2,0},
\end{aligned}\label{5.22}
\end{align}
where $\mathfrak{m}_i$ are some numbers, $G_i=G_i(x,\xi)$,
$f_i=f_i(x)$, $a=a(\xi)$, $V=V(x)$, $\t\psi_1:=0$,
$u_{0,0}:=\psi_0$, $u_{1,0}:=0$, $u_{i,j}\in
C^\infty(\mathbb{R})\cap\mathcal{V}$, $\supp f_i\subseteq
[-x_0,x_0]$, and the functions $\psi_{i,j}$ are 1-periodic and
obey the equalities
\begin{equation}\label{5.25}
\int\limits_{0}^1 \psi_{i,j}(\xi)\di\xi=0.
\end{equation}
The numbers $\l_i$ and $\tau_i$ are given by the formulas
\begin{align}
&\l_i=-\sum\limits_{j=4}^{i-4}\tau_j\tau_{i-j}+\mathfrak{a}_{i,+},\quad
\tau_i=\int\limits_\mathbb{R}\psi_0\t f_i\di x,
 \label{5.26}
\\
&
\begin{aligned}
\t
f_i&:=\sum\limits_{j=2}^{i-2}\left(2h_j^{(1)}
\frac{d}{dx}+h_j^{(2)}+ \l_j\right)u_{i-j,0}+\left(2\t
h_i^{(1)}\frac{d}{dx}+\t h_i^{(2)}+ \l_i\right)\psi_0
-\int\limits_0^1 a\t\psi_i\di\xi,
\end{aligned}\label{5.27}
\end{align}
where $\mathfrak{a}_{i,+}$ is defined in (\ref{5.23a}), $\t
f_i=\t f_i(x)$, $\supp\t f_i\subseteq[-x_0,x_0]$.
\end{lemma}

\begin{proof}
The conclusion of the lemma for $\psi_2$, $\psi_3$ follows from
(\ref{5.11b}). The formulas (\ref{5.26}) are valid for $\l_1$
and $\tau_1$, if we set $\l_1=\tau_1:=0$. We also note that the
equations (\ref{5.11}), (\ref{5.10}) hold for $i=0,1$. The
further proof is carried out by induction. Suppose the
equalities (\ref{5.16a}), (\ref{5.21}), (\ref{5.25}) hold for
$i\leqslant m+1$, and formulas (\ref{5.23}), (\ref{5.26}) be
valid for $i\leqslant m-1$. Let us prove that in this case the
conclusion of the lemma on $\psi_{m+2}$, $u_{m,0}$, $\l_m$,
$\tau_m$ is valid.

According the induction assumption, the functions $\psi_{i,j}$,
$i\leqslant m+1$, satisfy the equalities (\ref{5.25}). Taking
into account these equalities and (\ref{1.0}), we substitute the
representations (\ref{5.16a}) into the equation (\ref{5.10})
with $i=m$. As a result we get the following equation for
$u_{m,0}$:
\begin{equation}\label{5.30}
\left(-\frac{d^2}{dx^2}+V\right)u_{m,0}=f_m, \quad x\in
\mathbb{R},
\end{equation}
where $f_m$ is given by the formula (\ref{5.24}). By the
induction assumption, $u_{i,0}\in C^\infty(\mathbb{R})\cap
\mathcal{V}$, $i\leqslant m-1$, and $u_{m,j}\in
C^\infty(\mathbb{R})\cap \mathcal{V}$, $j\geqslant 1$, this is
why $f_m\in C^\infty(\mathbb{R})\cap \mathcal{V}$. Let us prove
that $\supp f_m\subseteq[-x_0,x_0]$. We denote
\begin{equation}\label{5.23a}
\begin{aligned}
&\mathfrak{f}_{m,\pm}:=\frac{f_m(x_0)}{2\b_+}\pm
\frac{f_m(-x_0)}{2\b_-}, \quad
\mathfrak{u}_i:=\frac{u_{i,0}(x_0)}{2\b_+}-
\frac{u_{i,0}(-x_0)}{2\b_-},
\\
&\h\psi_{i,\pm}(\xi):=\frac{\t\psi_i(x_0,\xi)}{2\b_+} \pm
\frac{\t\psi_i(-x_0,\xi)}{2\b_-},  \quad
\mathfrak{a}_{i,\pm}:=\int\limits_0^1
a(\xi)\h\psi_{i,\pm}(\xi)\di\xi.
\end{aligned}
\end{equation}
Clearly, the inclusion $\supp f_m\subseteq[-x_0,x_0]$ is
equivalent to the equalities $\mathfrak{f}_{m,\pm}=0$. Let us
calculate the numbers $\mathfrak{f}_{m,\pm}$. From the formula
(\ref{5.23}) for $f_m$, the equalities (\ref{5.12a}), and the
condition (\ref{5.16}) for $u_{i,0}$, $1\leqslant i\leqslant
m-1$ it follows:
\begin{equation}\label{5.85}
\mathfrak{f}_{m,+}=-\mathfrak{a}_{m,+}+\mathfrak{l}_m,\quad
\mathfrak{f}_{m,-}=-\mathfrak{a}_{m,-}+\sum\limits_{j=2}^{m-2}
\mathfrak{l}_{m-j}\mathfrak{u}_j,\quad
\mathfrak{l}_i:=\l_i+\sum\limits_{j=4}^{i-4}\tau_j\tau_{i-j}.
\end{equation}
The equality $\mathfrak{f}_{m,+}=0$ is a consequence of the
formula (\ref{5.26}) for $\l_m$.

Let us prove that $\mathfrak{f}_{m,-}=0$. By (\ref{5.11b}) and
the equality $\t\psi_1=0$ we obtain:
\begin{equation}\label{5.36}
\h\psi_{1,\pm}=0,\quad\h\psi_{2,+}=\psi_{2,1} ,\quad
\h\psi_{2,-} =0,\quad \h\psi_{3,\pm}=0.
\end{equation}
By the induction assumption it implies that the functions
$\h\psi_{i+2,\pm}$, $i\leqslant m-1$, are 1-periodic solutions
of the equations:
\begin{equation}\label{5.33}
\begin{aligned}
&\frac{d^2}{d\xi^2}\h\psi_{2,+}=a,\quad
\frac{d^2}{d\xi^2}\h\psi_{i+2,+}=a\h\psi_{i,+}
-\mathfrak{a}_{i,+}
+2\sum\limits_{j=4}^{i-1}\tau_j\frac{d}{d\xi} \h\psi_{i-j+1,-}
-\sum\limits_{j=2}^{i-2}\mathfrak{l}_j\h\psi_{i-j,+},
\\
&\frac{d^2}{d\xi^2}\h\psi_{i+2,-}=a\h\psi_{i,-} -
\mathfrak{a}_{i,-} +2\sum\limits_{j=4}^{i-1}\tau_j
\frac{d}{d\xi}\h\psi_{i-j+1,+} -\sum\limits_{j=2}^{i-2}
\mathfrak{l}_j\h\psi_{i-j,-} + a\mathfrak{u}_i,\quad \xi\in
\mathbb{R},
\end{aligned}
\end{equation}
where $i\geqslant 1$. Due to the equalities (\ref{5.26}) for
$\l_i$, $i\leqslant m-1$, we have
$\mathfrak{l}_i=\mathfrak{a}_{i,+}$, $i\leqslant m-1$. Bearing
in mind these equalities and (\ref{5.25}), we multiply the
equation for $\h\psi_{i+2,+}$ in (\ref{5.33}) by
$\h\psi_{m-i,+}$ in $L_2(0,1)$ and make a summation over
$i=2,\ldots,m-2$. This procedure results in
\begin{equation}\label{5.34}
\begin{aligned}
&\sum\limits_{i=2}^{m-2} \mathfrak{l}_{m-i} \mathfrak{u}_i =
\sum\limits_{i=2}^{m-2}\Big(\h\psi_{m-i,+},\frac{d^2}{d\xi^2}
\h\psi_{i+2,-}\Big)-\sum\limits_{i=2}^{m-2}
\left(\h\psi_{m-i,+},a\h\psi_{i,-}\right)-
\\
&-2\sum\limits_{i=2}^{m-2}\sum\limits_{j=4}^{i-1}\tau_j
\Big(\h\psi_{m-i,+},\frac{d}{d\xi}\h\psi_{i-j+1,+}\Big)
+\sum\limits_{i=2}^{m-2}\sum\limits_{j=2}^{i-2}
\mathfrak{l}_j\left(\h\psi_{m-i,+},\h\psi_{i-j,-}\right) .
\end{aligned}
\end{equation}
Here and everywhere till the end of the proof the symbol
$(\cdot,\cdot)$ indicates the inner product in $L_2(0,1)$. The
third summand in the right hand side of this equality is zero,
what follows from the following realtions:
\begin{equation}
\begin{aligned}
&2\sum\limits_{i=2}^{m-2}\sum\limits_{j=4}^{i-1}\tau_j
\Big(\h\psi_{m-i,+}, \frac{d}{d\xi}\h\psi_{i-j+1,+}\Big)
=\sum\limits_{j=4}^{m-3}\tau_j\sum\limits_{i=j+1}^{m-2}
2\Big(\h\psi_{m-i,+}, \frac{d}{d\xi}\h\psi_{i-j+1,+}\Big)=
\\
&=\sum\limits_{j=4}^{m-3}\tau_j
\left(\sum\limits_{i=j+1}^{m-2}\Big(\h\psi_{m-i,+},
\frac{d}{d\xi}\h\psi_{i-j+1,+}\Big)+\sum\limits_{\t
i=j+1}^{m-2}\Big(\h\psi_{\t i-j+1,+}, \frac{d}{d\xi}\h\psi_{m-\t
i,+}\Big)\right)=0,
\end{aligned}\label{5.34a}
\end{equation}
where $\t i=m-i+j-1$. The equalities obtained imply that the
third summand in the right hand side of (\ref{5.34}) is zero.
Changing the summation indexes $i\mapsto m-i+2$, $i\mapsto m-i$
in the first and second summands in the right hand side of
(\ref{5.34}), and then integrating by parts in the first summand
and employing (\ref{5.25}), (\ref{5.36}), and (\ref{5.33}), we
deduce:
\begin{align*}
&\sum\limits_{i=2}^{m-2}\mathfrak{l}_{m-i}\mathfrak{u}_i
=\sum\limits_{i=0}^{m-4}\Big(\h\psi_{m-i,-},
\frac{d^2}{d\xi^2}\h\psi_{i+2,+}\Big)-
\sum\limits_{i=2}^{m-2}\left(\h\psi_{i,+},
a\h\psi_{m-i,-}\right)+
\\
&+ \sum\limits_{i=2}^{m-2}\sum\limits_{j=2}^{i-2}\mathfrak{l}_j
\left(\h\psi_{m-i,+},\h\psi_{i-j,-}\right)= \mathfrak{a}_{m,-}
+2\sum\limits_{i=1}^{m-4}\sum\limits_{j=4}^{i-1}\tau_j
\Big(\h\psi_{m-i,-},\frac{d}{d\xi}\h\psi_{i-j+1,-}\Big)-
\\
&-\sum\limits_{i=1}^{m-4}\sum\limits_{j=2}^{i-2}\mathfrak{l}_j
\left(\h\psi_{m-i,-},\h\psi_{i-j,+}\right)
+\sum\limits_{i=2}^{m-2}\sum\limits_{j=2}^{i-2}\mathfrak{l}_j
\left(\h\psi_{m-i,+},\h\psi_{i-j,-}\right).
\end{align*}
By analogy with (\ref{5.34a}) it is not difficult to show that
the second summand in the right hand side of the last equality
is zero. Taking into account (\ref{5.36}) and applying
Lemma~\ref{lm2.8} with $p=m$, $s=2$, $A_0=A_1=0$,
$A_j=\mathfrak{l}_j$, $j\geqslant 2$,
$B_{i,j}=\big(\h\psi_{i,+},\h\psi_{j,-}\big)$ to the last
summand, we arrive at the equality:
$\sum\limits_{i=2}^{m-2}\mathfrak{l}_{m-i}\mathfrak{u}_i=
\mathfrak{a}_{m,-}$, what together with (\ref{5.85}) imply that
$\mathfrak{f}_{m,-}=0$. The inclusion $\supp
f_m\subseteq[-x_0,x_0]$ is proven.

By (\ref{5.11a}), (\ref{5.24}) the function $f_m$ can be
represented as
\begin{equation*}
f_m(x)=-\tau_m\left(\psi_0(x)\frac{d^2}{dx^2}+2\frac{d\psi_0(x)}{dx}
\frac{d}{dx}\right)\left(|x|(1-\chi(x))\right)+\t f_m(x),
\end{equation*}
where $\t f_m$ is defined by the equality (\ref{5.27}). The
function $\t f_m$ is compactly supported, since the functions
$f_m$ and $\left(\psi_0(x)\frac{\displaystyle d^2}{\displaystyle
dx^2}+2\frac{\displaystyle d\psi_0(x)}{\displaystyle dx}
\frac{\displaystyle d}{\displaystyle
dx}\right)\left(|x|(1-\chi(x))\right)$ are compactly supported.

In virtue of Lemma~\ref{lm5.1} the equation (\ref{5.30}) is
solvable in the space $\mathcal{V}$, if the solvability
condition
\begin{align*}
0=\int\limits_\mathbb{R}\psi_0(x) f_m(x)\di
x=&-\tau_m\int\limits_\mathbb{R}\frac{d}{dx}
\left(\psi_0^2(x)\frac{d}{dx}\left(|x|(1-\chi(x))\right)\right)\di
x+
\\
&+\int\limits_\mathbb{R}\psi_0(x)\t f_m(x)\di
x=-\tau_m+\int\limits_\mathbb{R}\psi_0(x)\t f_m(x)\di x
\end{align*}
holds, what implies the equality (\ref{5.26}) for $\tau_m$. In
accordance with Lemma~\ref{lm5.1} the solution $u_{m,0}$ of the
equation (\ref{5.30}) is given by the formula (\ref{5.23}).

Substituting the representations (\ref{5.16a}) for $i\leqslant
m+1$ into the equation (\ref{5.11}) with $i=m$ and taking into
account (\ref{5.30}), we get
\begin{equation}\label{5.30a}
-\frac{\partial^2}{\partial\xi^2}\psi_{m+2}=G_{m+2},\quad
(x,\xi)\in \mathbb{R}^2,
\end{equation}
where $G_{m+2}$ is defined by the formula (\ref{5.22}). Using
(\ref{5.21}), (\ref{5.25}) with $i\leqslant m+1$, by analogy
with (\ref{4.33})--(\ref{4.35}) it is not difficult to show that
the function $G_{m+2}$ is the finite series
\begin{equation*}
G_{m+2}(x,\xi)=\sum\limits_p G_{m+2,p}^{(1)}(x)
G_{m+2,p}^{(2)}(\xi),
\end{equation*}
where $G_{m+2,p}^{(1)}\in C^\infty(\mathbb{R})\cap\mathcal{V}$,
$G_{m+2,p}^{(2)}\in C^2(\mathbb{R})$ are 1-periodic functions,
satisfying the equalities
\begin{equation*}
\int\limits_0^1 G_{m+2,p}^{(2)}(\xi)\di\xi=0.
\end{equation*}
Substituting the obtained representation for $G_{m+2}$ into the
equation (\ref{5.30a}) and solving then this equation by
Lemma~\ref{lm2.3}, we arrive at the formulas (\ref{5.16a}),
(\ref{5.21}) for  $\psi_{m+2}$, where $u_{m+2,0}$ is a some
function, $u_{m+2,j}\in C^\infty(\mathbb{R})\cap \mathcal{V}$,
$j\geqslant 1$, and $\psi_{m+2,j}\in C^2(\mathbb{R})$ are
1-periodic functions satisfying the equalities (\ref{5.25}).
\end{proof}

We proceed to proving the formulas (\ref{1.11}). From
(\ref{2.18}), (\ref{5.11b}), (\ref{5.23a}), (\ref{5.36}) it
follows that
\begin{equation}\label{5.40}
\psi_{2,1}=\phi^+_{0,1},\quad
\mathfrak{a}_{2,+}=\int\limits_0^1a\phi_{0,1}^+\di\xi=\mu^+_{0,1},
\quad \mathfrak{a}_{3,+}=0,
\end{equation}
what together with (\ref{5.26}) imply the formulas (\ref{1.11})
for $\l_2$, $\l_3$. Employing the definition (\ref{5.24}) of the
functions $f_i$, the formulas (\ref{5.11b}), and the equality
$\l_2=0$, it is easy to check that $f_2=0$, what by (\ref{5.23})
yields the identity
\begin{equation}\label{5.43}
u_{2,0}(x)\equiv0.
\end{equation}
In view of (\ref{5.11b}), (\ref{5.24}), and the formulas
(\ref{1.11}) for $\l_2$ and $\l_3$ we get that
$f_3=-2\frac{\displaystyle d\psi_0}{\displaystyle
dx}\int\limits_0^1a\psi_{3,1}\di\xi$. The functions
$\psi_{2,1}$, $\psi_{3,1}$ defined in (\ref{5.11b}) coincide
with the functions $\psi_{2,1}$, $\psi_{3,1}$ from (\ref{4.7}),
(\ref{4.13a}), this is why by (\ref{4.13b}) we have $f_3=0$. Now
from (\ref{5.23}) it follows that
\begin{equation}\label{5.44}
u_{3,0}\equiv0.
\end{equation}
This equality, (\ref{5.11b}), (\ref{5.22}), (\ref{5.40}),
(\ref{5.43}), and the formulas (\ref{1.11}) for $\l_2$, $\l_3$
lead us to
\begin{equation*}
G_4(x,\xi)=4\frac{d^2}{dx^2}\psi_0(x)\frac{d}{d\xi}\psi_{3,1}(\xi)+
(-a(\xi)\phi_{0,2}^+(\xi)+\mu_{0,1}^+)\psi_0(x),
\end{equation*}
what due to (\ref{5.21}) and (\ref{2.18}) imply that
\begin{equation}\label{5.45}
\begin{aligned}
&\mathfrak{m}_4=2, \quad u_{4,1}=\psi_0,\quad
\psi_{4,1}=\phi^+_{0,2},\quad
u_{4,2}=4\frac{d^2\psi_0}{dx^2},\quad
\psi_{4,2}={L}_0\left[\frac{d\psi_{3,1}}{d\xi}\right].
\end{aligned}
\end{equation}
Hence,
\begin{equation}\label{5.46}
\h\psi_{4,+} =\phi_{0,2}^+ ,\quad \h\psi_{4,-} =0,\quad
\mathfrak{a}_{4,+}=\mu_{0,2}^+,
\end{equation}
and the formula (\ref{1.11}) for $\l_4$ holds true. The formulas
(\ref{1.11}) for $\l_2$, $\l_4$, (\ref{5.26}), (\ref{5.27}), and
the equalities (\ref{2.18}), (\ref{5.11b}), (\ref{5.43}),
(\ref{5.45}) allow us to determine $\tau_4$:
\begin{equation*}
\tau_4=-4\int\limits_\mathbb{R}\psi_0\frac{d^2\psi_0}{dx^2}dx
\int\limits_0^1a\psi_{4,2}\di\xi.
\end{equation*}
The function $\psi_{4,2}$ from (\ref{5.45}) coinciding with the
function $\psi_{4,2}$ in (\ref{4.13d}), by (\ref{4.20a}) we
deduce:
\begin{equation}\label{5.48}
\tau_4=4\int\limits_\mathbb{R}\Big|\frac{d\psi_0}{dx}\Big|^2\di
x\int\limits_0^1|\phi_{0,1}^+|^2\di\xi.
\end{equation}
From (\ref{5.36}), (\ref{5.40}) it follows that the right hand
side of the equation (\ref{5.33}) for $\h\psi_{5,+}$ is zero,
this is why $\h\psi_{5,+}=0$, what due to (\ref{5.23a}) yields
$\mathfrak{a}_{5,+}=0$. Similarly, taking into account the
equalities $\mathfrak{l}_i=\mathfrak{a}_{i,+}$ established in
the proof of Lemma~\ref{lm5.2}, one can check that
$\mathfrak{a}_{7,+}=0$. The equalities
$\mathfrak{l}_i=\mathfrak{a}_{i,+}$ and (\ref{2.18}),
(\ref{5.36}), (\ref{5.40}), (\ref{5.46}) imply that the equation
(\ref{5.33}) for $\h\psi_{6,+}$ coincide with the equation
(\ref{2.13}) for $\phi_{0,3}^+$, because of this
$\h\psi_{6,+}=\phi_{0,3}^+$, what by (\ref{2.18}) gives rise to
$\mathfrak{a}_{6,+}=\mu_{0,3}^+$. In the same way one can show
that $\mathfrak{a}_{8,+}=\mu_{0,4}^+$. The equalities obtained
and (\ref{5.26}), (\ref{5.48}) lead us to the formulas
(\ref{1.11}) for $\l_6$, $\l_7$, $\l_8$.

Let $m\geqslant 8$. We denote
\begin{gather*}
\tau_{\e,m}:=\sum\limits_{i=4}^m\e^i\tau_i,\quad
\l_{\e,m}:=\sum\limits_{i=2}^m\e^i\l_i,
\\
\psi_{\e,m}(x):=h(x,\tau_{\e,m})\left(\psi_0(x)+
\sum\limits_{i=2}^m\e^i\psi_i\left(x,\frac{x}{\e}\right)\right).
\end{gather*}
The equations (\ref{5.11}), the equalities (\ref{5.11b}),
(\ref{5.48}), and Lemma~\ref{lm5.2} yield the following
statement.

\begin{lemma}\label{lm5.3}
The function $\psi_{\e,m}\in C^2(\mathbb{R})$ converges to
$\psi_0$ in $\H^1(Q)$ for each $Q\in \mathfrak{C}$ as $\e\to0$.
The functions $\psi_{\e,m}$ and $\l_{\e,m}$ satisfy the equation
\begin{equation}\label{5.50}
\left(-\frac{d^2}{dx^2}+V(x)+a\left(\frac{x}{\e}\right)-\l_{\e,m}
\right)\psi_{\e,m}=h(x,\tau_{\e,m})f_{\e,m}(x),\quad
x\in\mathbb{R},
\end{equation}
and the estimate
\begin{equation}\label{5.51}
\max\limits_\mathbb{R}|f_{\e,m}(x)|=\Odr(\e^{m-1})
\end{equation}
holds.
\end{lemma}

We set $k_{\e,m}:=\sqrt{\mu_0^+(\e^2)-\l_{\e,m}}$. Due to
(\ref{1.3}), (\ref{1.11}) we have the inequality
$\mu_0^+(\e^2)>\l_{\e,m}$, this is why we suppose that
$k_{\e,m}>0$. Then the formulas (\ref{1.3}), (\ref{1.11}) imply
\begin{equation}\label{5.57}
k_{\e,m}=\e^4\tau_4+\Odr(\e^5),
\end{equation}
where $\tau_4>0$ (see (\ref{5.48})). Let $\tau_m(\e):=\e^{-1}\ln
\k(\e,k_{\e,m})$, where, we remind, $\k$ is from (\ref{3.14}).
Then (\ref{3.14a}) and (\ref{5.57}) give rise to
\begin{equation}\label{5.59b}
\tau_m(\e)=\e^4\tau_4+\Odr(\e^5).
\end{equation}

\begin{lemma}\label{lm5.4}
For each $m$ the equality
\begin{equation*}
\tau_m(\e)=\tau_{\e,m}+\Odr(\e^{m-5})
\end{equation*}
takes place.
\end{lemma}
\begin{proof}
From (\ref{5.4}), (\ref{5.11b}), and Lemma~\ref{lm5.2} it
follows that for $x\geqslant x_0$ the function $\psi_{\e,m}(x)$
can be represented as
\begin{equation}\label{5.64a}
\psi_{\e,m}(x)=\E^{-\tau_{\e,m}x}\psi_{\e,m}^{per}
\left(\frac{x}{\e}\right), \quad \psi_{\e,m}^{per}(\xi)=\b_+ +
\e^2\widetilde{\psi}_{\e,m}^{per}(\xi),
\end{equation}
where 1-periodic function $\widetilde{\psi}_{\e,m}^{per}$ is
bounded uniformly on $\e$ in the norm of the space $C[0,1]$.
Employing this representation, the definition of the function
$\Th^+$, the equality (\ref{3.14b}), and the equation
(\ref{5.50}), it is not difficult to check that the function
\begin{equation}\label{5.103}
\mathcal{U}(x,\e)
:=W_x\left(\Th^+_{per}\left(\frac{x}{\e}\right),
\psi_{\e,m}^{per}\left(\frac{x}{\e}\right)\right)+
\left(\tau_m(\e)-\tau_{\e,m}\right)
\Th^+_{per}\left(\frac{x}{\e}\right)
\psi_{\e,m}^{per}\left(\frac{x}{\e}\right),
\end{equation}
where $\Th^+_{per}(\xi):=\Th^+_{per}(\xi,\e,k_{\e,m})$, obeys
the equation
\begin{equation}\label{5.104}
\frac{d}{dx}\mathcal{U}(x,\e)-(\tau_{\e,m}+\tau_m(\e))
\mathcal{U}(x,\e)=
-\Th^+_{per}\left(\frac{x}{\e}\right)f_{\e,m}(x), \quad
x\geqslant x_0.
\end{equation}
Let us integrate this equation on $x$ from $x_0$ to $(x_0+\e)$
and take into account the $\e$-periodicity of the functions
$\psi_{\e,m}^{per}\left(\frac{x}{\e}\right)$ and
$\Th_{per}^+\left(\frac{x}{\e}\right)$. As a result we arrive at
the following equality:
\begin{equation}\label{5.66}
\left(\tau_{\e,m}^2-\tau^2_{m}(\e)\right)
\int\limits_{x_0}^{x_0+\e} \Th_{per}^+\left(\frac{x}{\e}\right)
\psi_{\e,m}^{per}\left(\frac{x}{\e}\right)\di x=
\int\limits_{x_0}^{x_0+\e} \Th_{per}^+\left(\frac{x}{\e}\right)
f_{\e,m}(x)\di x.
\end{equation}
It follows from (\ref{3.22}) and (\ref{5.64a}) that the integral
in the left hand side of the last equality meets the relation
\begin{equation*}
\int\limits_{x_0}^{x_0+\e} \Th_{per}^+\left(\frac{x}{\e}\right)
\psi_{\e,m}^{per}\left(\frac{x}{\e}\right)\di
x=\e\b_++\Odr(\e^3),
\end{equation*}
what together with (\ref{5.51}), (\ref{5.66}) lead us to
\begin{equation}\label{5.66a}
\tau_{\e,m}^2-\tau_m^2(\e)=\Odr(\e^{m-1}).
\end{equation}
The conclusion of the lemma follows from the equality obtained
and (\ref{5.57}), (\ref{5.59b}).
\end{proof}

Further we will make use of the following auxiliary statement.

\begin{lemma}\label{lm5.5}
There exists a function $R_{\e,m}(x)\in C^2(\mathbb{R})$, such
that the function $\h\psi_{\e,m}:=\psi_{\e,m}(x)-R_{\e,m}(x)$
obeys the equation (\ref{3.16}) with $\l=\l_{\e,m}$, $f(x)=\h
f_{\e,m}(x)$, $\supp \h f_{\e,m}\subseteq[-x_0,x_0]$, and the
constraint (\ref{3.17}) with $k=k_{\e,m}$, and the estimates
\begin{equation}\label{5.56}
\|R_{\e,m}\|_{C^2(\overline{Q})}=\Odr(\e^{m-8}), \quad \|\h
f_{\e,m}\|_{C[-x_0,x_0]}=\Odr(\e^{m-8})
\end{equation}
are valid for each interval $Q\in \mathfrak{C}$.
\end{lemma}

\begin{proof}
For the sake of brevity throughout the proof we denote
$\Th^\pm(\xi):=\Th^\pm(\xi,\e,k_{\e,m})$. Let us calculate the
wronskian of the functions $\Th^+\left(\frac{x}{\e}\right)$ and
$\Th^-\left(\frac{x}{\e}\right)$, taking into account the
equalities (\ref{3.16a}), (\ref{3.15}), (\ref{5.59b}), and
Lemma~\ref{lm3.6}:
\begin{equation}
\begin{aligned}
W(\e)&:=W_x\left(\Th^+\left(\frac{x}{\e}\right),
\Th^-\left(\frac{x}{\e}\right)\right)=\e^{-1}W_\xi
\left(\Th^+(\xi), \Th^-(\xi)\right)=
\\
&\hphantom{:}=\e^{-1}\Th_{2}(1,\e^2,k_{\e,m}^2)
\left(\E^{\tau_m(\e)}-\E^{-\tau_m(\e)}\right)=2\e^3\tau_4+
\Odr(\e^4).
\end{aligned}\label{5.58}
\end{equation}
Since $W(\e)\not=0$, it follows that the functions $\Th^\pm$
form the fundamental set of solutions for the equation
\begin{equation}\label{5.59}
\left(-\frac{d^2}{dx^2}+a\left(\frac{x}{\e}\right)-\l_{\e,m}\right)
u=f,
\end{equation}
with $f=0$. Let $R_{\pm}=R_{\pm}(x,\e)$ be the solutions of
these equations with$f(x)=h(x,\tau_{\e,m})f_{\e,m}(x)$ defined
by the formulas
\begin{equation}\label{5.101}
R_{\pm}(x)=\frac{1}{W(\e)}\int\limits_{\pm x_0}^x \left(
\Th^+\left(\frac{x}{\e}\right)\Th^-\left(\frac{t}{\e}\right)-
\Th^-\left(\frac{x}{\e}\right)\Th^+\left(\frac{t}{\e}\right)
\right) h(t,\tau_{\e,m}) f_{\e,m}(t)\di t.
\end{equation}
Clearly, $R_\pm\in C^2(\mathbb{R})$. In view of the equations
(\ref{5.50}), (\ref{5.59}) and $V$ being compactly supported the
function $(\psi_{\e,m}(x)-R_+(x,\e))$ as $x\geqslant x_0$ is a
solution of the equation (\ref{5.59}) with homogeneous right
hand side, this is why the equality
\begin{equation}\label{5.100}
\psi_{\e,m}(x)-R_{+}(x,\e)=C_+^+\Th^+\left(\frac{x}{\e}\right)
+C_+^-\Th^-\left(\frac{x}{\e}\right),\quad x\geqslant x_0,
\end{equation}
is valid. Let us estimate the coefficient $C_+^-$. Putting
$x\geqslant x_0$ in (\ref{5.101}) and replacing the function
$h(t,\tau_{\e,m})f_{\e,m}(t)$ by the left hand side of the
equation (\ref{5.50}), by integration by parts we obtain:
\begin{equation}\label{5.102}
C_+^-=\frac{W_x\left(\Th^+\left(\frac{x}{\e}\right),\psi_{\e,m}
(x)\right)}{W(\e)}\Bigg|_{x=x_0}=
\frac{\mathcal{U}(x_0,\e)\E^{-(\tau_{\e,m}+\tau_m(\e))x_0}
}{W(\e)},
\end{equation}
where the function $\mathcal{U}$ is defined in accordance with
(\ref{5.103}). From the equation (\ref{5.104}) it follows that
\begin{equation*}
\mathcal{U}(x,\e)=\E^{(\tau_{\e,m}+\tau_m(\e))x} \left(
\mathcal{U}(x_0,\e) -\int\limits_{x_0}^x
\E^{-(\tau_{\e,m}+\tau_m(\e))t}
\Th^+_{per}\left(\frac{t}{\e}\right)f_{\e,m}(t)\di t\right).
\end{equation*}
The function $\mathcal{U}$ is $\e$-periodic for $x\geqslant
x_0$, what gives rise to the equality
$\mathcal{U}(x_0+\e,\e)=\mathcal{U}(x_0,\e)$, which can be
rewritten as
\begin{equation*}
\mathcal{U}(x_0,\e)=\E^{\e(\tau_{\e,m}+\tau_m(\e))} \left(
\mathcal{U}(x_0,\e) -\int\limits_{x_0}^{x_0+\e}
\E^{-(\tau_{\e,m}+\tau_m(\e))t}
\Th^+_{per}\left(\frac{t}{\e}\right)f_{\e,m}(t)\di t\right).
\end{equation*}
Due to (\ref{5.102}) it follows that
\begin{equation*}%\l%abel{7.105}
C_+^-=-\frac{\E^{\e(\tau_{\e,m}+\tau_m(\e))}\int\limits_{x_0}^{x_0+\e}
\E^{-(\tau_{\e,m}+\tau_m(\e))t}
\Th^+_{per}\left(\frac{t}{\e}\right)f_{\e,m}(t)\di
t}{\left(1-\E^{\e(\tau_{\e,m}+\tau_m(\e))}\right)W(\e)}.
\end{equation*}
From (\ref{3.22}) we deduce that the function
$\Th^+_{per}\left(\frac{x}{\e}\right)$ is bounded uniformly on
$x$ and $\e$. Taking into account this fact, (\ref{5.51}),
(\ref{5.59b}) and (\ref{5.58}), we arrive at the estimate:
\begin{equation}\label{5.105}
C_+^-=\Odr(\e^{m-8}).
\end{equation}
The form of the function $R_+$, (\ref{3.14a}), (\ref{3.14b}),
(\ref{3.22}), (\ref{5.51}), (\ref{5.58}) imply that
\begin{equation}\label{5.106}
\|R_+\|_{C(\overline{Q})}=\Odr(\e^{m-4})
\end{equation}
for each $Q\in \mathfrak{C}$. As one can easily check,
\begin{align*}
\frac{d}{dx}R_+(x,\e)= \frac{1}{W(\e)}\Bigg(&
\frac{d}{dx}\Th^+\left(\frac{x}{\e}\right) \int\limits_{x_0}^x
\Th^-\left(\frac{t}{\e}\right)h(t,\tau_{\e,m}) f_{\e,m}(t)\di t-
\\
&-\frac{d}{dx} \Th^-\left(\frac{x}{\e}\right)
\int\limits_{x_0}^{x}
\Th^+\left(\frac{t}{\e}\right)h(t,\tau_{\e,m}) f_{\e,m}(t)\di
t\Bigg).
\end{align*}
Basing on this equality and (\ref{3.14a}), (\ref{3.14b}),
(\ref{3.22}), (\ref{5.51}), (\ref{5.58}), we prove the estimate
\begin{equation*}
\left\|\frac{d}{dx}R_{\e,+}\right\|_{C(\overline{Q})}=\Odr(\e^{m}),
\end{equation*}
which is valid for each $Q\in \mathfrak{C}$. From the estimate
obtained, the equation $R_+$, (\ref{5.51}), (\ref{5.106}) we
deduce that
\begin{equation}\label{5.62}
\left\|R_+\right\|_{C^2(\overline{Q})}=\Odr(\e^{m-4})
\end{equation}
for each $Q\in \mathfrak{C}$. By analogy with (\ref{5.100}) one
can prove that
\begin{equation*}
\psi_{\e,m}(x)-R_{-}(x,\e)=C_-^+\Th^+\left(\frac{x}{\e}\right)
+C_-^-\Th^-\left(\frac{x}{\e}\right),\quad x\leqslant -x_0.
\end{equation*}
The constant $C_-^+$ is estimated in the same way as this was
done for $C_+^-$; the only difference is that the function
$\mathcal{U}$ should be chosen as follows:
\begin{equation*}
\mathcal{U}(x,\e):=
W_x\left(\psi_{\e,m}^{per}\left(\frac{x}{\e}\right),
\Th^-_{per}\left(\frac{x}{\e}\right)\right)+
\left(\tau_m(\e)-\tau_{\e,m}\right)
\Th^-_{per}\left(\frac{x}{\e}\right)
\psi_{\e,m}^{per}\left(\frac{x}{\e}\right).
\end{equation*}
Similarly to (\ref{5.105}), (\ref{5.62}) it not difficult to
show that the estimates
\begin{equation*}
C_-^+=\Odr(\e^{m-8}), \quad
\left\|R_-\right\|_{C^2(\overline{Q})}=\Odr(\e^{m-4})
\end{equation*}
hold true for each $Q\in \mathfrak{C}$. Employing the obtained
estimates for $C_-^+$, $C_+^-$, $R_\pm$ and (\ref{3.14b}),
(\ref{3.22}), one can easily show that the function
\begin{align*}
&R_{\e,m}(x):=\left(1-\chi(x)\right)\t R_{\e,m}(x)+C_+^-
\Th^-\left(\frac{x}{\e}\right)+C_-^+
\Th^+\left(\frac{x}{\e}\right),
\\
&\t R_{\e,m}(x):=\left\{
\begin{aligned}
&R_{\e,+}(x),\quad \text{as $x>0$},
\\
&R_{\e,-}(x),\quad \text{as $x<0$},
\end{aligned}
\right.
\end{align*}
satisfies the lemma.
\end{proof}

In virtue of the proven lemma the function $\h\psi_{\e,m}$ is a
solution to the problem (\ref{3.16}), (\ref{3.17}). Therefore,
the Lemma~\ref{lm3.11} is applicable to the function
$F_{\e,m}(x):=\h
f_{\e,m}(x)/{\vp}\left(\frac{x}{\e},\e^2\right)$, namely, the
representation
\begin{equation}\label{5.71}
\h\psi_{\e,m}=\frac{T_8(\e,k_{\e,m})F_{\e,m}}{k_{\e,m}-k_\e}
\psi_\e^{(0)}+{\vp}\left(\frac{x}{\e},\e^2\right)
T_9(\e,k_{\e,m})F_{\e,m}
\end{equation}
holds true, where $\psi_\e^{(0)}$ is from Lemma~\ref{lm3.10}.
From (\ref{3.38b}) a uniform on $\e$ estimate
\begin{equation}\label{5.72}
\|\psi_\e^{(0)}\|_{\H^1(-x_0,x_0)}\leqslant C
\end{equation}
follows. Bearing in mind this estimate, (\ref{5.56}),
Lemma~\ref{lm3.5}, and uniform on $\e$ boundedness of the
functional $T_8(\e,k_{\e,m})$ and the operator
$T_9(\e,k_{\e,m})$ (see Lemma~\ref{lm3.11}), from (\ref{5.71})
we get
\begin{equation*}
\|\h\psi_{\e,m}-{\vp}T_9(\e,k_{\e,m})F_{\e,m}\|_{L_2(-x_0,x_0)}
\leqslant\frac{C\e^{m-8}}{|k_{\e,m}-k_\e|},
\end{equation*}
what due to the convergence of $\h\psi_{\e,m}$ to $\psi_0$ in
$L_2(-x_0,x_0)$ (see Lemmas~\ref{lm5.3},~\ref{lm5.5})
imply
\begin{equation}\label{5.73}
k_\e-k_{\e,m}=\Odr(\e^{m-8}).
\end{equation}
Putting now $m=13$, by (\ref{5.101}) we obtain that the solution
of the equation (\ref{3.32}) satisfies the equality
\begin{equation*}
k_\e=\e^4\tau_4+\Odr(\e^5),
\end{equation*}
and this is why $\RE k_\e>0$. By Lemma~\ref{lm3.10} it implies
that the operator $H_\e$ has the unique eigenvalue converging to
zero as $\e\to0$ which is given by the equality
$\l_\e^{(0)}=\mu_0^+(\e^2)-k_\e^2$ and is simple. It also
follows from (\ref{5.73}) and the definition of $k_{\e,m}$ that
the asymptotics of this eigenvalue is given by (\ref{1.10}),
(\ref{1.11}). The proof of Theorem~\ref{th1.2a} is complete.

In conclusion let us find out the asymptotics of the
eigenfunction associated with $\l_\e^{(0)}$. Multiplying the
representation (\ref{5.71}) by $\psi_\e^{(0)}$ in
$L_2(-x_0,x_0)$, on the base of
Lemmas~\ref{lm3.11}~and~\ref{lm5.3} and the convergence
(\ref{3.38b}) by analogy with the proof of the equality
(\ref{4.16}) it is not difficult to establish the existence of a
function $c(\e)$ such that
\begin{equation}\label{5.59a}
c(\e)=\frac{T_8(\e,k_{\e,m})F_{\e,m}}{k_{\e,m}-k_\e}+\Odr(\e^{m-8}),
\quad c(\e)=1+O(1), \quad \e\to0.
\end{equation}
Taking into account this formula, from the estimates
(\ref{5.56}), (\ref{5.72}), and the representation (\ref{5.71})
we get
\begin{equation*}
\|\h\psi_{\e,m}-c(\e)\psi_\e^{(0)}\|_{\H^2(-x_0,x_0)}=\Odr(\e^{m-8}).
\end{equation*}
The functions $\h\psi_{\e,m}$ and $\psi_\e$ obeying the
equalities (\ref{3.17}) with $k=k_{\e,m}$ and $k=k_\e$,
respectively, the last equality, (\ref{5.56}), (\ref{5.73}), and
Lemmas~\ref{lm3.6},~\ref{lm5.4} yield that
\begin{equation*}
\|\psi_{\e,m}-c(\e)\psi_\e^{(0)}\|_{\H^2(Q)}=\Odr(\e^{m-8})
\end{equation*}
for each $Q\in \mathfrak{C}$. Thus, we have just proved

\begin{theorem}\label{th5.1}
The eigenfunction associated with the eigenvalue $\l_\e^{(0)}$
of the operator $H_\e$ can be chosen so that it satisfies the
asymptotics expansion (\ref{5.3}) in the norm of $\H^2(Q)$ for
each $Q\in \mathfrak{C}$. The coefficients of this expansion are
determined in accordance with (\ref{5.11b}), (\ref{5.43}),
(\ref{5.44}), (\ref{5.45}), and Lemma~\ref{lm5.2}. As $\pm
x\geqslant x_0$ this eigenfunction satisfies the equalities
(\ref{3.17}), and the multipliers $\k_\pm=\k^{\pm 1}(\e,k_\e)$
corresponding to the functions $\Th^\pm$ meet the equality
$\k(\e,k_\e)=\E^{-\tau_\e}$, where $\tau_\e$ has the asymptotics
(\ref{5.5}) with the coefficients defined by the formula
(\ref{5.48}) and Lemma~\ref{lm5.2}.
\end{theorem}

\sect{Existence of the eigenvalues in a finite lacuna}

The aim of the present section is to prove Theorem~\ref{th1.3a}
and item~\ref{it2th1.3b} of Theorem~\ref{th1.3b}. Throughout the
section the hypothesis of Theorem~\ref{th1.3a} is supposed to
hold.

\begin{lemma}\label{lm6.1}
The solutions $\vp_{\pm,i}=\vp_{\pm,i}(\xi,\e^2)$ of the
equation (\ref{2.6}) with $M=\e^2\mu_n^\pm(\e^2)$ subject to the
initial conditions
\begin{equation*}
\vp_{\pm,1}(0,\e^2)=1,\quad \vp'_{\pm,1}(0,\e^2)=0,\quad
\vp_{\pm,2}(0,\e^2)=0,\quad \vp'_{\pm,2}(0,\e^2)=1,
\end{equation*}
are of the form
\begin{align*}
&\vp_{\pm,1}=\cos\pi n\xi+\sum\limits_{j=1}^\infty \e^{2j}
{P}_{\pm}^j(\e^2)[\cos\pi n \xi],
\\
&\vp_{\pm,2}=\frac{1}{\pi n}\sin\pi n \xi+\frac{1}{\pi n
}\sum\limits_{j=1}^\infty \e^{2j} {P}_{\pm}^j(\e^2)[\sin\pi n
\xi]
\end{align*}
for all sufficiently small $\e$. Here the operator
${P}_{\pm}(\e^2)$ is defined by the equality
\begin{equation*}
{P}_{\pm}(\e^2)[f](\xi,\e^2):=\frac{1}{\pi n }
\int\limits_0^\xi(a(\eta)-\mu_n^\pm(\e^2)+\pi^2 n^2 \e^{-2})\sin
\pi n(\xi-\eta) f(\eta)\di\eta
\end{equation*}
and is a linear bounded operator from $C[0,1]$ into $C^2[0,1]$
holomorphic on $\e^2$. The functions $\vp_{\pm,i}$ are
holomorphic on $\e^2$ in the norm of $C^2[0,1]$.
\end{lemma}

\begin{lemma}\label{lm6.2}
The solutions $\Th_{\pm,i}=\Th_{\pm,i}(\xi,\e^2,k^2)$  of the
equation (\ref{2.6}) $M=\e^2(\mu_n^\pm(\e^2)\mp k^2)$, where $k$
is a complex parameter, subject to the initial conditions
\begin{align*}
&\Th_{\pm,1}(0,\e^2,k^2)=1,\quad \Th'_{\pm,1}(0,\e^2,k^2)=0,
\\
&\Th_{\pm,2}(0,\e^2,k^2)=0,\quad \Th'_{\pm,2}(0,\e^2,k^2)=1,
\end{align*}
are of the form
\begin{equation*}
\Th_{\pm,i}=\vp_{\pm,i}+\sum\limits_{j=1}^\infty \e^{2j} k^{2j}
\widetilde{{P}}_{\pm}^j(\e^2)[\vp_{i}],
\end{equation*}
for all sufficiently small $\e$. Here the operator
$\widetilde{{P}}_{\pm}(\e^2)$ is defined by the equality
\begin{equation*}
\widetilde{{P}}_{\pm}(\e^2)[f](\xi,\e^2):=\pm \int\limits_0^\xi
\left(\vp_{\pm,2}(\xi,\e^2)\vp_{\pm,1}(\eta,\e^2)-
\vp_{\pm,2}(\eta,\e^2)\vp_{\pm,1}(\xi,\e^2)\right)
f(\eta)\di\eta
\end{equation*}
and is a linear bounded operator from $C[0,1]$ into $C^2[0,1]$
holomorphic on $\e^2$. The functions  $\Th_{\pm,i}$ are
holomorphic on $\e^2$ and $k^2$ in the norm of $C^2[0,1]$.
\end{lemma}

The proof of these lemmas is completely similar to the proof of
Lemmas~\ref{lm3.5},~\ref{lm3.6}.

Throughout this section we will not indicate explicitly the
dependence of the functions $\vp_{\pm,i}$ and $\Th_{\pm,i}$ on
$\e$ and $k$, putting for the sake of brevity
$\vp_{\pm,i}(\xi):=\vp_{\pm,i}(\xi,\e^2)$,
$\Th_{\pm,i}(\xi):=\vp_{\pm,i}(\xi,\e,k)$, $i=1,2$.

Directly from Lemmas~\ref{lm6.1},~\ref{lm6.2} we deduce
\begin{equation}
\begin{aligned}
&\vp_{\pm,1}(1)=(-1)^n\left(1-\e^2\frac{b_n}{2\pi
n}\right)+\Odr(\e^4),
\\
&\vp'_{\pm,1}(1)=(-1)^n\e^2\frac{a_n\mp\mu_{n,0}}{2}+\Odr(\e^4),
\\
&\vp_{\pm,2}(1)=(-1)^n\e^2\frac{a_n\pm\mu_{n,0}}{2\pi^2 n^2
}+\Odr(\e^4),
\\
&\vp'_{\pm,2}(1)=(-1)^n\left(1+\e^2\frac{b_n}{2\pi
n}\right)+\Odr(\e^4),
\end{aligned}\label{6.1}
\end{equation}
as well as
\begin{equation}
\begin{aligned}
&\Th_{\pm,1}(1)=(-1)^n\left(1-\e^2\frac{b_n}{2\pi
n}\right)+\Odr(\e^4
k^2),
\\
&\Th'_{\pm,1}(1)=(-1)^n\e^2\frac{a_n\mp\mu_{n,0} \pm
k^2}{2}+\Odr(\e^4 k^2),
\\
&\Th_{\pm,2}(1)=(-1)^n\e^2\frac{a_n\pm\mu_{n,0}\mp k^2 }{2\pi^2
n^2}+\Odr(\e^4 k^2),
\\
&\Th'_{\pm,2}(1)=(-1)^n\left(1+\e^2\frac{b_n}{2\pi
n}\right)+\Odr(\e^4 k^2),
\end{aligned}\label{6.1a}
\end{equation}
where $\mu_{n,0}:=\sqrt{a_n^2+b_n^2}>0$.

\begin{lemma}\label{lm6.3}
Let $c>0$ be an arbitrary constant independent on $\e$ and $k$.
The functions $\Th_{\pm,i}$ are holomorphic on $k$ in the norm
of $C^2[-c\e^{-1},c\e^{-1}]$. In the sense of the same norm the
functions $\Th_{\pm,i}$ together with all their derivatives on
$k$ are bounded uniformly on $\e$ and $k$.
\end{lemma}
\begin{proof}
Lemma~\ref{lm6.2} and equalities (\ref{6.1})
imply
\begin{equation*}
\|\vp_{\pm,i}\|_{C^2[0,1]}\leqslant C_0, \quad
|\vp_{\pm,1}(1)|+|\vp_{\pm,2}(1)|+|\vp'_{\pm,1}(1)|+
|\vp'_{\pm,2}(1)|\leqslant 1+c\e^2,
\end{equation*}
where the constants $C_0$ and $c$ are independent on $\e$. Let a
uniform on $\e$ estimate
\begin{equation*}
\|\vp_{\pm,i}\|_{C^2[m-1,m]}\leqslant C_m,\quad i=1,2,
\end{equation*}
holds true for an integer $m$ with a constant $C_m$. Then it
follows from the equality
\begin{equation}\label{6.11}
\vp_{\pm,i}(\xi+1)=\vp_{\pm,i}(1)\vp_{\pm,1}(\xi)+
\vp'_{\pm,i}(1)\vp_{\pm,2}(\xi)
\end{equation}
that
\begin{equation*}
\|\vp_{\pm,i}\|_{C^2[m,m+1]}\leqslant
\left(|\vp_{\pm,i}(1)|+|\vp'_{\pm,i}(1)|\right)
\|\vp_{\pm,i}\|_{C^2[m-1,m]}\leqslant C_m(1+c\e^2),
\end{equation*}
and the same estimate is valid for
$\|\vp_{\pm,i}\|_{C^2[m-2,m-1]}$. Applying the estimates
obtained by induction, we arrive at the inequality
\begin{equation*}
\|\vp_{\pm,i}\|_{C^2[m,m+1]}\leqslant C_0(1+c\e^2)^{m}, \quad
m\in \mathbb{Z},
\end{equation*}
which yields
\begin{equation*}
\|\vp_{\pm,i}\|_{C^2[-c\e^{-1},c\e^{-1}]}\leqslant
C_0(1+c\e^2)^{c\e^{-1}+1}\leqslant C,
\end{equation*}
where the constant $C$ is independent on $\e$. Therefore, the
operator $\t P_\pm(\e^2)$ from Lemma~\ref{lm6.2} satisfies a
uniform on $\e$ estimate
\begin{equation*}
\|\t P_\pm(\e^2)[f]\|_{C^2[-c\e^{-1},c\e^{-1}]}\leqslant
C\e^{-1}\|f\|_{C[-c\e^{-1},c\e^{-1}]}
\end{equation*}
for each fixed $c>0$. In view of Lemma~\ref{lm6.2} it implies
the conclusion of the lemma.
\end{proof}

Let us calculate the multipliers $\k_\pm^+=\k_\pm^+(\e,k)$ and
$\k_\pm^-=\k_\pm^-(\e,k)$ corresponding to the equation
(\ref{2.6}) with $M=\e^2(\mu_n^\pm(\e^2)\mp k^2)$. By the
formula (8.13) from \cite[Ch. 2, \S 2.8]{Sh} these multipliers
can be defined as follows:
\begin{align}
&\k_\pm^+=\k_\pm,\quad \k_\pm^-=\frac{1}{\k_\pm},\quad
\k_\pm=\frac{D_\pm+(-1)^n\sqrt{D_\pm^2-4}}{2},\label{6.2}
\\
&D_\pm=D_\pm(\e^2,k^2):=\Th_{\pm,1}(1,\e^2,k^2)+
\Th'_{\pm,2}(1,\e^2,k^2).\nonumber
\end{align}
Since $\mu_n^\pm(\e^2)$ are edges of the essential spectrum of
the operator $H_\e$ and the corresponding solutions $\phi_n^\pm$
of the equation (\ref{2.6}) obey either periodic or antiperiodic
boundary conditions (\ref{2.7}), (\ref{2.8}), according to
\cite[Ch. 2, \S 2.8]{Sh} it follows that the equality
\begin{equation}\label{6.3}
\vp_{\pm,1}(1)+\vp'_{\pm,2}(1)=2(-1)^n
\end{equation}
holds true. Using this equality, the relations (\ref{6.1a}), and
Lemmas~\ref{lm6.1},~\ref{lm6.2}, by direct calculate one can
check that
\begin{align}
&D_\pm(\e^2,k^2)=(-1)^n\left(2+\frac{\e^4 k^2}{4\pi^2 n^2}
\left(\t
D_\pm(\e^2)-k^2\right)\right)+\e^6k^4\widehat{D}_\pm(\e^2,k^2),\nonumber
\\
&\t D_\pm(\e^2):=\frac{4\pi^2n^2}{\e^2}\left(\widetilde{P}_\pm
[\vp_{\pm,1}](\xi,\e^2)+\frac{d}{d\xi}\widetilde{P}_\pm
[\vp_{\pm,2}](\xi,\e^2)\right)\bigg|_{\xi=1}=2\mu_{n,0}+\Odr(\e^2),
\label{6.3a}
\end{align}
where $\widehat{D}_\pm$ and $\t D_\pm$ are holomorphic function.
Substituting the equality obtained into (\ref{6.2}) leads us to
the following representation for $\k_\pm$:
\begin{equation}\label{6.4}
\begin{aligned}
&\k_\pm(\e,k)=(-1)^n\left(1+\frac{\e^2k}{2\pi n}\sqrt{\t
D_\pm(\e^2)-k^2}\right)+\e^4k^2\t\k_\pm(\e,k),
\\
&\ln\k_\pm(\e,k)=\ln\big((-1)^n\big)+\frac{\e^2k}{2\pi
n}\sqrt{2\mu_{n,0}-k^2}+\Odr(\e^4k),
\end{aligned}
\end{equation}
where $\t\k_\pm(\e,k)$ is a holomorphic function on two
variables and a logarithm branch is chosen by the condition $\ln
1=0$ and, therefore, $\ln(-1)=\pi\iu$. Here we also assume that
$|2\mu_{n,0}-k^2|\geqslant c>0$, where the constant $c$ is
independent on $\e$ and $k$. In accordance with Floquet-Lyapunov
theorem, the equation (\ref{2.6}) with
$M=\e^2\left(\mu_n^\pm(\e^2)\mp k^2\right)$ has a fundamental
set of solutions of the form
\begin{equation}\label{6.6}
\begin{aligned}
&\Th^+_\pm=\Th^+_\pm(\xi,\e,k)=\E^{-\xi\ln\k_\pm(\e,k)}
\Th_{per,\pm}^+(\xi,\e,k),
\\
& \Th^-_\pm=\Th^-_\pm(\xi,\e,k)=\E^{\xi\ln\k_\pm(\e,k)}
\Th_{per,\pm}^-(\xi,\e,k),
\end{aligned}
\end{equation}
where the functions $\Th_{per,\pm}^\pm(\xi,\e,k)$ are
1-periodic. Everywhere till the end of the section we will omit
the dependence on $\e$ and $k$ in the notations of the functions
$\Th^\pm_\pm$, putting for the sake of brevity:
$\Th_\pm^+(\xi):=\Th_\pm^+(\xi,\e,k)$,
$\Th_\pm^-(\xi):=\Th_\pm^-(\xi,\e,k)$.

Let us consider the equation (\ref{3.16}) with
$\l=\mu_n^\pm(\e^2)\mp k^2$. We will seek a solution of this
equation obeying the constraints:
\begin{equation}\label{6.17}
\begin{aligned}
&u(x,\e,k)=c_+(\e,k)\Th_\pm^+\left(\frac{x}{\e}\right),\quad
x\geqslant x_0,
\\
&u(x,\e,k)=c_-(\e,k)\Th_\pm^-\left(\frac{x}{\e}\right),\quad
x\leqslant -x_0,
\end{aligned}
\end{equation}
where $c_\pm(\e,k)$ are some constants. We denote
\begin{equation}\label{6.9a}
\begin{aligned}
&\t\Th_{\pm}^1(\xi,s):=\Th_{\pm,2}(1)\Th_{\pm,1}(\xi)+
\big(s-\Th_{\pm,1}(1)\big)\Th_{\pm,2}(\xi),
\\
&\t\Th_{\pm}^2(\xi,s):=-\Th'_{\pm,1}(1)\Th_{\pm,2}(\xi)+
\big(s-\Th_{\pm,1}(1)\big)\Th_{\pm,1}(\xi).
\end{aligned}
\end{equation}
Let $s=\k_\pm$ or $s=\k_\pm^{-1}$. Employing the equalities
\begin{equation}\label{6.7}
\begin{aligned}
&\Th_{\pm,i}(\xi+1)=\Th_{\pm,i}(1)\Th_{\pm,1}(\xi)+
\Th'_{\pm,i}(1)\Th_{\pm,2}(\xi),
\\
&W_\xi\left(\Th_{\pm,1}(\xi),\Th_{\pm,2}(\xi)\right)\equiv1,\quad
\k_\pm+\k_\pm^{-1}=\Th_{\pm,1}(1)+\Th'_{\pm,2}(1),
\end{aligned}
\end{equation}
it is easy to show that
$\t\Th_{\pm}^1(\xi+1,s)=s\t\Th_{\pm}^1(\xi,s)$,
$\t\Th_{\pm}^2(\xi+1,s)=s^{-1}\t\Th_{\pm}^2(\xi,s)$, what
implies
\begin{equation}\label{6.90}
\begin{aligned}
&\t\Th_\pm^1(\xi,\k_\pm)=c_1^\pm\Th_\pm^-(\xi),\quad
\t\Th_\pm^2(\xi,\k_\pm^{-1})=c_2^\pm\Th_\pm^+(\xi),
\\
&\t\Th_\pm^1(\xi,\k_\pm^{-1})=c_3^\pm\Th_\pm^+(\xi),\quad
\t\Th_\pm^2(\xi,\k_\pm)=c_4^\pm\Th_\pm^-(\xi),
\end{aligned}
\end{equation}
where $c_i^\pm$ are some constants depending on $\e$ and $k$
only.

On functions $g\in L_2(-x_0,x_0)$, $\supp g\subseteq[-x_0,x_0]$,
we define the operator
\begin{gather}
T_{11}^\pm(\e,k)g:=\frac{1}{\k_\pm(\e,k)-\k_\pm^{-1}(\e,k)}
T_{12}^\pm(\e,k)g,\label{6.20}
\\
\begin{aligned}
T_{12}^\pm g:=&\int\limits_{-\infty}^x
\left(\t\Th_\pm^1\left(\frac{x}{\e},\k_\pm^{-1}\right)
\Th_{\pm,1}\left(\frac{t}{\e}\right)+
\t\Th_\pm^2\left(\frac{x}{\e},\k_\pm\right)
\Th_{\pm,2}\left(\frac{t}{\e}\right)\right)g(t)\di t+
\\
&+\int\limits_x^{+\infty}
\left(\t\Th_\pm^1\left(\frac{x}{\e},\k_\pm\right)
\Th_{\pm,1}\left(\frac{t}{\e}\right)+
\t\Th_\pm^2\left(\frac{x}{\e},\k_\pm^{-1}\right)
\Th_{\pm,2}\left(\frac{t}{\e}\right)\right)g(t)\di t.
\end{aligned}\nonumber
\end{gather}

\begin{lemma}\label{lm6.7}
For each $Q\in \mathfrak{C}$ the operator $T_{11}^\pm$ is a
linear bounded operator from $L_2(-x_0,x_0)$ into $\H^2(Q)$. The
problem (\ref{3.16}), (\ref{6.17}) with $\l=\mu_n^\pm(\e^2)\mp
k^2$ is equivalent to the following operator equation in
$L_2(-x_0,x_0)$:
\begin{equation}
g+\e V T_{11}^\pm(\e,k)g=f,\label{6.18}
\end{equation}
where
\begin{equation}\label{6.16a}
g:=f-Vu,\quad u=\e T_{11}^\pm(\e,k)g.
\end{equation}
\end{lemma}

\begin{remark}\label{rm6.1}
In the lemma the functions from $L_2(-x_0,x_0)$ those the
operator $T_{11}^\pm$ is applied to are supposed to be extended
by zero outside the segment $[-x_0,x_0]$.
\end{remark}

\begin{proof}
The boundedness of the operator $T_{11}^\pm: L_2(-x_0,x_0)\to
\H^2(Q)$, $Q\in \mathfrak{C}$, follows directly from the
definition of this operator. Let $g$ be a solution of the
equation (\ref{6.18}). We define the function $u$ in accordance
with (\ref{6.16a}). In view of the definition of the functions
$t\Th_\pm^i$ one can easily check that the function  $u$ is a
solution of the equation
\begin{equation}\label{6.101}
\left(-\frac{d^2}{dx^2}+a\left(\frac{x}{\e}\right)-
\mu_n^\pm(\e^2)\pm k^2\right)u=g,\quad x\in \mathbb{R}.
\end{equation}
For $x\leqslant x_0$ the function $u$ takes the form:
\begin{align*}
&u(x)=\t\Th_{\pm}^1\left(\frac{x}{\e},\k_\pm^{-1}\right)
\int\limits_\mathbb{R} \Th_{\pm,1}\left(\frac{t}{\e}\right) g(t)
\di t+\t\Th_{\pm}^2\left(\frac{x}{\e},\k_\pm\right)
\int\limits_\mathbb{R} \Th_{\pm,2}\left(\frac{t}{\e}\right) g(t)
\di t,
\end{align*}
what together with (\ref{6.90}) yield that the function $u$
satisfies the former of the constraints (\ref{6.17}). In same
way one can prove that the function $u$ satisfies the latter of
the constraints (\ref{6.17}).

If $u$ is a solution to the problem (\ref{3.16}), (\ref{6.17}),
it is a solution of the equation (\ref{6.101}) with the right
hand side defined by (\ref{6.16a}). Taking into accound the
definition of the operator $T_{11}^\pm$, we get that $u=\e
T_{11}^\pm g$. Substituting this equality into (\ref{6.101}), we
arrive at the equation (\ref{6.18}).
\end{proof}

\begin{lemma}\label{lm6.4}
For each $\d>0$ the segment
$[\mu_n^-(\e^2)+\d,\mu_n^+(\e^2)-\d]$ of the real axis contains
no eigenvalues of the operator $H_\e$, if  $\e$ is small enough.
\end{lemma}

\begin{proof}
For $\l\in[\mu_n^-(\e^2)+\d,\mu_n^+(\e^2)-\d]$ we set
$k:=\sqrt{\mu_n^+(\e^2)-\l}>0$. Then it follows from
(\ref{1.4}), (\ref{1.6}) that for all sufficiently small $\e$
the estimate $0<c\leqslant k^2\leqslant 2\mu_{n,0}-c$ holds
where $c$ is a constant independent on $\e$ and $k$. This
estimate and (\ref{6.3a}), (\ref{6.4}) yield that the
denominator in (\ref{6.20}) obeys the inequality:
\begin{equation}\label{6.100}
|\k_\pm(\e,k)- \k_\pm^{-1}(\e,k)|\geqslant C\e^2,
\end{equation}
where the constant $C>0$ is independent on $\e$ and $k$. The
equalities (\ref{6.1a}), (\ref{6.3a}), (\ref{6.4}) and
Lemma~\ref{lm6.3} allow to estimate the norm of the operator
$T_{12}^\pm(\e,k): L_2(-x_0,x_0)\to L_2(-x_0,x_0)$ uniformly on
$\e$ and $k$ as follows: $\|T_{12}^\pm\|\leqslant C\e^2$. By
(\ref{6.100}) it implies that the operator $VT_{11}^\pm:
L_2(-x_0,x_0)\to L_2(-x_0,x_0)$ is bounded uniformly on $\e$ and
$k$. Therefore, for all sufficiently small $\e$ the operator
$(\I+\e VT_{11}^\pm(\e,k))$ is uniformly bounded, this is why
the equation (\ref{6.18}) with $f=0$ has a trivial solution
only. In virtue of Lemma~\ref{lm6.7} it follows that the problem
(\ref{3.16}), (\ref{6.17}) with $f=0$,
$\l\in[\mu_n^-(\e^2)+\d,\mu_n^+(\e^2)-\d]$ has no nontrivial
solution for all sufficiently small $\e$, what completes the
proof.
\end{proof}

It follows from the proven lemma that the eigenvalues of the
operator $H_\e$ located in lacuna
$(\mu_n^-(\e^2),\mu_n^+(\e^2))$ obey the equality (\ref{1.30}).
We will seek these eigenvalues as
$\l^{(n)}_{\e,\pm}:=\mu_n^\pm(\e^2)\mp k_{\e,\pm}^2$, where
$k_{\e,\pm}\to0$ as $\e\to0$. We study the existence of the
eigenvalues $\l^{(n)}_{\e,\pm}$ of the operator $H_\e$ by the
scheme which is similar to one employed in the third section in
studying the eigenvalue tending to zero as $\e\to0$.

Using Lemmas~\ref{lm6.2},~\ref{lm6.3}, the equalities
(\ref{6.1a}), (\ref{6.4}), and
\begin{equation*}
\k_\pm-\k_\pm^{-1}=\left(\k_\pm-(-1)^n\right)\left(1+(-1)^n\k_\pm^{-1}\right),
\end{equation*}
from (\ref{6.20}) and the definition of the functions
$\t\Th_\pm^i$ we deduce that the operator $T_{11}^\pm$ can be
represented as
\begin{gather}
T_{11}^\pm(\e,k)=\frac{1}{\k_\pm-\k_\pm^{-1}}
T_{13}^\pm(\e)+T_{14}^\pm(\e)+k T_{15}^\pm(\e,k),\label{6.21}
\\
\begin{aligned}
&T_{13}^\pm(\e) g=\left(
\vp_{\pm,2}(1)\vp_{\pm,1}\left(\frac{x}{\e}\right) +
\big((-1)^n-\vp_{\pm,1}(1)\big)\vp_{\pm,2}\left(\frac{x}{\e}\right)
\right)\int\limits_{\mathbb{R}}\vp_{\pm,1}\left(\frac{t}{\e}\right)
g(t)\di t+
\\
&+\left(-\vp'_{\pm,1}(1)\vp_{\pm,2}\left(\frac{x}{\e}\right) +
\big((-1)^n-\vp_{\pm,1}(1)\big)\vp_{\pm,1}\left(\frac{x}{\e}\right)
\right)\int\limits_{\mathbb{R}}\vp_{\pm,2}\left(\frac{t}{\e}\right)
g(t)\di t,
\end{aligned}\nonumber
\\
T_{14}^\pm(\e)g=\frac{1}{2}\int\limits_{\mathbb{R}} \left(
\vp_{\pm,1}\left(\frac{x}{\e}\right)
\vp_{\pm,2}\left(\frac{t}{\e}\right)
-\vp_{\pm,2}\left(\frac{x}{\e}\right)
\vp_{\pm,1}\left(\frac{t}{\e}\right)\right)\sgn(x-t)g(t)\di t,
\label{6.21a}
\end{gather}
where for each $Q\in \mathfrak{C}$ the linear operator
$T_{15}^\pm(\e,k): L_2(-x_0,x_0)\to \H^2(Q)$ together with its
derivative $\frac{\displaystyle d}{\displaystyle
dk}T_{15}^\pm(\e,k)$ are holomorphic on $k$ and bounded
uniformly on $\e$ and $k$. Since
$\vp_{\pm,i}(\xi,\e^2)=\Th_{\pm,i}(\xi,\e^2,0)$, it follows from
Lemma~\ref{lm6.3} that the linear operators $T_{13}^\pm(\e),
T_{14}(\e): L_2(-x_0,x_0)\to \H^2(Q)$ are bounded uniformly on
$\e$ for each $Q\in \mathfrak{C}$.

We set
\begin{equation*}
\vp_\pm(\xi,\e^2):=
\vp_{\pm,1}(\xi,\e^2)+\frac{(-1)^n-\vp_{\pm,1}(1,\e^2)}
{\vp_{\pm,2}(1,\e^2)}\vp_{\pm,2}(\xi,\e^2),
\end{equation*}
if $b_n\not=0$ or $b_n=0$, $\sgn a_n=\pm 1$, and
\begin{equation*}
\vp_\pm(\xi,\e^2):=
\vp_{\pm,2}(\xi,\e^2)-\frac{(-1)^n-\vp_{\pm,1}(1,\e^2)}
{\vp'_{\pm,1}(1,\e^2)}\vp_{\pm,1}(\xi,\e^2),
\end{equation*}
if $b_n=0$, $\sgn a_n=\mp 1$. The function $\vp_\pm$ is well
defined, since the denominators $\vp_{\pm,2}(1)$ and
$\vp'_{\pm,1}(1)$ in its definition are nonzero due to
(\ref{6.1}). The equalities (\ref{6.11}) imply that the function
$\vp_\pm$ is 1-periodic on $\xi$ for even $n$ and 1-antiperiodic
for odd $n$. Taking into account (\ref{6.1}) and
Lemma~\ref{lm6.1} it is easy to make sure that the function
$\vp_\pm$ is holomorphic on $\e^2$ in the norm of  $C^2[0,1]$.
For the sake of brevity everywhere till the end of the section
we denote $\vp_\pm(\xi):=\vp_\pm(\xi,\e^2)$.

The function $\vp_\pm$ is a solution of the boundary value
problem (\ref{2.6}), (\ref{2.7}) for even $n$ and the boundary
value problem (\ref{2.6}), (\ref{2.8}) for odd $n$ with
$M=\e^2\mu_n^\pm(\e^2)$. Therefore,
$\vp_\pm=C_\pm(\e^2)\phi_n^\pm$, where $C_\pm$ is a constant,
and $\phi_n^\pm$, we remind, is defined in accordance with
(\ref{2.20}), (\ref{2.9c}) and Lemma~\ref{lm2.7}. By the
formulas (\ref{6.1}) and Lemma~\ref{lm6.1} the equality
\begin{equation*}
\vp_\pm(\xi,0)=\left\{
\begin{aligned}
&\cos\pi n\xi+\frac{b_n\sin\pi n\xi}{a_n\pm\mu_{n,0}},
&&\text{if $b_n\not=0$ or $b_n=0$, $\sgn a_n=\pm 1$},
\\
&\frac{\sin\pi n\xi}{\pi n}-\frac{b_n\cos\pi n\xi}{\pi n
(a_n\mp\mu_{n,0})},&&\text{if $b_n=0$, $\sgn a_n=\mp 1$},
\end{aligned}
\right.
\end{equation*}
holds true. In virtue of holomorphy on $\e^2$ of the functions
$\vp_\pm$ from the last equality we deduce that $C_\pm$ is
holomorphic on $\e^2$ and $C_\pm(\e^2)=C_\pm(0)+\Odr(\e^2)$,
where $C_\pm(0)$ is a nonzero real constant.

From the relations (\ref{6.3}) and
$W_\xi\left(\vp_{\pm,1}\left(\frac{x}{\e}\right),
\vp_{\pm,2}\left(\frac{x}{\e}\right)\right)\equiv1$ one can
easily obtain the equality
$\left((-1)^n-\vp_{\pm,1}\right)^2=-\vp'_{\pm,1}(1)\vp_{\pm,2}(1)$,
which is being taken into account, by direct calculations we
check that
\begin{align}
&T_{13}^\pm(\e) g=\e^2
c_\pm(\e^2)\phi_n^\pm\left(\frac{x}{\e},\e^2\right)
T_{16}^\pm(\e)g, \quad T_{16}^\pm(\e)g:=
\int\limits_{\mathbb{R}}\phi_n^\pm\left(\frac{t}{\e},\e^2\right)
g(t)\di t,\label{6.24}
\\
&c_\pm(\e^2)=\left\{
\begin{aligned}
&\frac{\vp_{\pm,2}(1,\e^2)C_\pm^2(\e^2)}{\e^2},&&\text{if
$b_n\not=0$ or $b_n=0$, $\sgn a_n=\pm 1$},
\\
-&\frac{\vp'_{\pm,1}(1,\e^2)C_\pm^2(\e^2)}{\e^2},&&\text{if
$b_n=0$, $\sgn a_n=\mp 1$}.
\end{aligned}
\right.\nonumber%\l%abel{6.24a}
\end{align}
The functional $T_{16}^\pm(\e): L_2(-x_0,x_0)\to \mathbb{C}$ is
bounded uniformly on $\e$, since the function
$\phi_n^\pm(\xi,\e^2)$ is 1-periodic and holomorphic on $\e^2$
in the norm $C^2[0,1]$. The equalities  (\ref{6.21}),
(\ref{6.24}) allow to rewrite the equation (\ref{6.18}) as
\begin{equation}\label{6.25}
g+\frac{\e^3 c_\pm V\phi_n^\pm}{\k_\pm-\k_\pm^{-1}}T_{16}^\pm
g+\e V T_{14}^\pm g+\e k V T_{15}^\pm g=f.
\end{equation}
The operators $T_{14}^\pm, T_{15}^\pm: L_2(-x_0,x_0)\to
L_2(-x_0,x_0)$ being bounded uniformly on $\e$ and $k$, there
exists a bounded inverse operator $T_{17}^\pm(\e,k):=\big(\I+\e
VT_{14}(\e)+\e k VT_{15}^\pm(\e,k)\big)^{-1}: L_2(-x_0,x_0)\to
L_2(-x_0,x_0)$. The operator $T_{17}^\pm$ is holomorphic on $k$
and obey a uniform on $k$ equality
\begin{equation}\label{6.26}
T_{17}^\pm(\e,k)=\I+\Odr(\e),\quad \e\to0.
\end{equation}
Applying the operator $T_{17}^\pm$ to the equation (\ref{6.25}),
we get:
\begin{equation}\label{6.27} g+\frac{\e^3 c_\pm
T_{16}^\pm g}{\k_\pm-\k_\pm^{-1}} T_{17}^\pm V\phi_n^\pm
=T_{17}^\pm f.
\end{equation}
Acting now by the functional $T_{16}^\pm$ on this equation, we
obtain
\begin{equation}\label{6.28}
\left(1+\frac{\e^3 c_\pm}{\k_\pm-\k_\pm^{-1}}T_{16}^\pm
T_{17}^\pm V\phi_n^\pm\right)T_{16}^\pm g=T_{16}^\pm T_{17}^\pm
f.
\end{equation}
By analogy with the way we employed to deduce the equation
(\ref{3.32}) from (\ref{3.30}), (\ref{3.31}), from (\ref{6.27}),
(\ref{6.28}) we deduce that the values $k$ those the equation
(\ref{6.18}) with $f=0$ has a nontrivial solution for are
defined by the equation
\begin{gather}
k=\e\mathfrak{g}_\pm(\e,k),\label{6.29}
\\
\mathfrak{g}_\pm(\e,k):=-\frac{k \e^2 c_\pm(\e^2)T_{16}^\pm(\e)
T_{17}^\pm(\e,k)V(x)\phi_n^\pm\left(\frac{x}{\e},\e^2\right)}
{\k_\pm(\e,k)-\k_\pm^{-1}(\e,k)}.\nonumber
\end{gather}
From Lemma~\ref{lm6.1}, equalities (\ref{6.1}) and holomorphy of
$C_\pm(\e^2)$ it follows that $c_\pm(\e^2)$ is a holomorphic
function and
\begin{equation*}%\l%abel{6.29b}
c_\pm(0)=\left\{
\begin{aligned}
&(-1)^nC_\pm^2(0)\frac{a_n\pm\mu_{n,0}}{2\pi^2n^2},&&\text{if
$b_n\not=0$ or $b_n=0$, $\sgn a_n=\pm 1$},
\\
&(-1)^{n+1}C_\pm^2(0)\frac{a_n\mp\mu_{n,0}}{2},&&\text{if
$b_n=0$, $\sgn a_n=\mp 1$}.
\end{aligned}
\right.
\end{equation*}
Clearly, $c_\pm(0)\not=0$, and $(-1)^{n+1}c_+(0)<0$. Bearing in
mind (\ref{6.4}), we conclude that the function
$\mathfrak{g}_\pm$ is holomorphic on $k$ and together with its
derivative $\frac{\displaystyle d
\mathfrak{g}_\pm}{\displaystyle dk}$ are uniformly bounded on
$\e$. For all sufficiently small $\e$ the equation (\ref{6.29})
has a unique root $k=k_{\e,\pm}$ tending to zero as $\e\to0$,
what can be proved completely by analogy with the proof of the
unique solvability of the equation (\ref{3.32}). The nontrivial
solution of the equation (\ref{6.18}) with $f=0$ and
$k=k_{\e,\pm}$ is defined up to a multiplicative constant and is
of the form
\begin{equation}\label{6.30}
g_{\e,\pm}=-T_{17}(\e,k_{\e,\pm})V\phi_n^\pm.
\end{equation}
The solution $\psi_{\e,\pm}$ of the problem (\ref{3.16}),
(\ref{6.17}) associated with $g_{\e,\pm}$, in view of
(\ref{6.16a}) is connected with $g_{\e,\pm}$ by the equalities:
\begin{equation}\label{6.30e}
\psi_{\e,\pm}=\e T_{11}^\pm(\e,k_{\e,\pm}) g_{\e,\pm},\quad
g_{\e,\pm}=-V\psi_{\e,\pm}.
\end{equation}
If  $\RE k_{\e,\pm}>0$, then the function $\psi_{\e,\pm}(x)$
decays exponentially as $x\to\pm\infty$ (see
(\ref{6.4})-(\ref{6.17})). Therefore, in this case
$\psi_{\e,\pm}$ belongs to $L_2(\mathbb{R})$ and due to this
reason it is the eigenfunction of the operator $H_\e$ associated
with the eigenvalue $\l_{\e,\pm}:=\mu_n^\pm(\e^2)\mp
k_{\e,\pm}^2$. The eigenvalue $\l_{\e,\pm}$ is simple. If $\RE
k_{\e,\pm}\leqslant 0$, it follows that the function
$\psi_{\e,\pm}$ is not an element of $L_2(\mathbb{R})$, and in
this case the operator $H_\e$ has no eigenvalues obeying the
equality (\ref{1.30}).

From (\ref{6.26}), (\ref{6.30}) we deduce that in the norm of
$L_2(-x_0,x_0)$ the equality
\begin{equation}\label{6.30a}
g_{\e,\pm}(x)=-V(x)\phi_n^\pm\left(\frac{x}{\e},\e^2\right)
+\Odr(\e)
\end{equation}
holds. Due to (\ref{6.21}), (\ref{6.24}), (\ref{6.30a}) and
holomorphy of the function $\phi_n$ on $\e^2$ it follows that
\begin{equation}\label{6.30b}
\begin{aligned}
&\psi_{\e,\pm}(x)=-\frac{\e^3 c_\pm(\e^2)
T_{16}^\pm(\e)g_{\e,\pm}}{\k_\pm(\e,k_{\e,\pm})-
\k_\pm^{-1}(\e,k_{\e,\pm})}\phi_n^\pm\left(\frac{x}{\e},\e^2\right)
+\Odr(\e^2),
\\
&g_{\e,\pm}(x)=\frac{\e^3 c_\pm(\e^2)
T_{16}^\pm(\e)g_{\e,\pm}}{\k_\pm(\e,k_{\e,\pm})-
\k_\pm^{-1}(\e,k_{\e,\pm})}V(x)\phi_n^\pm\left(\frac{x}{\e},\e^2\right)
+\Odr(\e^2),
\end{aligned}
\end{equation}
where the former of the equalities holds in the norm of
$\H^2(Q)$ for each $Q\in \mathfrak{C}$, and the latter does in
the norm of $L_2(-x_0,x_0)$. Since by Riemann-Lebesgue lemma
\begin{equation*}
\int\limits_{-x_0}^{x_0} V^2(x)\cos \frac{2\pi n x}{\e}\di x
=o(1),\quad\e\to0,
\end{equation*}
from (\ref{2.20}), (\ref{2.9c}) it follows that
\begin{equation}\label{6.30c}
\left\|V\phi_{n}^\pm \right\|^2_{L_2(-x_0,x_0)}=
\left\|V\phi_{n,0}^\pm\right\|^2_{L_2(-x_0,x_0)}+\Odr(\e^2)
=\|V\|^2_{L_2(-x_0,x_0)}+o(1),
\end{equation}
where $\phi_n^\pm=\phi_{n}^\pm\left(\frac{x}{\e},\e^2\right)$,
$\phi_{n,0}^\pm=\phi_{n,0}^\pm\left(\frac{x}{\e}\right)$. We
multiply the equality (\ref{6.30a}) and the latter of the
equalities (\ref{6.30b}) by
$\phi_n^\pm\left(\frac{x}{\e},\e^2\right)$ in $L_2(-x_0,x_0)$
and obtain as a result that
\begin{equation*}
\Big\|V(x)\phi_n^\pm\left(\frac{x}{\e},\e^2\right)\Big\|_{L_2(-x_0,x_0)}
\left(1+\frac{\e^3 c_\pm(\e^2)
T_{16}^\pm(\e)g_{\e,\pm}}{\k_\pm(\e,k_{\e,\pm})-
\k_\pm^{-1}(\e,k_{\e,\pm})}\right)=\Odr(\e),
\end{equation*}
what by (\ref{6.30c}) implies
\begin{equation*}
-\frac{\e^3 c_\pm(\e^2)
T_{16}^\pm(\e)g_{\e,\pm}}{\k_\pm(\e,k_{\e,\pm})-
\k_\pm^{-1}(\e,k_{\e,\pm})}=1+\Odr(\e).
\end{equation*}
By the equality obtained, (\ref{2.9c}), and the former of the
equalities (\ref{6.30b}) we get that
\begin{equation}\label{6.30d}
\psi_{\e,\pm}(x)=\phi_{n,0}^\pm\left(\frac{x}{\e}\right)+\Odr(\e),
\end{equation}
in the norm of $\H^2(Q)$ for each $Q\in \mathfrak{C}$. Thus, we
have just proved
\begin{lemma}\label{lm6.6}
Theorem~\ref{th1.3a} is valid. The operator $H_\e$ has the
eigenvalue $\l_{\e,\pm}$, if and only if $\RE k_{\e,\pm}>0$,
which is given in this case by the equality
$\l_{\e,\pm}=\mu_n^\pm(\e^2)\mp k_{\e,\pm}^2$, and the
associated eigenfunction $\psi_{\e,\pm}$ is determined by the
formulas (\ref{6.30}), (\ref{6.30e}). The equality (\ref{6.30d})
holds true for the function $\psi_{\e,\pm}$.
\end{lemma}

The form of the function $\mathfrak{g}_+(\e,k)$ and (\ref{6.4})
yield
\begin{equation*}%\l%abel{6.29a}
\mathfrak{g}_+(\e,k)=(-1)^{n+1}\frac{c_+(\e^2)\pi n} {\sqrt{\t
D_+(\e^2)-k^2}}T_{16}^+(\e)(\I+\e V T_{14}^+(\e))^{-1}
V(x)\phi_n^+\left(\frac{x}{\e},\e^2\right)+\Odr(\e k).
\end{equation*}
We substitute this equality into the equation (\ref{6.29}) with
$k=k_{\e,+}$ and obtain
\begin{equation*}
k_{\e,+}\big(1+\Odr(\e^2)\big)=(-1)^{n+1}\frac{\e c_+(\e^2)\pi
n}{\sqrt{\t D_+(\e^2)-k_{\e,+}^2}}T_{16}^+(\e)(I+\e V
T_{14}(\e))^{-1}V(x)\phi_n^+\left(\frac{x}{\e},\e^2\right),
\end{equation*}
what together with (\ref{6.3a}), (\ref{6.24}), holomorphy of
$c_+(\e^2)$ on $\e^2$, realness of the function $\phi_n^+$, and
the inequality $(-1)^{n+1}c_+(0)<0$ imply that the condition
$\RE k_{\e,+}>0$ is equivalent to (\ref{1.16}). The proof of
item~\ref{it2th1.3b} of Theorem~\ref{th1.3b} is complete.

In the next section we will make use of an auxiliary lemma, and
it is convenient to formulate it in this section.

\begin{lemma}\label{lm6.5}
For all sufficiently small $\e$ and $k$ the representation
\begin{equation*}
(\I+\e V T_{11}^\pm(\e,k))^{-1}f=\frac{\e
T_{18}^\pm(\e,k)f}{k-k_{\e,\pm}} g_{\e,\pm}+T_{19}^\pm(\e,k)f,
\end{equation*}
holds true, where $T_{18}^\pm: L_2(-x_0,x_0)\to\mathbb{C}$ and
$T_{19}^\pm:L_2(-x_0,x_0)\to L_2(-x_0,x_0)$  are linear
functional and operator bounded uniformly on $\e$ and $k$.
\end{lemma}

The proof of this lemma is carried out completely by analogy
with the proof of Lemma~\ref{lm3.11} on the base of the
equations (\ref{6.27})-(\ref{6.29}), equality (\ref{6.26}), and
formulas (\ref{2.9c}), (\ref{6.3a}), (\ref{6.4}), (\ref{6.30}).

\sect{Asymptotics for the eigenvalues in a finite lacuna}

In this section we will prove Theorems~\ref{th1.4},~\ref{th1.5}
and item~\ref{it1th1.3b} of Theorem~\ref{th1.3b}. The proofs
will be based on the asymptotics expansions for the numbers
$\l_{\e,\pm}^{(n)}:=\mu_n^\pm(\e^2)\mp k_{\e,\pm}^2$, where
$k_{\e,\pm}$, we remind, are the solutions of the equations
(\ref{6.29}). As in the fifth section, first we formally
construct the asymptotics expansions for these numbers and the
associated nontrivial solutions $\psi_{\e,\pm}$ of the problem
(\ref{3.16}), (\ref{6.17})  from (\ref{6.30e}), and then we
justify them rigorously.

We construct the asymptotics of the number $\l_{\e,\pm}$ as the
corresponding series from (\ref{1.17}), (\ref{1.20}), and the
asymptotics for the associated function $\psi_{\e,\pm}$ is
sought as follows
\begin{equation}\label{7.1}
\psi_{\e,\pm}(x)=h(x,\tau_{\e}^{\pm})\left(\phi_{n,0}^\pm(\xi)+
\sum\limits_{i=1}^\infty \e^i\psi_{i}^{\pm}(x,\xi)\right),
\end{equation}
where, we remind, the function $h$ is defined by the equality
(\ref{5.4}). The asymptotics of the function $\tau_{\e}^{\pm}$
is constructed as follows
\begin{equation}\label{7.2}
\tau_{\e}^{\pm}=\sum\limits_{i=2}^\infty \e^i\tau_{i}^{\pm}.
\end{equation}
The aim of the formal constructing is to determine the functions
$\psi_{i}^{\pm}$ and the numbers
$\l_{i}^{\pm}:=\l_{i,\pm}^{(n)}$ and $\tau_{i}^{\pm}$.

The functions $\psi_{i}^{\pm}$ are sought as 1-periodic on
$\xi$, of $n$ is even and 1-antiperiodic, if $n$ is odd. We also
postulate the functions $\psi_i^\pm$ to belong to the space
$\mathcal{V}$ as functions on $x$. As in the fifth section, the
ansatze (\ref{7.1}) is chosen so that it has the same structure
for $|x|\geqslant x_0$ as the functions $\Th^\pm_\pm$ in
(\ref{6.6}). In particular, the requirement for the functions
$\psi_i^\pm$ to be 1-periodic (1-antiperiodic) is explained by
1-periodicity (1-antiperiodicity) of the functions
$\E^{-\xi\ln((-1)^n)}\Th_{per,\pm}^+(\xi,\e,k)$ and
$\E^{\xi\ln((-1)^n)}\Th_{per,\pm}^-(\xi,\e,k)$  (see
(\ref{6.4}), (\ref{6.6}), (\ref{7.2})).

We proceed to the formal constructing of the asymptotics. We
substitute (\ref{7.1}), (\ref{7.2}) and the corresponding series
from (\ref{1.17}), (\ref{1.20}) into the equation (\ref{3.16})
with $f=0$, divide the equation obtained by
$h(x,\tau_{\e}^{\pm})$, expand it in a power asymptotic series
on $\e$ and collect the coefficients of the same powers of $\e$.
As a result we arrive at the following equations for the
functions $\psi_{i}^{\pm}$:
\begin{equation}\label{7.3}
\begin{aligned}
-&\left(\frac{\partial^2}{\partial\xi^2}+\pi^2
n^2\right)\psi_{i+2}^{\pm} =2\frac{\partial^2}{\partial
x\partial\xi}\psi_{i+1}^{\pm}+\left(\frac{\partial^2}{\partial
x^2 }-a-V\right)\psi_{i}^{\pm}+
\\
&+\sum\limits_{j=0}^{i} \left(2h_j^{(1)}\frac{\partial}{\partial
x}+h_j^{(2)}+\l_{j}^{\pm}\right)
\psi_{i-j}^{\pm}+2\sum\limits_{j=2}^{i+1}h_j^{(1)}
\frac{\partial}{\partial\xi}\psi_{i-j+1}^{\pm},\quad (x,\xi)\in
\mathbb{R}^2,
\end{aligned}
\end{equation}
where $i\geqslant -1$, $a=a(\xi)$, $V=V(x)$, $\psi_{-1}^\pm:=0$,
$\psi_{0}^{\pm}(x,\xi):=\phi_{n,0}^\pm(\xi)$,
$h_i^{(1)}=h_i^{(2)}:=0$, $i=0,1$. The functions
$h_i^{(1)}=h_i^{(1)}(x,\boldsymbol{\tau}_{i}^{\pm})$ and
$h_i^{(2)}=h_i^{(2)}(x,\boldsymbol{\tau}_{i}^{\pm})$,
$\boldsymbol{\tau}_{i}^{\pm}:=(\tau_{2}^{\pm},\ldots,\tau_{i}^{\pm})$,
$i\geqslant 2$, are the coefficients of the expansions of the
functions $h'(x,\tau_{\e}^{\pm})/h(x,\tau_{\e}^{\pm})$ and
$h''(x,\tau_{\e}^{\pm})/h(x,\tau_{\e}^{\pm})$ in the power
asymptotic series on $\e$, respectively. These coefficients obey
the equalities (\ref{5.11a}) with
$\boldsymbol{\tau}_i=\boldsymbol{\tau}_{i}^{\pm}$ for
$i\geqslant 2$, where $\t h_i^{(j)}\in
C^\infty(\mathbb{R})\cap\mathcal{V}$, and, in particular,
\begin{equation}\label{7.4}
\t h_2^{(1)}(x)=\t h_2^{(2)}(x)=0.
\end{equation}
For $\pm x\geqslant x_0$ the equalities
\begin{equation}\label{7.5}
h_i^{(1)}(x,\boldsymbol{\tau}_{i})=\mp\tau_{i},\quad
h_i^{(2)}(x,\boldsymbol{\tau}_i)=
\sum\limits_{j=2}^{i-2}\tau_j\tau_{i-j}
\end{equation}
hold true, where $\tau_p=\tau_{p}^+$ or $\tau_p=\tau_{p}^{-}$.
These equalities follow from (\ref{5.8a}) and (\ref{7.2}).

In view of Lemma~\ref{lm2.5} the equations (\ref{7.3}) are
solvable in the class of 1-periodic on $\xi$ functions for even
$n$ and 1-antiperiodic on $\xi$ functions for odd $n$, if the
following solvability conditions
\begin{equation}\label{7.6}
\begin{aligned}
&2\frac{d}{dx}\int\limits_0^1\phi^\pm_{n,0}
\psi_{i+1}^{\pm}\di\xi=c_0^\pm\left(\frac{d^2}{dx^2}-V\right)
\int\limits_0^1 \phi_{n,0}^\mp\psi_{i}^{\pm}\di\xi-
2\sum\limits_{j=2}^{i+1} h_j^{(1)}
\int\limits_0^1\phi_{n,0}^\pm\psi_{i-j+1}^{\pm}\di\xi+
\\
&+c_0^\pm\sum\limits_{j=0}^i \left(2h_j^{(1)}\frac{d}{dx}+
h_j^{(2)}+\l_{j}^{\pm}\right) \int\limits_0^1\phi_{n,0}^\mp
\psi_{i-j}^{\pm}\di\xi-c_0^\pm\int\limits_0^1
a\phi_{n,0}^\mp\psi_{i}^{\pm}\di\xi,\quad x\in \mathbb{R},
\\
&2\frac{d}{dx}\int\limits_0^1\phi^\mp_{n,0}
\psi_{i+1}^{\pm}\di\xi=-c_0^\pm\left(\frac{d^2}{dx^2}-V\right)
\int\limits_0^1
\phi_{n,0}^\pm\psi_{i}^{\pm}\di\xi-2\sum\limits_{j=2}^{i+1}
h_j^{(1)} \int\limits_0^1\phi_{n,0}^\mp\psi_{i-j+1}^{\pm}\di\xi-
\\
&-c_0^\pm\sum\limits_{j=0}^i \left(2h_j^{(1)}\frac{d}{dx}+
h_j^{(2)}+\l_{j}^{\pm}\right) \int\limits_0^1\phi_{n,0}^\pm
\psi_{i-j}^{\pm}\di\xi+c_0^\pm\int\limits_0^1
a\phi_{n,0}^\pm\psi_{i}^{\pm}\di\xi ,\quad x\in \mathbb{R},
\end{aligned}
\end{equation}
take place, where $c_0^\pm:=\pm\frac{1}{\pi n}$, $i\geqslant 0$.
In deducing these equations we have also made use of the
equality
\begin{equation}\label{7.6a}
\frac{d}{d\xi}\phi_{n,0}^\pm(\xi)=-\frac{1}{c_0^\pm}
\phi_{n,0}^\mp(\xi)
\end{equation}
and integrated once by parts in some of the integrals, assuming
that the functions $\psi_j$ are 1-periodic (1-antiperiodic) on
$\xi$.

In order to study the solvability of the equations (\ref{7.6})
we will employ the following obvious statement.

\begin{lemma}\label{lm7.1}
Let $f\in C(\mathbb{R})$. Then the equation
\begin{equation*}%\l%abel{7.7}
2\frac{du}{dx}=f,\quad x\in\mathbb{R},
\end{equation*}
has a solution $u\in C^1(\mathbb{R})\cap\mathcal{V}$, if and
only if $\supp f\subseteq[-x_0,x_0]$. In this case there exists
a unique solution of this equation, satisfying the equality
\begin{equation}\label{7.8}
u(x_0)+u(-x_0)=0,
\end{equation}
which is of the form
\begin{equation*}
u(x)=\int\limits_{\mathbb{R}}\sgn(x-t)f(t)\di t.
\end{equation*}
\end{lemma}

The equation (\ref{7.3}) for the function $\psi_{1}^{\pm}$
implies that this function is of the form:
\begin{equation}\label{7.9}
\psi_{1}^{\pm}(x,\xi)=u_{1,0}^{\pm}(x)\phi_{n,0}^\pm(\xi)+
u_{1,1}^{\pm}(x)\phi_{n,0}^\mp(\xi),
\end{equation}
where $u_{1,j}^{\pm}(x)$ are some functions. This representation
for the function $\psi_1^\pm$ and the equalities (\ref{2.30})
for $N=1$ allow us to rewrite the equations (\ref{7.6}) for
$i=0$ as follows
\begin{equation}\label{7.9a}
2\frac{du_{1,0}^\pm}{dx}=0,\quad
2\frac{du_{1,1}^\pm}{dx}=f_{1,1}^\pm,\quad x\in \mathbb{R},
\end{equation}
where $f_{1,1}^\pm:=c_0^\pm
\left(V+\mu_{n,0}^\pm-\l_0^\pm\right)$. Clearly, the inclusion
$\psi_{1}^\pm(\cdot,\xi)\in \mathcal{V}$ is valid, if
$u_{1,j}^\pm\in \mathcal{V}$, $j=0,1$. By Lemma~\ref{lm7.1} the
equations (\ref{7.9a}) are solvable in the class $\mathcal{V}$,
if $\l_0^\pm$ is chosen in accordance with (\ref{1.18}),
(\ref{1.21}), and the solutions of these equations are of the
form
\begin{equation}\label{7.10}
u_{1,0}^\pm=0,\quad u_{1,1}^\pm=\t u_{1}^\pm(x)+c_{1}^\pm,\quad
\t u_{1}^\pm(x)= c_0^\pm\int\limits_{\mathbb{R}}\sgn(x-t)V(t)\di
t,
\end{equation}
where $c_1^\pm$ is some constant, and the function $\t u_{1}$
satisfies the equality (\ref{7.8}).

\begin{remark}\label{rm7.1}
The solution to the former of the equations (\ref{7.9a}) is
determined up to an additive constant. Without loss of
generality we set this constant equal zero, since it can be
always achieved, multiplying the asymptotics (\ref{7.1}) by a
suitable number.
\end{remark}

The equations (\ref{7.6}) being valid for $i=0$, the equation
(\ref{7.3}) with $i=0$ is solvable in the class of 1-periodic
(1-antiperiodic) functions. Using (\ref{7.6a}), (\ref{7.9}),
(\ref{7.10}), and formulas (\ref{1.18}), (\ref{1.21}) for
$\l_0^\pm$, it is easy to check that the equation (\ref{7.3})
with $i=0$ reads as follows
\begin{equation*}
-\left(\frac{\partial^2}{\partial\xi^2}+\pi^2
n^2\right)\psi_2^\pm=
\left(-a+\mu_{n,0}^\pm\right)\phi_{n,0}^\pm,\quad (x,\xi)\in
\mathbb{R}^2.
\end{equation*}
In accordance with Lemma~\ref{lm2.5} the solution of this
equation is given by the formula
\begin{equation}\label{7.11}
\begin{aligned}
&\psi_2^\pm(x,\xi)=\t\psi_2^\pm(x,\xi)+
u_{2,0}^\pm(x)\phi_{n,0}^\pm(\xi)+u_{2,1}^\pm(x)\phi_{n,0}^\mp(\xi),
\\
&\t\psi_2^\pm(x,\xi)=L_n[-a\phi_{n,0}^\pm+
\mu_{n,0}^\pm\phi_{n,0}^\pm](\xi)=\t\phi_{n,1}^\pm(\xi),
\end{aligned}
\end{equation}
where the latter equality follows from  (\ref{2.31}) with $N=1$.
Now we substitute (\ref{7.9}), (\ref{7.10}), (\ref{7.11}) into
the equations (\ref{7.6}) with $i=1$ and take into account the
relations (\ref{5.11a}), (\ref{7.4}), (\ref{2.30}) with $N=1$,
and the orthogonality of the function $\t\phi_{n,1}^\pm$ to the
functions $\phi_{n,0}^+$ and $\phi_{n,0}^-$ in $L_2(0,1)$. As a
result we obtain
\begin{gather}
2\frac{du_{2,0}^{\pm}}{dx}=f_{2,0}^\pm,\quad
2\frac{du_{2,1}^\pm}{dx}=f_{2,1}^\pm=-c_0^\pm\l_1^\pm,\quad
x\in\mathbb{R}, \nonumber
\\
f_{2,0}^\pm=c_0^\pm\left(\frac{d^2}{dx^2}-V\right)
u_{1,1}^\pm+2c_0^\pm\mu_{n,0}^\pm u_{1,1}^\pm + 2\tau_2^\pm
\frac{d}{dx}\left(|x|(1-\chi(x))\right),\label{7.12a}
\end{gather}
We seek the solutions of these equations in the class
$\mathcal{V}$, what is explained by the equality (\ref{7.11})
and the constraints $\psi_2^\pm(\cdot,\xi)\in\mathcal{V}$.
Obviously, $f_{2,j}^\pm\in \mathcal{V}$, $j=0,1$, this is why
the inclusion  $\supp f_{2,1}^\pm\subseteq[-x_0,x_0]$ is
equivalent to the formulas (\ref{1.18}), (\ref{1.21}) for
$\l_1^\pm$, and the inclusion $\supp
f_{2,0}^\pm\subseteq[-x_0,x_0]$ is equivalent to the equalities
$f_{2,0}^\pm(x_0)+f_{2,0}^\pm(-x_0)=0$,
$f_{2,0}^\pm(x_0)-f_{2,0}^\pm(-x_0)=0$. Taking into account the
condition (\ref{7.8}) for the function $\t u_1^\pm$, it is easy
to check that the last two equalities hold, if we set
\begin{equation}\label{7.12}
c_1^\pm=0,\quad \tau_2^\pm=-c_0^\pm\mu_{n,0}^\pm\t u_1^\pm(x_0)
=\mp\frac{\mu_{n,0}}{\pi^2 n^2}\int\limits_{\mathbb{R}} V(x)\di
x,
\end{equation}
where, we remind, $\mu_{n,0}=\sqrt{a_n^2+b_n^2}$. Therefore,
\begin{equation}\label{7.13}
u_{2,0}^\pm(x)=\int\limits_{\mathbb{R}}\sgn(x-t)f_{2,0}^\pm(t)\di
t,\quad u_{2,1}^\pm(x)=\t u_2^\pm(x)+c_2^\pm,\quad \t
u_2^\pm(x)\equiv0,
\end{equation}
where $c_2^\pm$ is a some constant.

\begin{lemma}\label{lm7.2}
There exist the solutions to the equations (\ref{7.3}),
(\ref{7.6}) of the form
\begin{align}
&\psi_i^\pm(x,\xi)=\t\psi_i^\pm(x,\xi)+u_{i,0}^\pm(x)
\phi_{n,0}^\pm(\xi)+u_{i,1}^\pm(x)\phi_{n,0}^\mp(\xi),\label{7.14}
\\
&\t\psi_i^\pm(x,\xi)=L_n[G_i^\pm(x,\cdot)](\xi)=
\sum\limits_{j=2}^{\mathfrak{m}_i^\pm}u_{i,j}^\pm(x)\psi_{i,j}^\pm(\xi)
\label{7.15}
\\
&
\begin{aligned}
&u_{i,0}^\pm(x)=\int\limits_{\mathbb{R}}\sgn(x-t)f_{i,0}^\pm(x)\di
x,\quad u_{i,1}^\pm(x)=\t u_i^\pm(x)+c_i^\pm,
\\
&\t
u_i^\pm(x)=\int\limits_{\mathbb{R}}\sgn(x-t)f_{i,1}^\pm(x)\di x,
\end{aligned}
\label{7.16}
\end{align}
\begin{align}
&
\begin{aligned}
G_i^\pm=&2\frac{\partial^2}{\partial
x\partial\xi}\t\psi_{i-1}^\pm+ \left(\frac{\partial^2}{\partial
x^2}-V-a\right)\t\psi_{i-2}^\pm+ \phi_{n,0}^\pm\int\limits_0^1
a\phi_{n,0}^\pm \t\psi_{i-2}^\pm\di\xi+
\\
&+ \phi_{n,0}^\mp\int\limits_0^1
a\phi_{n,0}^\mp\t\psi_{i-2}^\pm\di\xi
+2\sum\limits_{j=2}^{i-3}h_j^{(1)}
\frac{\partial}{\partial\xi}\t\psi_{i-j-1}^\pm
-\big(a-\mu_{n,0}^\pm\big)\phi_{n,0}^\pm u_{i-2,0}^\pm+
\\
&+\sum\limits_{j=0}^{i-4}
\left(2h_j^{(1)}\frac{\partial}{\partial
x}+h_j^{(2)}+\l_j^\pm\right) \t\psi_{i-j-2}^\pm-
\big(a+\mu_{n,0}^\pm\big)\phi_{n,0}^\mp u_{i-2,1}^\pm,
\end{aligned}\label{7.17}
\\
&
\begin{aligned}
f_{i,0}^\pm=&c_0^\pm\left(\frac{d^2}{dx^2}-V\right)
u_{i-1,1}^\pm-c_0^\pm\int\limits_0^1
a\phi_{n,0}^\mp\t\psi_{i-1}^\pm\di\xi-2 \sum\limits_{j=2}^i
h_j^{(1)} u_{i-j,0}^\pm+
\\
&+c_0^\pm\sum\limits_{j=2}^{i-2} \left(2h_j^{(1)}\frac{d}{dx}+
h_j^{(2)}+\l_{j}^\pm\right) u_{i-j-1,1}^\pm+
2c_0^\pm\mu_{n,0}^\pm u_{i-1,1}^\pm,
\\
f_{i,1}^\pm=&-c_0^\pm\left(\frac{d^2}{dx^2}-V\right)
u_{i-1,0}^\pm+c_0^\pm\int\limits_0^1
a\phi_{n,0}^\pm\t\psi_{i-1}^\pm\di\xi-2 \sum\limits_{j=2}^{i-1}
h_j^{(1)} u_{i-j,1}^\pm+
\\
&-c_0^\pm\sum\limits_{j=2}^{i-1} \left(2h_j^{(1)}\frac{d}{dx}+
h_j^{(2)}+\l_{j}^\pm\right) u_{i-j-1,0}^\pm,
\end{aligned}\label{7.18}
\end{align}
where $\mathfrak{m}_i^\pm$ are some numbers,
$G_i^\pm=G_i^\pm(x,\xi)$, $f_{i,j}^\pm=f_{i,j}^\pm(x)$,
$V=V(x)$, $a=a(\xi)$, $\phi_{n,0}^\pm:=\phi_{n,0}^\pm(\xi)$,
$\t\psi_{-1}^\pm=\t\psi_0^\pm:=0$, $u_{0,0}^\pm(x)\equiv 1$,
$u_{-1,j}^\pm=u_{0,1}^\pm:=0$, $u_{i,j}^\pm\in
C^\infty(\mathbb{R})\cap \mathcal{V}$, $\supp
f_{i,p}^\pm(x)\subseteq[-x_0,x_0]$, $p=0,1$, the functions
$\psi_{i,j}^\pm$ are 1-periodic for even $n$ and 1-antiperiodic
for odd $n$ and obey the equalities
\begin{equation}\label{7.19}
\int\limits_0^1 \phi_{n,0}^+(\xi)\psi_{i,j}^\pm(\xi)\di\xi=
\int\limits_0^1 \phi_{n,0}^-(\xi)\psi_{i,j}^\pm(\xi)\di\xi=0.
\end{equation}
The functions $u_{i,0}^\pm$ and $\t u_{i}^\pm$ meet the
condition (\ref{7.8}). The numbers $\l_i^\pm$, $\tau_i^\pm$ and
$c_i^\pm$ are given by the formulas
\begin{align}
&\l_i^\pm=-\sum\limits_{j=2}^{i-2}\tau_j^\pm\tau_{i-j}^\pm+
\mathfrak{a}^{\pm}_{i,+}+\frac{2}{c_0^\pm}\sum\limits_{j=2}^i
\tau_j^\pm \widetilde{\mathfrak{u}}_{i-j+1}^\pm,\label{7.20}
\\
&\tau_i^\pm=\frac{c_0^\pm}{2}
\left(\widetilde{\mathfrak{a}}^\pm_{i-1,-}-2\mu_{n,0}^\pm
\widetilde{\mathfrak{u}}_{i-1}^\pm-\sum\limits_{j=2}^{i-2}
\mathfrak{l}_j^\pm \widetilde{\mathfrak{u}}_{i-j-1}^\pm
\right)\label{7.21}
\\
& c_i^\pm=\frac{1}{2\mu_{n,0}^\pm}
\left(\widetilde{\mathfrak{a}}^\pm_{i,+} -
\sum\limits_{j=2}^{i-2} \mathfrak{l}_j^\pm
c_{i-j}^\pm-\frac{2}{c_0^\pm}\sum\limits_{j=2}^{i-1} \tau_j^\pm
\mathfrak{u}_{i-j+1}^\pm \right),\label{7.22}
\end{align}
where
$\mathfrak{l}_i^\pm:=\l_i^\pm+\sum\limits_{j=2}^{i-2}\tau_j^\pm
\tau_{i-j}^\pm$, and the numbers $\mathfrak{a}^\pm_{i,\pm}$,
$\widetilde{\mathfrak{a}}^\pm_{i,\pm}$ $\mathfrak{u}_{i}^\pm$,
$\widetilde{\mathfrak{u}}_{i}^\pm$ are defined in (\ref{7.25}).
\end{lemma}

\begin{proof}
The conclusion of the lemma on $\psi_1^\pm$, $\t\psi_2^\pm$,
$u_{i,j}^\pm$, $i=1,2$, $j=0,1$, $\l_0^\pm$, $\t u_{2,1}^\pm$,
$\l_1^\pm$, $c_1^\pm$, and $\tau_2^\pm$ follows from
(\ref{7.9}), (\ref{7.10}), (\ref{7.11}), (\ref{7.12}),
(\ref{7.13}) and the formulas (\ref{1.18}), (\ref{1.21}) for
$\l_0^\pm$ and $\l_1^\pm$. The further proof is carried out by
induction. Suppose the formulas (\ref{7.14}), (\ref{7.15}),
(\ref{7.19}) be valid for $i\leqslant m+1$, the formulas
(\ref{7.16}), (\ref{7.21}) be valid for $i\leqslant m$, the
formulas (\ref{7.20}), (\ref{7.22}) be valid for $i\leqslant
m-1$. Let us prove that in this case the conclusion of the lemma
on $\psi_{m+2}^\pm$, $\t\psi_{m+2}^\pm$, $u_{m+1,0}^\pm$,
$u_{m+1,1}^\pm$, $\t u_{m+1}^\pm$, $\l_m^\pm$, $\tau_{m+1}^\pm$
and $c_m^\pm$ is true.

Throughout the proof the symbol $(\cdot,\cdot)$ indicates the
inner product in $L_2(0,1)$. We also denote
\begin{equation}\label{7.25}
\begin{aligned}
&\h\psi_{i,+}^\pm(\xi):=
\frac{\t\psi_i^\pm(x_0,\xi)+\t\psi_i^\pm(-x_0,\xi)}{2},\quad
\h\psi_{i,-}^\pm(\xi):=
\frac{\t\psi_i^\pm(x_0,\xi)-\t\psi_i^\pm(-x_0,\xi)}{2},
\\
&\mathfrak{a}_{i,+}^\pm:=(\phi_{n,0}^\pm,a\h\psi_{i,+}^\pm),\quad
\mathfrak{a}_{i,-}^\pm:=(\phi_{n,0}^\pm,a\h\psi_{i,-}^\pm),
\quad\widetilde{\mathfrak{a}}_{i,+}^\pm:=
(\phi_{n,0}^\mp,a\h\psi_{i,+}^\pm),
\\
&\widetilde{\mathfrak{a}}_{i,-}^\pm:=
(\phi_{n,0}^\mp,a\h\psi_{i,-}^\pm), \quad \mathfrak{u}_i^\pm:=
u_{i,0}(x_0),\quad \widetilde{\mathfrak{u}}_i^\pm:=
u_{i,1}(x_0).
\end{aligned}
\end{equation}
It follows from the formulas (\ref{1.18}), (\ref{1.21}) for
$\l_0^\pm$, $\l_1^\pm$, and the relations (\ref{7.9}),
(\ref{7.10}), (\ref{7.11}), (\ref{7.12}), (\ref{7.13}) that
\begin{equation}\label{7.27}
\begin{aligned}
& \mathfrak{u}_{1}^\pm=\widetilde{\mathfrak{u}}_{2}^\pm=0,\quad
\h\psi_{j,+}^\pm=\h\psi_{j,-}^\pm=\h\psi_{2,-}^\pm=0,\quad
j=0,1,
\\
&\mathfrak{a}_{1,+}^\pm=\mathfrak{a}_{1,-}^\pm=
\widetilde{\mathfrak{a}}_{1,+}^\pm=
\widetilde{\mathfrak{a}}_{1,-}^\pm=0,\quad
\mathfrak{l}_0^\pm=\mu_{n,0}^\pm,\quad \mathfrak{l}_1^\pm=0 .
\end{aligned}
\end{equation}
According to the induction assumption, the functions
$\psi_{i,j}^\pm$, $j\geqslant 2$, $i\leqslant m+1$, meet the
equalities (\ref{7.19}). Bearing in mind these equalities,
(\ref{7.27}), and (\ref{2.30}) with $N=1$ and substituting the
representations (\ref{7.14}), (\ref{7.15}) into the equations
(\ref{7.6}) with $i=m+1$, we obtain:
\begin{equation}\label{7.24}
2\frac{d}{dx}u_{m+1,j}^\pm=f_{m+1,j}^\pm,\quad x\in
\mathbb{R},\quad j=0,1,
\end{equation}
where the functions $f_{m+1,j}^\pm\in
C^\infty(R)\cap\mathcal{V}$ are given by the formulas
(\ref{7.18}). The solutions of these equations are sought in the
space $\mathcal{V}$ in order to ensure the belonging
$\psi_{m+1}^\pm(\cdot,\xi)\in \mathcal{V}$. In accordance with
Lemma~\ref{7.1} the condition of solvability of the equations
(\ref{7.24}) in the space $\mathcal{V}$ is the inclusion $\supp
f_{m+1,j}^\pm\subseteq[-x_0,x_0]$. Clearly, these inclusions are
equivalent to the equalities $\mathfrak{f}_{m+1,j,+}^\pm=
\mathfrak{f}_{m+1,j,-}^\pm=0$, where
\begin{equation*}
\mathfrak{f}_{m+1,j,+}^\pm:=
\frac{f_{m+1,j}^\pm(x_0)+f_{m+1,j}^\pm(-x_0)}{2},\quad
\mathfrak{f}_{m+1,j,-}^\pm:=
\frac{f_{m+1,j}^\pm(x_0)-f_{m+1,j}^\pm(-x_0)}{2}.
\end{equation*}
It follows from the formulas (\ref{7.18}) for
$f_{m+1,j,\pm}^\pm$, (\ref{7.16}) with $i\leqslant m$, the
condition (\ref{7.8}) for the functions $u_{i,0}^\pm$, $\t
u_i^\pm$, $i\leqslant m$, equalities (\ref{7.5}), (\ref{7.12}),
(\ref{7.27}), and the definition of the numbers
$\mathfrak{f}_{m+1,j,\pm}^\pm$ that
\begin{align}
&\mathfrak{f}_{m+1,0,+}^\pm=-c_0^\pm
\widetilde{\mathfrak{a}}_{m,+}^\pm+ 2\sum\limits_{j=2}^{m-1}
\tau_j^\pm \mathfrak{u}_{m-j+1}^\pm+
c_0^\pm\sum\limits_{j=2}^{m-2}\mathfrak{l}_j^\pm
c^\pm_{m-j}+2c_0^\pm\mu_{n,0}^\pm c_m^\pm,\nonumber
\\
&\mathfrak{f}_{m+1,0,-}^\pm=-c_0^\pm
\widetilde{\mathfrak{a}}_{m,-}^\pm+2\tau_{m+1}^\pm
+c_0^\pm\sum\limits_{j=2}^{m-1} \mathfrak{l}_j^\pm
\widetilde{\mathfrak{u}}_{m-j}^\pm+ 2c_0^\pm\mu_{n,0}^\pm
\widetilde{\mathfrak{u}}_m^\pm,\nonumber
\\
&\mathfrak{f}_{m+1,1,+}^\pm=c_0^\pm
\mathfrak{a}_{m,+}^\pm+2\sum\limits_{j=2}^m \tau_j^\pm
\widetilde{\mathfrak{u}}_{m-j+1}^\pm-
c_0^\pm\mathfrak{l}_m^\pm,\nonumber
\\
&\mathfrak{f}_{m+1,1,-}^\pm=c_0^\pm
\mathfrak{a}_{m,-}^\pm+2\sum\limits_{j=2}^{m-1}\tau_j^\pm
c_{m-j+1}^\pm - c_0^\pm\sum\limits_{j=2}^{m-2}\mathfrak{l}_j^\pm
\mathfrak{u}_{m-j}^\pm.\label{7.26}
\end{align}
The numbers $\mathfrak{f}_{m+1,0,+}^\pm$,
$\mathfrak{f}_{m+1,0,-}^\pm$, $\mathfrak{f}_{m+1,1,+}^\pm$ are
zero, if we define $c_m^\pm$, $\tau_{m+1}^\pm$ and $\l_m^\pm$ in
accordance with (\ref{7.20}), (\ref{7.21}), (\ref{7.22}). Let us
prove that $\mathfrak{f}_{m+1,1,-}^\pm=0$.

Formulas (\ref{7.15}) and (\ref{7.27}) imply that the functions
$\h\psi_{i,\pm}^\pm$ are 1-periodic for even $n$ and
1-antiperiodic for even $n$ and satisfy the equations
\begin{equation}\label{7.29}
\begin{aligned}
&\left(\frac{d^2}{d\xi^2}+\pi^2n^2\right) \h\psi_{2,+}^\pm=
(a-\mu_{n,0}^\pm)\phi_{n,0}^\pm,\quad \xi\in \mathbb{R},
\\
&\left(\frac{d^2}{d\xi^2}+\pi^2n^2\right) \h\psi_{i+2,+}^\pm= 2
\sum\limits_{j=2}^{i-2}\tau_j^\pm\frac{d}{d\xi}\h\psi_{i-j+1,-}^\pm
+a\h\psi_{i,+}^\pm-\mathfrak{a}_{i,+}^\pm\phi_{n,0}^\pm-
\widetilde{\mathfrak{a}}_{i,+}^\pm\phi_{n,0}^\mp-
\\
&\hphantom{\left(\frac{d^2}{d\xi^2}+\pi^2n^2\right)\h\psi_{i+2,+}^\pm}
-
\sum\limits_{j=0}^{i-2}\mathfrak{l}_j^\pm\h\psi_{i-j,+}^\pm+
c_i^\pm(a+\mu_{n,0}^\pm)\phi_{n,0}^\mp,\quad \xi\in
\mathbb{R},\quad i\geqslant 1,
\\
&\left(\frac{d^2}{d\xi^2}+\pi^2n^2\right) \h\psi_{i+2,-}^\pm= 2
\sum\limits_{j=2}^{i-1}\tau_j^\pm\frac{d}{d\xi}\h\psi_{i-j+1,+}^\pm
+a\h\psi_{i,-}^\pm-\mathfrak{a}_{i,-}^\pm\phi_{n,0}^\pm-
\widetilde{\mathfrak{a}}_{i,-}^\pm\phi_{n,0}^\mp-
\\
&\hphantom{\Bigg(a}
-
\sum\limits_{j=0}^{i-3}\mathfrak{l}_j^\pm\h\psi_{i-j,-}^\pm+
\mathfrak{u}_i^\pm(a-\mu_{n,0}^\pm)\phi_{n,0}^\pm+
\widetilde{\mathfrak{u}}_i^\pm(a+\mu_{n,0}^\pm)\phi_{n,0}^\mp,\quad
\xi\in \mathbb{R},\quad i\geqslant 1.
\end{aligned}
\end{equation}
In deducing these equations we also took into account the
equalities (\ref{7.27}). Using the equations (\ref{7.29}) and
the equalities (\ref{7.19}), (\ref{7.27}), let us calculate the
following sum:
\begin{align*}
&\mathfrak{a}_{m,-}^\pm+\sum\limits_{j=2}^{m-1} c_j^\pm
\widetilde{\mathfrak{a}}_{m-j,-}^\pm=
\big(\h\psi_{m,-}^\pm,(a-\mu_{n,0}^\pm)\phi_{n,0}^\pm\big)+
\sum\limits_{j=1}^{m-1}c_j^\pm\big(\h\psi_{m-j,-}^\pm,
(a+\mu_{n,0}^\pm)\phi_{n,0}^\mp\big)=
\\
&=\sum\limits_{j=0}^{m-3}\left(\h\psi_{m-j,-}^\pm,
\left(\frac{d^2}{d\xi^2}+\pi^2
n^2\right)\h\psi_{j+2,+}^\pm\right)
-2\sum\limits_{j=1}^{m-1}\sum\limits_{p=2}^{j-2}\tau_p^\pm
\left(\h\psi_{m-j,-}^\pm,\frac{d}{d\xi}\h\psi_{j-p+1,-}^\pm\right)
-
\\
&- \sum\limits_{j=2}^{m-1}\big(\h\psi_{m-j,-}^\pm,
a\h\psi_{j,+}^\pm\big)+\sum\limits_{j=1}^{m-3}\sum\limits_{p=0}^{m-2}
\mathfrak{l}_p^\pm\big(\h\psi_{m-j,-}^\pm,\h\psi_{j-p,+}^\pm\big).
\end{align*}
Completely by analogy with (\ref{5.34a}) one can easily show
that the second summand in the right hand side of the last
equality is zero. Changing the summation index $j\mapsto m-j-2$
and integrating by parts twice in the first summand in the right
hand side of the last equality and changing the summation index
$j\mapsto m-j$ in the third summand, in view of (\ref{7.27}),
(\ref{7.29}) we obtain:
\begin{align*}
&\mathfrak{a}_{m,-}^\pm+\sum\limits_{j=2}^{m-1} c_j^\pm
\widetilde{\mathfrak{a}}_{m-j,-}^\pm=
2\sum\limits_{j=1}^{m-2}\sum\limits_{p=2}^{j-1}\tau_p^\pm
\left(\h\psi_{m-j,+}^\pm,\frac{d}{d\xi}\h\psi_{j-p+1,+}^\pm\right)
+\sum\limits_{j=2}^{m-2}\mathfrak{u}_j^\pm
\mathfrak{a}_{m-j,+}^\pm+
\\
&+\sum\limits_{j=1}^{m-2} \widetilde{\mathfrak{u}}_j^\pm
\widetilde{\mathfrak{a}}_{m-j,+}^\pm
-\sum\limits_{j=1}^{m-2}\sum\limits_{p=0}^{j-3}
\mathfrak{l}_p^\pm \big(\h\psi_{m-j,+}^\pm,\h\psi_{j-p,-}^\pm
\big)+\sum\limits_{j=1}^{m-3}\sum\limits_{p=0}^{j-2}
\mathfrak{l}_p^\pm\big(\h\psi_{m-j,-}^\pm,\h\psi_{j-p,+}^\pm\big).
\end{align*}
Similarly to (\ref{5.34a}) one can prove that the first summand
in the right hand side of the last equality is zero. The sum of
the fourth and fifth summand is zero, what follows from the
equality (\ref{7.27}) and Lemma~\ref{lm2.8} with $p=m$, $s=1$,
$A_j=\mathfrak{l}_j^\pm$,
$B_{i,j}=\big(\h\psi_{i,+}^\pm,\h\psi_{j,-}^\pm\big)$.
Therefore,
\begin{equation*}
\mathfrak{a}_{m,-}^\pm+\sum\limits_{j=2}^{m-1} c_j^\pm
\widetilde{\mathfrak{a}}_{m-j,-}^\pm=
\sum\limits_{j=2}^{m-2}\mathfrak{u}_j^\pm
\mathfrak{a}_{m-j,+}^\pm+\sum\limits_{j=1}^{m-2}
\widetilde{\mathfrak{u}}_j^\pm
\widetilde{\mathfrak{a}}_{m-j,+}^\pm.
\end{equation*}
Employing this equality, (\ref{7.12}), (\ref{7.21}),
(\ref{7.22}), (\ref{7.27}) and applying Lemma~\ref{lm2.8} with
$p=m$, $s=1$, $A_0=A_1=0$, $A_j=\mathfrak{l}_j^\pm$, $j\geqslant
2$, $B_{i,j}=c_i^\pm\widetilde{\mathfrak{u}}_j^\pm $, we get
\begin{align*}
&\mathfrak{a}_{m,-}^\pm +\frac{2}{c_0^\pm}
\sum\limits_{j=2}^{m-1} c_j^\pm\tau_{m-j+1}^\pm =
\sum\limits_{j=2}^{m-2}\mathfrak{u}_j^\pm
\mathfrak{a}_{m-j,+}^\pm+\sum\limits_{j=1}^{m-2}
\widetilde{\mathfrak{u}}_j^\pm
\widetilde{\mathfrak{a}}_{m-j,+}^\pm-2\mu_{n,0}\sum\limits_{j=2}^{m-1}
\widetilde{\mathfrak{u}}_{m-j}^\pm c_j^\pm-
\\
&-\sum\limits_{j=2}^{m-3}\sum\limits_{p=2}^{m-j-1}
\mathfrak{l}_p^\pm c_j^\pm
 \widetilde{\mathfrak{u}}_{m-j-p}^\pm=
\sum\limits_{j=2}^{m-2}\mathfrak{u}_j^\pm
\mathfrak{a}_{m-j,+}^\pm+\sum\limits_{j=2}^{m-1}
\widetilde{\mathfrak{u}}_{m-j}^\pm
\big(\widetilde{\mathfrak{a}}_{j,+}^\pm-2\mu_{n,0}^\pm
c_j^\pm\big)-
\\
&-\sum\limits_{j=3}^{m-2}\sum\limits_{p=2}^{j-1}
\mathfrak{l}_p^\pm c_{m-j}^\pm
\widetilde{\mathfrak{u}}_{j-p}^\pm=
\sum\limits_{j=2}^{m-2}\mathfrak{u}_j^\pm
\mathfrak{a}_{m-j,+}^\pm
+\frac{2}{c_0^\pm}\sum\limits_{j=2}^{m-1}\sum\limits_{p=2}^{j-1}
\tau_p^\pm
\mathfrak{u}_{j-p+1}^\pm\widetilde{\mathfrak{u}}_{m-j}^\pm.
\end{align*}
Let us transform the last summand in the right hand side of the
equality obtained:
\begin{align*}
&\sum\limits_{j=2}^{m-1}\sum\limits_{p=2}^{j-1} \tau_p^\pm
\mathfrak{u}_{j-p+1}^\pm\widetilde{\mathfrak{u}}_{m-j}^\pm=
\sum\limits_{p=2}^{m-2}\sum\limits_{j=p+1}^{m-1} \tau_p^\pm
\mathfrak{u}_{j-p+1}^\pm\widetilde{\mathfrak{u}}_{m-j}^\pm=
\\
&=\sum\limits_{p=2}^{m-2}\sum\limits_{j=2}^{m-p} \tau_p^\pm
\mathfrak{u}_j^\pm \widetilde{\mathfrak{u}}_{m-j-p+1}^\pm
=
\sum\limits_{j=2}^{m-2}\mathfrak{u}_j^\pm\sum\limits_{p=2}^{m-j}
\tau_p^\pm \widetilde{\mathfrak{u}}_{m-j-p+1}^\pm,
\end{align*}
The equalities obtained and (\ref{7.20}) allow to continue the
calculations:
\begin{equation*}
\mathfrak{a}_{m,-}^\pm +\frac{2}{c_0^\pm}
\sum\limits_{j=2}^{m-1}
c_j^\pm\tau_{m-j+1}^\pm=\sum\limits_{j=2}^{m-2}
\mathfrak{u}_j^\pm
\mathfrak{a}_{m-j,+}^\pm+\sum\limits_{j=2}^{m-2}
\mathfrak{u}_j^\pm \sum\limits_{p=2}^{m-j}\tau_p^\pm
\widetilde{\mathfrak{u}}_{m-j-p+1}^\pm=\sum\limits_{j=2}^{m-2}
\mathfrak{u}_j^\pm \mathfrak{l}_{m-j}^\pm,
\end{equation*}
what by (\ref{7.26}) gives rise to the equality
$\mathfrak{f}_{m+1,1,-}=0$. Thus, the functions $f_{m+1,0}^\pm$
are compactly supported. By Lemma~\ref{lm7.1} it follows that
the equations (\ref{7.24}) are solvable in $\mathcal{V}$ and
their solutions are given by the formulas (\ref{7.16}), where
$c_{m+1}^\pm$ are some constants and the functions $u_{i,0}, \t
u_{i}\in C^\infty(\mathbb{R})\cap \mathcal{V}$ meet the
condition (\ref{7.8}). Substituting now the representations
(\ref{7.14})-(\ref{7.16}) for the functions $\psi_{i}^\pm$,
$i\leqslant m+1$, into the equation (\ref{7.3}) with $i=m$, we
get that the function $\psi_{m+2}^\pm$ is a solution of the
equation
\begin{equation*}
-\left(\frac{\partial^2}{\partial\xi^2}+\pi^2
n^2\right)\psi_{m+2}^\pm= G_{m+2}^\pm,\quad (x,\xi)\in\mathbb{
R}^2,
\end{equation*}
where the right hand side is given by the formula (\ref{7.17}).
Due to the form of the functions $G_{m+2}^\pm$, the equalities
(\ref{7.19}) for $i\leqslant m+1$ and (\ref{2.30}) with $N=1$,
and Lemma~\ref{lm2.5}, similarly to (\ref{4.33})--(\ref{4.35})
it is not difficult to show that the representation
(\ref{7.14}), (\ref{7.15}) for the function $\psi_{m+2}^\pm$ is
valid, where $u_{m+2,j}^\pm$, $j=0,1$ are some functions and the
equalities (\ref{7.19}) with $i=m+2$ take place.
\end{proof}

Let us calculate the numbers $\tau_3^\pm$ and $\tau_4^\pm$. From
the formulas (\ref{7.21}), (\ref{7.27}) it follows that
\begin{equation}\label{7.29a}
\tau_3^\pm=0.
\end{equation}
Suppose that $\tau_2^\pm=0$, i.e., $\int\limits_\mathbb{R}
V(x)\di x=0$. In this case $\widetilde{\mathfrak{u}}_1^\pm=0$.
In virtue of the assumption made from (\ref{7.21}) we deduce:
$\tau_4^\pm=c_0^\pm\left(
\widetilde{\mathfrak{a}}_{3,-}^\pm-2\mu_{n,0}^\pm
\widetilde{\mathfrak{u}}_3^\pm\right)/2$. The equation
(\ref{7.29}) for $\h\psi_{3,-}^\pm$ has a zero right hand side
in this case, this is why $\h\psi_{3,-}^\pm=0$. Therefore,
$\widetilde{\mathfrak{a}}_{3,-}^\pm=0$ and
\begin{equation}\label{7.30}
\tau_4^\pm=-\mu_{n,0}^\pm c_0^\pm
\widetilde{\mathfrak{u}}_3^\pm.
\end{equation}
From (\ref{7.11}) and (\ref{7.20}) it follows that
$\l_2^\pm=\mathfrak{a}_{2,+}^\pm$,
$\h\psi_{2,+}^\pm=\t\psi_2^\pm$, what due to (\ref{5.11a}),
(\ref{7.4}), (\ref{7.18}) leads us to
$f_{3,1}^\pm=-c_0^\pm\left(\frac{\displaystyle
d^2}{\displaystyle dx^2}-V\right)u_{2,0}^\pm$. The last equality
and (\ref{7.9a}), (\ref{7.10}), (\ref{7.12a}), (\ref{7.12})
imply:
\begin{align*}
\widetilde{\mathfrak{u}}_3^\pm&=\int\limits_\mathbb{R}
f_{3,1}^\pm\di x=c_0^\pm\int\limits_\mathbb{R} V u_{2,0}^\pm \di
x=2\int\limits_\mathbb{R} u_{2,0}^\pm\frac{d\t u_1^\pm}{dx}=
-2\int\limits_\mathbb{R} \t u_1^\pm\frac{d u_{2,0}^\pm}{dx}\di
x=
\\
&=-c_0^\pm\int\limits_\mathbb{R} \t u_1^\pm
\left(\frac{d^2}{dx^2}-V+2\mu_{n,0}^\pm\right)\t u_1^\pm\di
x=
\\
&= c_0^\pm\left(\int\limits_\mathbb{R} \left(\frac{d\t
u_1^\pm }{dx}\right)^2\di x-2\mu_{n,0}^\pm\int\limits_\mathbb{R}
\left(\t u_1^\pm\right)^2 \di x\right)+
\\
&\hphantom{=}+2\int\limits_\mathbb{R}\left(\t u_1^\pm\right)^2
\frac{d\t u_1^\pm}{dx}\di x=c_0^\pm\left(\int\limits_\mathbb{R}
\left(\frac{d\t u_1^\pm }{dx}\right)^2\di
x-2\mu_{n,0}^\pm\int\limits_\mathbb{R} \left(\t u_1^\pm\right)^2
\di x\right),
\end{align*}
what together with (\ref{7.10}), (\ref{7.30}) yield
\begin{equation}\label{7.30a}
\tau_4^\pm=-\frac{2\mu_{n,0}^\pm}{\pi^4 n^4}\left(
2\int\limits_\mathbb{R} V^2(x)\di
x-\mu_{n,0}^\pm\int\limits_\mathbb{R} \left(
\int\limits_\mathbb{R} \sgn(x-t) V(t)\di t\right)^2 \di x
\right),
\end{equation}
if $\int\limits_\mathbb{R} V(x)\di x=0$.

Let $m\geqslant 2$. We denote
\begin{gather*}
\tau_{\e,m}^\pm:=\sum\limits_{i=2}^m\e^i\tau_i^\pm,\quad
\l_{\e,m}^\pm:=\frac{\pi^2n^2}{\e^2}+\sum\limits_{i=0}^m\e^i\l_i^\pm,
\\
\psi_{\e,m}^\pm(x):=h(x,\tau_{\e,m}^\pm)\left(\phi_{n,0}^\pm
\left(\frac{x}{\e}\right)+
\sum\limits_{i=1}^m\e^i\psi_i^\pm\left(x,\frac{x}{\e}\right)\right).
\end{gather*}
From (\ref{7.9})--(\ref{7.13}) and Lemma~\ref{lm7.2} it follows
\begin{lemma}\label{lm7.3}
As $\e\to0$ the function $\psi_{\e,m}^\pm\in C^2(\mathbb{R})$
satisfies the equality
\begin{equation*}
\left\|\psi_{\e,m}^\pm(x)-
\phi_{n,0}^\pm\left(\frac{x}{\e}\right)\right\|_{L_2(Q)}=\Odr(\e)
\end{equation*}
for each $Q\in \mathfrak{C}$. The functions $\psi_{\e,m}^\pm$
and $\l_{\e,m}^\pm$ obey the equation (\ref{5.50}) with
$\psi_{\e,m}=\psi_{\e,m}^\pm$, $\l_{\e,m}=\l_{\e,m}^\pm$,
$f_{\e,m}=f_{\e,m}^\pm$, where the functions $f_{\e,m}^\pm$ meet
the estimates (\ref{5.51}).
\end{lemma}

Suppose that there exist the numbers $\mathfrak{n}_\pm$ such
that $\tau_i^\pm=0$, $i\leqslant \mathfrak{n}_\pm-1$,
$\tau_{\mathfrak{n}_\pm}\not=0$. Observe, it follows from the
formulas (\ref{7.12}), (\ref{7.30a}) that at least one of the
numbers $\tau_2^-$, $\tau_4^-$ is nonzero, this is why
$\mathfrak{n}_-\leqslant 4$. We take $m\geqslant
4\mathfrak{n}-2$, where $\mathfrak{n}=\mathfrak{n}_-$ or
$\mathfrak{n}=\mathfrak{n}_+$. We set
$k_{\e,m}^\pm:=\sqrt{\pm\big(\mu_n^\pm(\e^2)-\l_{\e,m}^\pm\big)}$.
The branch of the root in this definition is chosen as follows.
If the radicand is positive, we take $k_{\e,m}^\pm$ of the same
sign as $\tau_{\mathfrak{n}_\pm}$. If the radicand is negative,
we choose $k_{\e,m}^\pm$ from the condition the imaginary part
of a number being positive. We denote
$\tau_m^\pm(\e):=\e^{-1}\left(\ln\k_\pm(\e,k_{\e,m}^\pm)-
\ln((-1)^n)\right)$.

\begin{lemma}\label{lm7.4}
For each $m$ the relations
\begin{equation}\label{7.31}
\tau_m^\pm(\e)=\tau_{\e,m}^\pm+\Odr(\e^{m-2\mathfrak{n}_\pm-1}),
\quad \l_{\e,m}^+\leqslant \mu_{n}^+(\e^2),\quad
\l_{\e,m}^-\geqslant \mu_{n}^-(\e^2).
\end{equation}
hold. The equalities
\begin{equation}\label{7.32}
\l_{2j}^\pm=\mu_{n,j}^\pm,\quad \l_{2j+1}^\pm=0,\quad j\leqslant
\mathfrak{n}_\pm-2,\quad
\l_{2\mathfrak{n}_\pm-2}^\pm=\mu_{n,\mathfrak{n}_\pm-1}^\pm\mp
\frac{2\pi^2n^2\left(\tau_{\mathfrak{n}_\pm}^\pm\right)^2}
{\mu_{n,0}}
\end{equation}
are valid.
\end{lemma}

\begin{proof}
In view of (\ref{6.7}) we choose the functions $\Th_\pm^\pm$ in
(\ref{6.6}) as follows:
\begin{equation}\label{6.8}
\Th_\pm^+(\xi)=\t\Th_\pm^1(\xi,\k_\pm^{-1}), \quad
\Th_\pm^-(\xi)=\t\Th_\pm^1(\xi,\k_\pm),
\end{equation}
where the function, we remind, is from (\ref{6.9a}). These
functions are linear independent, since their wronskian is of
the form:
\begin{equation}\label{6.9}
W_\xi\left(\Th_\pm^+(\xi),\Th_\pm^-(\xi)\right)
=\left(\k_\pm-\k_\pm^{-1}\right)\Th_{\pm,2}(1)
\end{equation}
and due to (\ref{6.1a}), (\ref{6.4}) it is nonzero for all
sufficiently small $\e$ and $k\not=0$. We denote
$\t\Th^+_\pm(\xi):=\E^{-\xi\ln((-1)^n)}
\Th_{per,\pm}^+(\xi,\e,k_{\e,m}^\pm)$,
$\t\Th^-_\pm(\xi):=\E^{\xi\ln((-1)^n)}
\Th_{per,\pm}^-(\xi,\e,k_{\e,m}^\pm)$. The functions
$\t\Th^+_\pm(\xi)$ and $\t\Th^-_\pm(\xi)$ are 1-periodic for
even $n$ and 1-antiperiodic for odd $n$, what follows from the
periodicity of the functions $\Th_{per,\pm}^\pm$. By
(\ref{6.1a}) we deduce that the equalities
\begin{equation*}
\t\Th_\pm^+(\xi)=\phi_{n,0}^+(\xi)+\Odr(\e^2),\quad
\t\Th_\pm^-(\xi)=\phi_{n,0}^-(\xi)+\Odr(\e^2)
\end{equation*}
hold true in the norm of $C^2[0,1]$. In virtue of
Lemma~\ref{lm7.2} for $x\geqslant x_0$ the function
$\psi_{\e,m}$ is of the form
\begin{equation*}%\l%abel{7.32a}
\psi_{\e,m}^\pm(x)=\E^{-\tau_{\e,m}^\pm
x}\psi_{\e,m}^{per,\pm}\left(\frac{x}{\e}\right),\quad
\psi_{\e,m}^{per,\pm}(\xi)=\phi_{n,0}^\pm(\xi)+\e\t
\psi_{\e,m}^{per,\pm}(\xi),
\end{equation*}
where the function $\t \psi_{\e,m}^{per,\pm}(\xi)$ is 1-periodic
for even $n$ and 1-antiperiodic for odd $n$. Using the
properties of the functions $\psi_{\e,m}$ and $\t\Th_\pm^\pm$
listed above, Lemma~\ref{lm7.3} and choosing the function
$\mathcal{U}$ as
\begin{equation*}
\mathcal{U}(x,\e)
:=W_x\left(\t\Th^+_{\pm}\left(\frac{x}{\e}\right),
\psi_{\e,m}^{per,\pm}\left(\frac{x}{\e}\right)\right)+
\left(\tau_m^\pm(\e)-\tau_{\e,m}^\pm\right)
\t\Th^+_{\pm}\left(\frac{x}{\e}\right)
\psi_{\e,m}^{per,\pm}\left(\frac{x}{\e}\right)
\end{equation*}
instead of (\ref{5.103}), completely by analogy with the
deducing of the equation (\ref{5.66a}) in the proof of
Lemma~\ref{lm5.4} it is not difficult to show that the equality
(\ref{5.66a}) is valid with $\tau_{\e,m}=\tau_{\e,m}^\pm$,
$\tau_m(\e)=\tau_m^\pm(\e)$. From (\ref{1.4}), (\ref{1.6}) and
the formulas (\ref{1.18}), (\ref{1.21}) for $\l_0^\pm$,
$\l_1^\pm$ we obtain that $k_{\e,m}^\pm\to0$, $\e\to0$. By
(\ref{6.4}) it follows that
\begin{equation*}
\tau_m^\pm(\e)=\frac{\e k_{\e,m}^\pm\sqrt{\mu_{n,0}}}
{\sqrt{2}\pi n}\left(1+\Odr(\e)\right).
\end{equation*}
Substituting of the equality obtained into (\ref{5.66a}) gives
rise to
\begin{equation*}
\e^2\frac{\mu_{n,0}\left(k_{\e,m}^\pm\right)^2}{2\pi^2 n^2}
\left(1+\Odr(\e)\right)=\left(\tau_{\e,m}^\pm\right)^2+
\Odr(\e^{m-1}).
\end{equation*}
Setting now $m=2\mathfrak{n}_\pm+2$, we get:
\begin{equation}\label{7.32b}
\left(k_{\e,2\mathfrak{n}_\pm+2}\right)^2=\e^{2\mathfrak{n}^\pm-2}
\frac{2\pi^2 n^2\left(\tau_{\mathfrak{n}_\pm}^\pm\right)^2}
{\mu_{n,0}}+\Odr(\e^{2\mathfrak{n}_\pm-1})
\end{equation}
In view of the definition of $k_{\e,m}^\pm$ from the last
equality we get
\begin{equation*}
\l_{\e,2\mathfrak{n}_\pm-2}^\pm=\mu_{n,0}^\pm(\e^2)\mp
\e^{2\mathfrak{n}_\pm-2}\frac{2\pi^2 n^2
\left(\tau_{\mathfrak{n}_\pm}^\pm\right)^2}{\mu_{n,0}}+
\Odr(\e^{2\mathfrak{n}_\pm-1}),
\end{equation*}
what together with (\ref{1.4}) imply the formulas (\ref{7.32})
and the inequalities (\ref{7.31}) for $\l_{\e,m}^\pm$. These
inequalities and (\ref{6.4}) yield that $k_{\e,m}^\pm$ is real,
and $\tau_{m}^\pm(\e)$ and $\tau_{\e,m}^\pm$ are of the same
sign. The equality (\ref{5.66a}) for $\tau_m^\pm(\e)$ and
$\tau_{\e,m}^\pm$ lead us to the equality (\ref{7.31}) for
$\tau_m(\e)$.
\end{proof}

From the proven lemma and (\ref{7.13}) the formulas
(\ref{1.18}), (\ref{1.21}) follow.

\begin{lemma}\label{lm7.5}
There exists a function $R_{\e,m}^\pm(x)\in C^2(\mathbb{R})$
such that the function
$\h\psi_{\e,m}^\pm:=\psi_{\e,m}^\pm(x)-R_{\e,m}^\pm(x)$
satisfies the equation (\ref{3.16}) with $\l=\l_{\e,m}^\pm$,
$f(x)=\h f_{\e,m}^\pm(x)$, $\supp \h
f_{\e,m}^\pm\subseteq[-x_0,x_0]$, and the constraints
(\ref{6.17}) with $k=k_{\e,m}^\pm$, and the estimates
\begin{equation*}%\l%abel{7.33}
\|R_{\e,m}^\pm\|_{C^2(\overline{Q})}=
\Odr(\e^{m-4\mathfrak{n}_\pm+3}), \quad \|\h
f_{\e,m}^\pm\|_{C[-x_0,x_0]}=\Odr(\e^{m-4\mathfrak{n}_\pm+3})
\end{equation*}
hold true for each interval $Q\in \mathfrak{C}$.
\end{lemma}

The proof of the lemma is carried out completely by analogy with
the proof of Lemma~\ref{lm5.5}. The functions $\mathcal{U}$ and
$\Th_\pm^\pm$ should be defined as in the proof of
Lemma~\ref{lm7.4}. Instead of the estimate (\ref{5.58}) one
should make use of the following inequality
\begin{equation*}
\left|W_x\left(\Th^+_\pm\left(\frac{x}{\e}\right),
\Th^-_\pm\left(\frac{x}{\e}\right)\right)\right|\geqslant
C\e^3\left(k_{\e,m}^\pm\right)^3\geqslant
C\e^{3\mathfrak{n}_\pm},
\end{equation*}
where the positive constant $C$ is independent on $\e$. This
inequality is a consequence of the relations (\ref{6.1a}),
(\ref{6.8}), (\ref{6.9}).

We denote $g_{\e,m}^\pm:=\h f_{\e,m}^\pm-V\h\psi_{\e,m}^\pm$.
From Lemmas~\ref{lm7.3},~\ref{lm7.5} we deduce that
\begin{equation}\label{7.35a}
\left\|g_{\e,m}^\pm(x)-V(x)\phi_{n,0}^\pm
\left(\frac{x}{\e}\right)\right\|_{L_2(-x_0,x_0)}=\Odr(\e).
\end{equation}
According to Lemma~\ref{lm7.5}, the functions
$\h\psi_{\e,m}^\pm$ is a solution of the problem (\ref{3.16}),
(\ref{6.17}) with $\l=\l_{\e,m}^\pm$, $k=k_{\e,m}^\pm$, $f=\h
f_{\e,m}^\pm$, this is why by Lemma~\ref{lm6.7},~\ref{lm6.5}
$g_{\e,m}^\pm=(\I+\e V T_{11}^\pm(\e,k_{\e,m}^\pm))^{-1}\h
f_{\e,m}^\pm$ and the representation
\begin{equation}\label{7.43}
g_{\e,m}^\pm=\frac{T_{18}^\pm(\e,k_{\e,m}^\pm)\h f_{\e,m}^\pm}
{k_{\e,m}^\pm-k_{\e,\pm}}g_{\e,\pm}+T_{19}^\pm(\e,k_{\e,m}^\pm)
\h f_{\e,m}^\pm
\end{equation}
holds true. By analogy with the way the estimate (\ref{5.73})
was obtained from (\ref{5.71}), from the last equality on the
base of (\ref{6.30c}), (\ref{7.35a}) and
Lemmas~\ref{lm6.5},~\ref{lm7.5} one can easily deduce that
\begin{equation}\label{7.44}
k_\e^\pm-k_{\e,m}^\pm=\Odr(\e^{m-4\mathfrak{n}_\pm+4}).
\end{equation}
Due to (\ref{6.3a}), (\ref{6.4}) and (\ref{7.32b}) it follows
that
\begin{equation}\label{7.44a}
k_{\e,\pm}=\e^{\mathfrak{n}_\pm-1} \frac{\sqrt{2}\pi
n\tau_{\mathfrak{n}_\pm}^\pm}
{\sqrt{\mu_{n,0}}}+\Odr(\e^{\mathfrak{n}_\pm}).
\end{equation}
In view of Lemma~\ref{lm6.6} now we conclude that the operator
$H_\e$ has an eigenvalue $\l_{\e,\pm}$ meeting the equality
(\ref{1.30}) with $\mu=\mu_n^\pm$, if and only if
$\tau_{\mathfrak{n}_\pm}^\pm>0$. As it follows from
(\ref{7.44}), in this case the asymptotics of the eigenvalue
$\l_{\e,\pm}$ coincides  with the corresponding series from
(\ref{1.17}), (\ref{1.20}) with the coefficients defined in this
section. From formulas (\ref{1.6}), (\ref{7.10}) we deduce that
in the case $\int\limits_\mathbb{R} V \di x>0$ the inequality
$\tau_2^->0$ holds. If $\int\limits_\mathbb{R} V \di x=0$, then
$\tau_2^-=0$, $\tau_3^-=0$, and according to (\ref{7.30a})
$\tau_4^->0$. Therefore, the criterion for the existence of the
eigenvalue $\l_{\e,-}$ is the inequality $\int\limits_\mathbb{R}
V \di x\geqslant 0$. The proof of item~\ref{it1th1.3b} of
Theorem~\ref{th1.3b} and Theorems~\ref{th1.4},~\ref{th1.5} are
complete.

Suppose the eigenvalue $\l_{\e,\pm}$ exist. Let us find out the
asymptotics expansions of the associated eigenfunction
$\psi_{\e,\pm}$.

Similarly to the proof of (\ref{4.16}), from  (\ref{6.30a}),
(\ref{6.30c}), (\ref{7.35a}), and
Lemmas~\ref{lm6.5},~\ref{lm7.3},~\ref{lm7.5} one can easily
deduce the existence of the function $c_\pm(\e)$ satisfying the
equalities:
\begin{equation*}%\l%abel{7.41}
c_\pm(\e)=\frac{\e T_{18}^\pm(\e,k_{\e,m}^\pm)\h f_{\e,m}^\pm}
{k_{\e,m}^\pm-k_{\e,\pm}}+\Odr(\e^{m-4\mathfrak{n}_\pm+3}),\quad
c_\pm(\e)=1+o(1),\quad \e\to0,
\end{equation*}
for all $m$. By (\ref{6.30a}), (\ref{6.30c}), (\ref{7.43}),
Lemma~\ref{lm7.5} and uniform boundedness of the operator
$T_{20}^\pm$ it implies the equality:
\begin{equation}\label{7.45}
\e g_{\e,m}^\pm= c_\pm(\e)g_{\e,\pm}+
\Odr(\e^{m-4\mathfrak{n}_\pm+3}),
\end{equation}
which is valid in the norm of $L_2(-x_0,x_0)$. By (\ref{6.16a})
the function $\h\psi_{\e,m}^\pm$ can be expressed by
$g_{\e,m}^\pm$ as follows: $\h\psi_{\e.m}^\pm=\e
T_{11}^\pm(\e,k_{\e,m}^\pm)g_{\e,m}^\pm$. From (\ref{6.4}),
(\ref{6.21}) we obtain that for each $Q\in \mathfrak{C}$ the
operator $T_{11}^\pm(\e,k): L_2(-x_0,x_0)\to \H^2(Q)$ obeys the
uniform on $\e$ and $k$ estimates:
\begin{equation*}
\|T_{11}^\pm\|\leqslant C\e^{-2}k,\quad
\left\|\frac{d}{dk}T_{11}^\pm\right\|\leqslant C\e^{-2}k^{-2}.
\end{equation*}
Taking into account these estimates, (\ref{7.44}),
Lemma~\ref{lm7.5} and applying the operator
$T_{11}^\pm(\e,k_{\e,m}^\pm)$ to (\ref{7.45}), we get:
\begin{align*}
\h\psi_{\e,m}^\pm&=c_\pm(\e)\psi_{\e,\pm}+c_\pm(\e)
\left(T_{11}^\pm(\e,k_{\e,m}^\pm) -
T_{11}^\pm(\e,k_{\e,\pm})\right)g_{\e,\pm}+
\Odr(\e^{m-5\mathfrak{n}_\pm+2})=
\\
&=c_\pm(\e)\psi_{\e,\pm}+\Odr(\e^{m-6\mathfrak{n}_\pm+4}),
\end{align*}
where the equality holds in the norm of $\H^2(Q)$ for each $Q\in
\mathfrak{C}$. Thus, we have proved
\begin{theorem}\label{th7.1}
Suppose the eigenvalue $\l_{\e,\pm}$ exists. The associated
eigenfunction $\psi_{\e,\pm}$ can be chosen so that it satisfies
the asymptotics expansion  (\ref{7.1}) in the norm of $\H^2(Q)$
for each $Q\in \mathfrak{C}$. The coefficients of this expansion
are determined according to (\ref{7.9}), (\ref{7.10}),
(\ref{7.11}), (\ref{7.12}), and Lemma~\ref{lm7.2}. For $\pm
x\geqslant x_0$ this eigenfunction meet the equalities
(\ref{6.17}), and the multipliers $\k_\pm$ and $\k_\pm^{-1}$
corresponding to the functions $\Th_\pm^+$ and $\Th_\pm^-$ obey
the equality $\k_\pm=(-1)^n\E^{-\tau_\e^\pm}$, where the
function $\tau_\e^\pm$ has the asymptotics (\ref{7.2}) with the
coefficients determined by the formulas (\ref{7.10}),
(\ref{7.30a}) and Lemma~\ref{lm7.2}.
\end{theorem}

\sectn{Acknowledgments}

The author thanks R.R.~Gadyl'shin for attention to the work and
valuable remarks.

The work is partially supported by RFBR (05-01-97912-r\_agidel)
and by the programs ''Leading scientific schools''
(NSh-1446.2003.1) and ''Universities of Russia'' (UR.04.01.484).

\renewcommand{\refname}
{

\end{document}